\documentclass{article}
\usepackage[italian,english]{babel} 
\usepackage{enumerate}
\usepackage{graphicx}
\usepackage{latexsym}
\usepackage{url}
\usepackage{natbib}
\usepackage[ansinew]{inputenc} 
\usepackage[OT2,T1]{fontenc}

\usepackage{listings}
\usepackage{tabularx}
\usepackage{booktabs}
\usepackage{multirow}
\usepackage[para]{threeparttable}

\let\origquotation\quotation
\let\endorigquotation\endquotation
\renewenvironment{quotation}{%
  \vspace{-0.5\parskip}
  \origquotation
  \footnotesize
}%
{\endorigquotation}

\title{Design Ltd.: Renovated Myths for the Development of Socially Embedded Technologies\footnote{This manuscript is the extended draft of the Chapter entitled ``Building Socially Embedded Technologies: Implications on Design.'' that has been submitted for review and publication in \emph{Designing Socially Embedded Technologies: A European Challenge}, a book edited by David Randall, Kjeld Schmidt, and Volker Wulf, forthcoming.
}}
\author{Federico Cabitza and Carla Simone\\Universit\'{a} degli Studi di Milano-Bicocca\\Viale Sarca 336, 20126 Milano, Italy}
\date{November 2012}

\begin{document}

\maketitle

\section{Motivations and Background(s)}

\label{sec:intro}
As often reported in the specialist literature in the last thirty years or so, and even recently~\citep[e.g., ][]{lyytinen_learning_1999,shapiro_participatory_2005,pan_information_2008,warkentin_analysis_2009} with a tinge of disgruntled resignation, approximately half to two-thirds (if not more) Information Systems (IS) projects fail. Of course, there is little comfort in that this seems largely due to organizational and social, rather than technical factors~\citep{pan_information_2008,kaplan_health_2009}. Indeed, this strikes even more in light of the almost universal recognition that the practice of information systems development has undergone in this lapse of time a radical transformation and has abandoned naive strictly structured life cycle methods of development toward more flexible, dynamic and multidisciplinary approaches; if this is true it is probably also because some principles and sensibilities typical within the HCI, CSCW and PD fields have so to say ``trickled down'' in the ``consciousness'' of IT practitioners in the ``real'' world~\citep[cf. e.g.,~][]{shapiro_participatory_2005,fitzpatrick_review_2012}).

Among the papers that try to go beyond both the typical optimism of the Titanic designers'\footnote{To this regard, we would like to notice that one of those designers, Edward Wilding, embarked at the Titanic Quarter, the place where the famous liner had been built in Belfast, and disembarked at Southampton 28 hours later, before the liner actually embarked its 3,000 passengers and began its maiden voyage; while another one, Thomas Andrews, was among those passengers and went down with the Titanic trying to heroically rescue as many passengers as he could (and probably refusing to flee to safety himself). These two stories, once they have been moved to a purely symbolic level, in terms of Wildingism and Andrewism, can be taken as archetypes of different attitudes of designers in relation to ``their'' systems: who abandons the system soon after the ``sea trials''; and who succumbs in the vain attempt to save its users.} and the fatalistic attitudes of the technological Cassandras, we like to mention a relatively underrated one, where~\citet{harris_better_1999} make the point that ``while our hardware technology has improved by orders of magnitude, and our software has grown comparably more complex, the relationship between people (individually or in groups) and computers has only improved incrementally. In some cases, \emph{it has even deteriorated}'' (p. 88, our emphasis). They suggestively address the reason why it is so difficult ``to translate [research insights] into comparable improvements in the usability (and more generally, the social integration) of computers'' by advocating the adoption of a ``better mythology for System Design'' in alternative to the standard mythology. This latter encompasses a set of ``myths''\footnote{Here and in the following, the word myth is not opposed to any truth fact, but it is rather used as synonym of ``archetypical story'' to indicate one possible \emph{stance}, among many other as much as legitimate and reasonable ones. On the other hand, we keep using the term mythology for its powerful and evocative connotation, although probably the most indicated term would be ``metanarrative'', in the sense after~\citet{lyotard_postmodern_1986}.}, i.e., stories that emphasize particular aspects of the practice of IT development and all together do not drive practice in any strong sense; rather ``shape it by helping each participant construct and frame their account of their practice''. The ``standard mythology''~\citet{harris_better_1999} outline is rooted in the assumptions that:

\begin{quotation}
\begin{itemize}
\item The parts of the system must interact according to a pre-established harmony defined during its design.
\item The job of a designer is to discover, clarify, and when necessary invent the rules that define that harmony, and then embed them into the computer system.
\item The users must interact with the system in terms of the language or ontology that these rules create.
\end{itemize}
\end{quotation}

These are in a nutshell the main assumptions underlying the traditional ``mythology of professional design''. This mythology sustains the very idea of a conceptual process that is carried out by \emph{experts} (in designing) with the participation of \emph{experts} (in their own practices) in order to \emph{represent} and \emph{direct} the unfolding of the production of computer-based information systems in an orderly manner in face of \emph{Chaos}~\citep{vv._aa._extreme_2001}. The main merit of this kind of contributions is to let some given-for-granted assumptions surface once again and be object of further review and discussion. In fact, only when ``the limited and inaccurate perspective on work and technology imposed by the standard myths of both organization and system design [have been recognized], we can start to search
for more effective approaches and write better myths around them''.

In this chapter we also will try to submit a sort of \emph{paralogy}~\citep{lyotard_postmodern_1986}, or little contrarian mythology, in which to write about other myths that could help us tackle the wicked problem of system design~\citep[p. 4]{fitzpatrick_locales_2003}. Although we also agree with the tenets~\citet{harris_better_1999} proposed within their ``mythology for the long term'' almost fifteen years ago\footnote{Although the interested reader can refer to the original paper, we here summarize the main high-level recommendations contained in the mythology proposed by~\citet{harris_better_1999}: that we should i) honor every particularity, even those that do not fit the regularities imposed by the organizational rules; ii) honor accomodation, i.e., the ``ad hoc elaboration of rules in use'' and iii) honor change, which is a intrinsic and unavoidable feature of real world system.}, in our little contrarian mythology we will go a step further (right in virtue of the time passed in the meanwhile) by arguing around the idea that the very conception of design that we are all well used to (and many of us also are fond of) 
should be challenged, and indeed considered one of the most decisive factors leading to manifest failure\footnote{We will certainly not try to prove this assumption, as we could never get over the causality fallacy that such a prove would entail (\emph{post hoc, propter hoc}). A similar argument, far from being a mere provocation, was also put forward by~\citet{bryant_its_2000}.}. 

If this hypothesis is true, there can be many reasons to account for it, which we will not investigate as not related with the chapter's aims. Our conjecture is that this idea of design is too disconnected from practice, although it heavily relies on even complicated representations of this latter, and it is nourished as the central element of a reductionistic framework where the process of system \emph{development} --~a word that until the industrial 19th century denoted a continuous unfolding of things till their maturity~-- is rationally \emph{phased down} in sub-components that are ontologically distinct, and where responsibilities are assigned on the basis of nominal competencies somehow reified in terms of specialized roles. We are aware that a rational and engineering approach is not bad \emph{per se}, but we submit that it reflects a conceptualization of ``what complex is'' and ``what to cope with complexity means'' that, as we will see in Section~\ref{sec:complexity} can tap in false assumptions and bring to ungrounded expectations\footnote{Moreover, Ivan Illich was among the first thinkers to denote a similar phenomenon as ``principle of (paradoxically) counterproductivity'': once most practices are institutionalized and engineered, they \emph{backfire} on some of the stakeholders.}. 

Thus, however imprudent this hypothesis may seem, we will take it seriously in order to submit the idea that such a conception (and the related professional activity) is \emph{not} really necessary to build any successful computational, material artifact \emph{with which users have to interact to have their work done}, and to propose an alternative approach that could do without formal or conceptual design and indeed any distinction between design and use (especially when this distinction come along with a supremacy of the former).  

As we discussed also in~\citep{cabitza_remain_2011}, we limit ourselves to contesting the necessity and primacy of conceptual design in the development of computational interactive applications and information systems to be embedded in a social cooperative setting. With design we then denote the specific phase of the larger process in which professional analysts meet some (or many) user representatives and/or their managers to draw more or less formal models of how work is and should be accomplished, produce detailed specifications of the needs of the various stakeholders involved, and of how the computational system will support work to fulfil needs and expectations. Thus, although we take this term in a quite broadly meaning that encompasses business analysis, requirement elicitation, conceptual modelling, process (re)design, specification formalization and analysis, in what follows we will use the general term ``design'' for brevity's sake. Other authors have cautioned about the fictional and ritualistic nature of this activity~\citep[e.g. ][]{robey_rituals_1984,robinson_questioning_1991,nandhakumar_fiction_1999}. We contest the myths in which this ritual is considered necessary~\citep[as also maintained within the CSCW, e.g.~][]{shipman_formality_1999}, given for granted and therefore substantially unquestionable~\citep[see also~][]{blackwell_abstract_2008}. Conversely, we will argue in favor of alternative myths according to which all the layers pertaining to human-computer interaction in the broadest sense, --~what is often denoted as the \emph{trimurti} of the Model, the Control and the View~-- can be realized by composition of elementary components without any rational design, and be put to work by end users alone, eventually (but not necessarily) flanked by IT professionals that are explicitly called to play the role of catalysts of a ``reaction'' that is substantially under no control, since it pertains to the dynamics of complex (socio-technical) systems.

In this lies the reason why we also speak of a mythology, as we are aware that also our argumentation encompasses some myths, the most notable of which is that of the ``end user that can develop her own artifacts'' \emph{somehow}. Moreover, this new ``mythology'' we are after is not ``original'' in any strong sense; rather it can be brought into focus where three different but complementary recent discourses meet, and seen as the outcome of a bricolage made of some indications and suggestions coming from the three strands we will outline in the following sections. This presentation of ideas is done towards the foundation of an alternative way to build socially embedded systems. This chapter regards this alternative way.

The rest of chapter will be articulated as follows: in Section~\ref{sec:nodesign}, we will address the given-for-grantedness and indisputability of conceptual design in system development by recalling its roots in recent history. In the next three sections (i.e., Section~\ref{sec:complexity},~\ref{sec:performance} and ~\ref{sec:bricolage}) we will briefly gather suggestions from three distinct discourses on which to ground our different mythology: complexity thinking (in Section~\ref{sec:complexity}) will invite us to distrust any formal model of social practices for their intrinsic opacity with respect to emerging phenomena and their unexpected evolution, and distrust design-based methods for system development for their role in conceiving the requirement of flexibility mainly (if not totally) in terms of exception handling~\citep{cabitza_computational_2013}.
Performativity thinking (see Section~\ref{sec:performance}) will help us reappraise the value for IT development of the karstic river that connects many influential thinkers from Nietzsche to Suchman, and will provide us the conceptual space to think system development differently from a design-driven process. Lastly, the metaphor of the \emph{bricoleur} (see Section~\ref{sec:bricolage}), i.e., who performs the activity of bricolage, will suggest us a new strategy for building computer-based support \emph{in the wild}. This pathway will lead us toward an alternative proposal, that we will articulate in Section~\ref{sec:environments} and discuss to some extent in  Section~\ref{sec:remarks}; Section~\ref{sec:conclusions} will end the contribution with a quick look at a research agenda coherent with this alternative mythology for IT system development.

\section{For a genealogy of the idea of Design}
\label{sec:nodesign}
\begin{quotation}
Good design is good business.\\
(Thomas Watson, Jr., 1971\footnote{Excerpt from a talk given at the Wharton School of Business.})
\end{quotation}
An alternative mythology of system development can be ``constructed'' only downstream of a process of deconstruction of the design-based development; this process would be aimed at identifying the ideas that in such a mythology are taken from granted; recognizing if any of these ideas are in a vibrant relationship with other often opposite ones (hence the idea of opposition); trying to understand where those ideas come from and how those oppositions got consolidated over time and are currently used in the social construction of meaning and values.~\citet{beath_contradictory_1994} began following this kind of approach, after~\citet{derrida_positions_1981} to unravel some of the incompatible assumptions about the role of designers and users during development. We advocate that further similar initiatives could be undertaken, for instance to make sense of often-cited oppositions like those between performative vs. ostensive modelling~\citep{poltrock_modeling_2009}, plan vs. planning~\citep{suchman_human-machine_2006}, development vs. growing~\citep{truex_growing_1999}, global vs. local~\citep{rolland_balancing_2002} and the two related design vs.~use and designer vs.~user~\citep{bowers_janus_1991} ones. We believe these deconstructive analyses could be useful, for instance, to understand if there is something in the rethorics of the proposals that stem from the communities that are closer to the EUSSET forum (i.e., HCI, CSCW, PD, etc.) that undermines their potential to really mitigate the risk of failure in IT projects, as they would not aim to ``overturn'' the above mentioned oppositions, nor merge and surpass them, but rather to recognize them being in a continuous interplay~\citep{derrida_positions_1981} that both asserts their irreversible \emph{difference}, and their necessity to produce sense in our fields. 

For brevity's sake our contribution will pretend to come after that such an undertaking has been attempted to shed light on the surreptitious hierarchy that acts in the design-use opposition, and that we could try to find some new way to look at this dyad. In particular, we will pretend that such a deconstruction has shown that ``the design of a thing'' governs (i.e., affects, conditions) its use, and that such an activity, which regards the creation of new things, is considered sort of intellectually superior, or just deserving more interest, than the latter, i.e. the mere and opportunistic use of what has been produced to this aim\footnote{We are aware that this point could be as easily as harshly objected with respect to IT system design, at least within the fields that are closer to the EUSSET community. But if things were \emph{actually} different, why the undeniable user-related orientations of approaches like the user-centered \emph{design}, interaction \emph{design}, contextual \emph{design}, participatory/cooperative \emph{design}, seem only to affect ``design'' in terms of attributive specifications without ever challenging the common concept of design? Is this linguistic phenomenon one of the ``semantic pathologies'' or ``frame conflicts'' (cf. Reddy) that also recently have been mentioned in CSCW when~\citet{dugdale_trouble_2012} recognized their potential for subtly harmful bias in IT-related discourses? In that case we could probably trace back this phenomenon to a specific case of the so called ``Sapir-Whorf hypothesis'', i.e., the conjecture according to which linguistic categories and usage do influence thought and attitudes in complex ways. Obviously we leave the questions mentioned above open to discussion by all means.}. 

Granted this preliminary insight for the sake of argument, in what follows we will tell a story towards a possible ``genealogy'' 
of the tension between design and use, in which we will recall that this dyad is not a timeless construct, but rather it emerged out of contingent turns of history, not as the outcome of rationally inevitable trends or part of any grand scheme of progressive history, but rather at the point in history where complex, mundane, i.e., social, cultural and economic, and not necessarily glorious conditions got entangled\footnote{Here suffice to say that ancient engineers involved in the building of impressive civil facilities, stately buildings and reliable ships, did not feel the need to have a term indicating the \emph{full specification} --~i.e., the \emph{de-sign} (attested since the 16th century)~-- of a \emph{future accomplishment} --~i.e., the \emph{project} (attested since the 15th century). No specific term indicating the concept of design can be found in the whole multi-volume work De Architectura ("On Architecture"), authored by Vitruvius, a Roman architect and engineer and one of the most theoretical writers of his times, who rather used the terms `species dispositionis', `descriptio' or `compositio' (made by `cogitatio' and `inventio') with different meanings. This is not a merely nominalistic point: what modern architect would not consider `design' as part and parcel of Architecture as a general topic? Conversely, Vitruvius wrote: ``Partes ipsius architecturae sunt tres : aedificatio, gnomonice, machinatio''; that is: architecture consists of three components: construction, scheduling and machine deployment. Here we can see how what the modern stance sees as the realization of the representation of an intellectual and creative achievement is dis-articulated in a connotation of design that relates it more to the enablement and enactment of the coordinative tasks that different practitioners involved in the construction articulate; something that resonates with the ethnographic work of architectural practices described by~\citet{schmidt_ordering_2004}.}.

\subsection{A little tale of design-as-we-know-it}
\label{subsec:story}

This story begins in the 1950s\footnote{The following, very short and partial indeed, account is based on the informative and insightful historiographical studies done by~\citet{campbell-kelly_computer:_2004} and, more recently, by~\citet{haigh_software_2002,haigh_crisis_2010,haigh_inventing_2011}. We also consulted the videos published by IBM Corporation for its 100th anniversary on its page on Youtube (www.youtube.com/user/IBM).} when a bold idea of design, which was purporting itself as solution for a very wide class of problems, met a very specific demand for an effective solution that was fed by the emergence of the so retrospectively called ``software crisis''\citep{haigh_crisis_2010}. This crisis can be related to two related phenomena: first, private companies that were customers of mainframe-based information systems were continually running out of memory of their storage systems: they continuously needed to process more and more data over time and for this reason they were slowly but inexorably moving from punched card- and sequential access tape-based storage systems to much more efficient random access storage systems; this caused these companies to shift from having small teams of in house  programmers writing very specific code for their particular needs (usually in Assembly) to buying new and more powerful hardware \& software bundles; in this move hardware systems were typically very little compatible with preexisting legacy systems, while software was getting increasingly more and more difficult to write (and usually in Cobol) and maintain for the corresponding low level intricacies that the nascent relational model was introducing for the more efficient management of random access storage. Second, mainframe vendors were experiencing as many difficulties, in coping with an ever increasing complexity of their storage and processing systems and of the closely related commercial offers, in face of at least three main factors: customer intrinsic heterogeneity, market competition, and what epitomizes both, i.e., continuous change\footnote{We recall that one chapter of the influential book that Brooks wrote in 1975 as a summa of the lessons learned in the sixties (cf. ``The mythical man-month: essays on software'') was aptly entitled ``\emph{Plan the system} for change'' (our emphasis).}. 

The particular idea of design that could be met in these hard times for the information storing and processing industry was spontaneously emerging from the second half of the 1950s (cf. the foundation of the SHARE association\footnote{http://www.share.org}, still existing) and got a first historical legitimization within the cultural milieu where researchers gathered at the first two UK conferences on design methods in 1962 and 1965.  As reconstructed by~\citet[Appendix 1][]{love_social_2007}, in those intellectual ambits, design was asserted as ``the use of scientific principles, technical information and imagination in the definition of a mechanical structure, machine or system to perform pre-specified functions with the maximum economy and efficiency''~\citep{eder_definitions_1966}, and as the activity whose ``effect is to initiate change in man made things''~\citep{jones_design_1970}; these are all definitions that resonate with the irresistible ontology elaborated by~\citet{simon_sciences_1981}, in which an ``artificial world'' is given to the Promethean attitude of system designers to be recreated and managed by these latter mainly through their rational activity of problem solving and scientific planning.

When such an idea met the elusive (and then just nascent) world of software abstract architectures (abstract with respect to all possible situated implementations) and abstract data models (abstract with respect to any situated nuance and ambiguity that business concepts could still preserve)--~during the great endeavour of IBM to build the full compatible system, the System/360~-- conceptual design became \emph{both} (part of) the solution to cope with the increasing complexity~\citep{noble_america_1979}, \emph{and} part of what could be delivered (and sold) to customers all together with the ``big iron''~\citep{haigh_software_2002}\footnote{It is also noteworthy that the so called ``hardware'', in those same years, was looking more and more like cold ``black boxes'', due to the coeval microelectronics revolution that was irreversibly substituting the large amount of hand-crafted wiring and ``tangible'' logic of a computer with new and inscrutable integrated circuits.}. As reported by~\citet{haigh_inventing_2011}, the conviction that ``to be effective, information systems must be designed --~engineered if you prefer'' (p. 27) justified the transformation from ``a tightly knit family of `computer people' into a diverse and highly fragmented collection of programmer/coders, systems analysts, and information technology specialists''~\citep{ensmenger_black_2001}, the concurrent ``elevation of systems men [or designers] over both accountants and data-processing technicians''~\citep{haigh_inventing_2011} and the shift from the art of programming~\citep{ensmenger_black_2001} to a set of methodological devices, whose gist (irrespective of the extent the flow of activities is pipelined, looped, spiralled, and the like) is ``rational'' (or better yet, reductionistic) \emph{phasing}, i.e., the idea that there \emph{are} distinct phases, and the implied ideas of either predictive planning and development parcelization that phase thinking
made possible\footnote{This parcelization is nowadays so obvious that offshore outsourcing has become a practice that is conceptually legitimate even before than technologically feasible.}. 

In virtue of this ``new'' (or just renovated) labor force, as soon as the need emerged to really \emph{embed} a computational system within an organization (and not only put it in one of its facilities as mere punched card and magnetic tape equipment), design and analysis left the realm of the \emph{builders of hardware} to be projected in the realm of the commercial and consulting services~\citep{haigh_inventing_2011} that were oriented to isolate the customer's needs, understand how to satisfy them\footnote{As Brooks noticed ``a programmer delivers satisfaction of a user need rather than any tangible product''. Yet Cosgrove, already in 1971, recognized that ``both the actual need and the user's perception of [its] needs will change as programs are built, tested, and used''.}, and understand how to deploy more and more sophisticated applications in their settings to integrate those systems with their organizational procedures\footnote{To this respect, we deem the quotation mentioned at the beginning of this section by the President of IBM relevant and indicative of those commercial strategies, beyond the conventional (and perhaps not so true) connotation of that sentence by which good design ``naturally'' leads to good systems, and these latter to good sales.}. 

It is while analysis and design became part and parcel of a deliverable commodity that had to be attuned to the strategic goals of the customer~\citep{haigh_software_2002}\footnote{We recall the definition of software given by Bauer in 1965: ``\emph{systems analysis and design}, programming and computer-based services, accomplished by users, computer manufacturers and others.'' (our emphasis), or the definition given by Head in 1968 ``\emph{the entire process of systems analysis and design}, programming, testing and implementation, as well as the documentation that accompanies this process.'' (our emphasis). All citations taken from~\citep{haigh_software_2002}.}, that these activities took the prominently representational and visual flavour they have still today\footnote{From~\citep{haigh_crisis_2010}: ``the systems men of the 1950s had also paid attention to flowcharting and analysis, having invented many of the charting techniques modified by the new generation of software engineering gurus''.}, since, in so doing, they could also more faithfully \emph{mirror} the conceptual ideas (cf. the Vitruvian \emph{cogitatio}) they reified. As both a commodity task and a social ritual~\citep{robey_rituals_1984}, design consolidated itself as the phase in which to generate representations that could \emph{duly and precisely specify the system} (that is the root meaning of the word \emph{de-sign}) and shed light (that is the root meaning of \emph{project}) on the future activities of factual realization of the system, while at the same time also ``displaying'' the necessary expertise needed in both these activities.
In such a ritual, visual languages, specification formalisms, both diagrammatic and textual representations that characterize current design have since then played a role in creating and maintaining the divide between ``who can read'' and ``who can't'', between the hired pundit and the ordinary user~\citep{illich_disabling_1977}. In so doing, the divide between these two roles came to its current ratification; the practice of design became a sort of play~\citep{nandhakumar_fiction_1999} where ostentation of expertise both justified and legitimized the economic exchange going on between the pundit/supplier and the customer, i.e., between who designs and who buys the outcome of design, also in light of the fact that this explicit and representationally paraded expertise was often ``bounced off'', so that the programs of digitization could act as symbols of an organization's rationality and reliability, both for the inside and the outside. 


From this very succinct account, the wrongest conclusion to draw would be that before the 1960s people did not design computational systems, even very complicated ones. Yet, we argue that before that particular period of time, when the word ``software'' did not even exist~\citep{haigh_software_2002}, the concept of design was a semantic bipole, so to say, vibrating in a state of balance: on the one hand, one would detect a representational pole, from which design is seen as the conception of ``clear and distinct'' representations of what a system (i.e., an orderly something made of parts) \emph{is} (i.e., its structure) and of what it \emph{does} (i.e., its behavior); on the other hand, a performative stance, from where design is related to a collaborative process of construction where material representations are used for the sake of reaching a mutual understanding and stipulating a tangible agreement of what is to be built and how. Every time some material arrangement must be done, design will also come with both these flavours\footnote{Indeed, if one is to look carefully at the world of machines, things would look to be much more performative than representational, even if we tend to think that a representation of intended action, i.e., the algorithm within a structured procedural program, really informs and affects its execution (the machine's behavior). Indeed, we here recall that it was in fact the wind and the water that made the mill within the Benedictine monastery \emph{to mill} (cf. the marvellous account by Mumford contained in The Myth of the Machine, 1967), not the plan that described the articulation of its parts (if any), much in the same way that it is electrical power that makes electronic artifacts act (if we agree that no such a thing like a ghost in the machine exists), in virtue of an orderly articulation of a pattern of solids and voids, of present and absent magnetic fields, of ``positive and negative spaces'' (like the cogs and slots in the cogwheels of a mill machine, like the holes of a paper punched card) that in history have progressively become \emph{softer and softer}.}. Yet when 
software architectures and high-level behaviors of a computational system are at stake, representations and their generative power~\citep{bowers_politics_1992} get the upper hand. For instance, it is true that a UML Collaboration Diagram makes an idea of interaction between (more or less abstract) components \emph{explicit}, and hence something that could be collaboratively discussed both within and across communities of practitioners~\citep[e.g., ][]{mcleod_documents_2010}: this is its descriptive role; but such a diagram is also intended to formally \emph{specify} the (abstract) structure and behavior of a computational machine whose components are supposed to interact as \emph{design}-ated: in being a formalism, such a design object is able to generate other ``representations through the operation of rules over some vocabulary''~\citep{bowers_politics_1992} and to prescribe things. When this generative power is trans-lated again in the arena of human collaboration, a series of small and sometimes imperceptible (but not for this reason more harmless) ontological drifts described by~\citet{robinson_questioning_1991} occur, and among those drifts the most dangerous one occurs when the prescriptive power of models, their being \emph{ostensive} in the denotation of~\citet{poltrock_modeling_2009}, is confused with their descriptive power, i.e., their \emph{performative} nature~\citep{poltrock_modeling_2009}. Since this tension has received a lot of consideration within the CSCW community, it is not necessary to linger over it any longer. 

In summary, the idea of design we have outlined above got crescent legitimization and strength within an \emph{engineering} response to the software crisis that broke software development into phases denoted with terms such as `analysis', `design',
`implementation', `testing', and so forth, and that also brought forth the idea that such activities could be performed in a circumscribed and orderly way (not to mention the idea that these could be controlled in a strict sequential fashion, as it happens in other ``hard'' engineering fields, like civil and aeronautics engineering).
Till nowadays, all current software development methodologies are imbued with the idea of \emph{method}, that is of orderly ``path'' between the ``idea'' of something and its realization, i.e., the separation and distinction (that is ontological, temporal and causal) between the ``design'' of a thing and the designed thing; between the plan of doing something (and its representation, of course) and the ``real'' thing. Differently from other concepts borrowed from ``hard engineering'' that were almost immediately denoted as naive and inadequate with regard to software (e.g., the notorious waterfall model), the idea of phasing, hence the opposition design-use, has become pervasive and it has been molded over time up to its current versions, i.e., the iterative, participatory and evolutionary methods that characterize, among others, the agile approaches (e.g., Scrum, XP programming). The idea (and practice!) of phasing brought in the dichotomy between design and subsequent phases; the idea of design, in its turn, brought in (with authority) the legitimacy of the practice of modeling a software artifact. This contributed in digging an increasingly deeper rift between the category of user and the one of the designers, who have differentiated themselves from laymen also in virtue of their jargon, representational formalisms and modeling technicalities.

\subsection{Sketches from a different future}
\label{subsec:story}

Could things have gone different from how they did? Does reconstructing how the current idea of design got consolidated over time tell us something in regard to whether a different idea of it could ever stand out in the future? Andrew Pickering, in a series of interesting articles~\citep{pickering_science_2004,pickering_beyond_2008} and even a book~\citep{pickering_cybernetic_2010}, has thoroughly focused on how some early British scientists and engineers (who were happy to call themselves ``cyberneticians''~~\citep[p. 223]{pickering_cybernetic_2010}), and notably Gordon Pask and Stafford Beer, were addressing the same goal of IBM of improving the Western organization and decision making of its management in the same years but, instead of leveraging representational and conceptual means, by developing a radically alternative approach, through the first proposals and experimentations in biological computing and cybernetic feedback theories. Their approach, according to Pickering, was dispensed entirely with representation and devoted to the concrete experimenting of ``productive and performative relations via open-ended and intimate engagements'': Pask and Beer, in the years in which Cobol was introducing the seminal concept of input and output as a first class statement, claimed that managers were not actually interested in getting an answer in symbolic and digital format (that is in outputs), but rather in \emph{using} the answer, that is in making sense of the state of affairs in any form this could come to their attention\footnote{Just to make this point more clear, we recall that their first experiments to this regard regarded the responses of \emph{pond ecosystems} to environmental stimuli as alternative information processing systems than electronic mainframes!}.

This research program was not pursued for an intrinsic eccentricity of isolated researchers\footnote{One could object that the cybernetic, embodied, biological approach of the early cyberneticians that Pickering speaks about just did not stand the test of time, while the representational stance has brought to us incredibly fast and reliable logistics management as well as pervasive systems for the provision of goods and services worldwide. As a matter of fact, no one can really guess how the viable system model developed by Beer would have changed the Western firm (or how) with its biological and systemic metaphors and holistic interventions. Little help to this respect can come from post-mortem analyses of the Cybersyn Project, which tried to build a comprehensive decision support system for the governance of Chile economy during the government of President Salvador Allende (1971-1973), as any interpretation of the most controversial aspects of that socio-economic experiment could be easily made an instrument of political propaganda. Rather, we refer again to the contributions mentioned at the beginning of Section~\ref{sec:intro} arguing about the alleged successes of modern IT production methodologies, as well of the actual impact of these systems on general productivity and affluence~\citep[see also][]{carr_does_2004}).}, but rather as consequence of the profound realization that ``in a world of exceedingly complex systems, for which any representation can only be provisional, performance is what we need to care about. The important thing is that the firm adapts to its ever-changing environment, not that we find the right representation of either entity''~\citep[p. 235]{pickering_cybernetic_2010}. It is the realization that ``the world is populated by a multiplicity of interacting exceedingly complex systems'' that urges us to be fascinated by alternative mythologies other than the representational, symbolic one sponsored by IBM exactly fifty years ago.

\section{It's a complex (new) world}
\begin{quotation}
The captain may, by simply moving an electrical switch, instantly close the doors throughout and make the vessel practically unsinkable\footnote{Shipbuilder, Special Edition, 1911}. [Olympic and Titanic] are designed to be unsinkable\footnote{White Star promotional flyer, 1912}.
\end{quotation}
\label{sec:complexity}
In the first decades of the 21st century many disciplines in the human, social and organizational studies have discovered complexity. In finding (or acknowledging) their realm of interest being \emph{complex}, in some cases, authors from those traditions have promoted a debate in their communities about the very foundations of their practices and methods~\citep[see e.g.,~][]{greenhalgh_response_2010}; more often, elements or suggestions from complexity theory are included in more or less traditional frameworks to reinforce their ramparts against the gales of unpredictability~\citep[e.g.,~][]{norman_engineering_2004,pavard_contribution_2006}; in these sallies in the world of complexity, authors usually proceed by recalling the definition of complex system that they deem more informative for their aims and pick up some of the properties that such systems exhibit to concentrate on how those could impact on their fields of research: in this jumble of definitions and claims, the casual reader could get confused, also for the fact that complexity seems to be the carrefour where hard science and softer ones, and within these latter where pragmatist/culturalist and rationalist/functionalist approaches, all meet and blend together creatively if not promiscuously~\citep{kaghan_out_2001,kim_interpreting_2006}\footnote{For this reason, trying to make even a brief account for the vast plethora of contributions that introduce complexity-related in IT system design would be out of scope.}. 

More specifically, in the IT system development discourse the usual line of argumentation is: the social and the technical components constituting an organizational `overvall' system are highly inter-dependent (this leading to the concept of socio-technical systems by the the Tavistock School); these systems encompass many components, of different types, which are linked in various ways, whose patterns can change unexpectedly and often~\citep{schneberger_complexity_2003}; \emph{hence} they are complex systems, \emph{hence} ``our knowledge and understanding of how different components work and interact, and accordingly how the system as a whole work, will be incomplete''~\citep[p. 5]{hanseth_risk_2007}, \emph{hence} ``the more complex a system is, the more incomplete our knowledge will be, and the more unintended effects our interventions will produce.''. We substantially subscribe this line of thought, which is founded on the catchy idea that ``where there are strong interactions among elements''~\citep{axelrod_harnessing_1999}, there you find a complex system (made of those elements, of course); yet we also think a word of caution is necessary. 

We recognize that the simplest way to see complexity is to take it as an \emph{explanatory concept}, that is to acknowledge that it is invoked or referred to ``to proffer a form of explanation''~\citep{paley_complexity_2011} for a situation, a pattern of behavior that it is thought to derive (the right term here would be ``emerge'') in non-linear and almost unpredictable ways from a multitude of interacting agents. Yet, for complexity to \emph{explain} something, one has also to assume that the agents involved follow, or can be interpreted as following, relatively simple rules\footnote{This point deserves to continue quoting~\citet{paley_complexity_2011}: ``however, a complex system is an abstraction, and so are [its generative] `rules' of behaviour. In all non-computer cases, the `rule' as a semantic expression is an inference, partly inductive and partly hypothetico-deductive. Inductively, it is inferred from the inveterate behaviour of individuals; at the same time, it functions as a hypothesis tested by its ability, in combination with the other rules in the set, to produce or predict the global patterns of the complex system''}. If we retain this as the main principle underlying complex systems, and how the ``theory'' conceive of them, then we should also agree with those authors who claim that most of the recent contributions coming from the human studies (including the organizational domain) often end up by producing references to complexity theory that are not dissimilar from ``mere retellings of old tales, [which use] complexity terminology tacked on retrospectively, gratuitously and, in many cases, quite awkwardly''~\citep{maguire_complexity_1999}; more precisely,~\citet{paley_complex_2007} makes the point that many researches that declare a focus in complex systems do actually refer to the open systems thinking, which between the 1960s and the 80s was aimed at replacing the Tayloristic organisation-as-machine metaphor with the metaphor of organisation-as-organism. 

For this reason, we prefer to take a pragmatic stance and refrain from relying too much on complexity as an explanatory term; rather in what follows we adopt it as a descriptive term, to account for post-hoc analyses. As even Simon acknowledged in 1968: ``the main route to the development and improvement of time-sharing systems was to build them and see how they behaved''~\citep[p. 20]{simon_sciences_1981}. These systems turned out to be exceedingly complex systems, \emph{that is} systems that, to be understood, ``had to be constructed, and their behavior observed'' (ibidem). Therefore we like to start the line of argumentation mentioned above \emph{from the end}, that is from the realization that our knowledge of how a complex (socio-technical) system will behave is necessarily incomplete (if not wrong to some extent) and that ``unintended consequences'' (either positive or negative) are \emph{unavoidable and certain} irrespective of the unremitting diligentness we put in analysing and planning how a specific system (encompassing people and things) ``behaves''. Indeed, an increasing number of researchers, in the line of that strand of research that focuses on unintended consequences~\citep[e.g.,~][]{merton_unanticipated_1936,tenner_why_1997,ash_unintended_2004} have recently come to speaking of ``law of unintended consequences''~\citep[e.g., ][]{mansfield_nature_2010} and agree that the only way to genuinely eliminate risks of (socio-technical) complex systems is to follow the recommendation ``do not build them!''\citep[p. 10]{hanseth_risk_2007}\footnote{This resonates with both the half-serious epigram of the ``Murphy's Law'' and with the full-serious theory of Normal Accident by Charles Perrow.}. We continuously observe that any claim to know how a technology-in-practice will impact on the setting in which it is embedded, that is, to know in advance that a system that was specifically designed for enabling a set of functions, supporting a set of tasks and fulfilling a set of needs will actually reach the goals established at design time, is frustrated by evidence and history~\citep{jones_patterns_1996,rochlin_trapped_1998}. 

In complex systems change must be expected: this has been an inductively supported common place in system design since its foundation, and we can related it to the vast research undergone about exception handling in software engineering so far; yet the effect of change must be also recognized as intrinsically unpredictable, and we believe that this insight comes to be less recognized than it should. Thus,~\citet[p. 25]{mansfield_nature_2010} argues that `if the majority of computer-based socio-technical systems fail to meet the expectations of their sponsors, perhaps that is due to their architecture''; consequently he submits design-oriented principles that call for the employment of small components that are mutually loosely coupled, so as to avoid that the break down of one of those could propagate to the others in complex ways (cf. the butterfly effect); and the adoption of a layered architecture where some layers are allowed to change (and adapt to changes, or evolve in response to them) at different rates to account for the varying rate of impact of the events affecting them. 

In addition to these recommendations, we perceive the undertaking of designing ``to meet specified expectations'' to be a ``wicked problem''\citep{fitzpatrick_locales_2003}, i.e., something for which there is no definitive formulation, no stopping rule, and such that any attempt to treat it as it were not a wicked problem would only worsen things\footnote{To this respect, then, rather than speaking of complex systems, and their fuzzy definition, we like to mention the taxonomy proposed by Ackoff, according to which there are three types of system: messes, problems and puzzles. A mess is a complex arrangement that defies attempts to define it precisely; a problem, conversely, does have a specific structure that can be specified at some level of detail, but not necessarily any single, clear-cut solution; the puzzle is a well-defined and well-structured problem with a specific solution. Accounts from the shop floor and the field of work show that ``when embarking on the creation of a large socio-technical system, many sponsors believe they have a problem, when they actually have a mess, while engineers believe they have a puzzle when, if they are lucky, they actually have a problem''~\citep[p. 118]{mansfield_nature_2010}. In the same vein, Pidd concluded that ``the greatest mistake that can be made when dealing with a mess is to carve off part of the mess, treat it as a problem and then solve it as a puzzle~--~ignoring its links with other aspects of the mess.'' (ibidem).}. For this reason, we submit that the simpler recipe to mitigate the opportunity of failure is to change \emph{both} the architecture and the \emph{method}, or better yet, the attitude: therefore, to abandon any method that deals with the future, that \emph{projects} an idea of future in the realization of the tool that is supposed to reach it, and embrace the possibility that the components of the socio-technical system would adapt and co-evolve in ways that no designer would imagine, not necessarily for the worse. On a practical side, this would mean to ``follow the actors''\citep{bannon_human_1992,latour_interobjectivity_1996}, but also to turn to their traditional artifacts, not only as sources of inspiration but also as ground upon which to build new tools~\citep{cabitza_remain_2011} and, above all, to let actors support themselves. 

Thus, the discourse on complexity turned out to become (or unveil) a discourse on unpredictability (cf. the iceberg struck by the Titanic); and this, in its turn, leads us to the seemingly dullest of the design principles, the one expressed in the French phrase \emph{laissez-faire}, literally ``let [them] do''. Although this recommendation could look shallow, in the next three sections we will see what ``letting do'' means (and why it is important), who must be let do (i.e., end users), and how system development should be oriented towards the building of an environment to build systems that is specifically \emph{designed} to allow \emph{for doing without design} and to allow for unexpected things to just happen.

\section{The rediscovery of performativity}
\label{sec:performance}
\begin{quotation}
Many things difficult to design prove easy to performance.\\
(Samuel Johnson, 1759\footnote{Line pronounced by Imlac within the apologue `Rasselas'.})
\end{quotation}
%

In this section we will first consider what we mean when we advocate that the alternative mythology for IT system development we are envisioning should make a ``performative turn''. This expression is getting more and more importance in fields that have historically had always an influence on design-oriented discourse. Then, we will consider how the performative discourse has already come into design-related mythologies, in order to highlight the specific strand we aim to renovate for our proposal.

\subsection{The performative turn} 

The expression ``performative turn'' is usually used to indicate two strictly related things that we nevertheless prefer to distinguish for clarity's sake. On the one hand, it indicates a historically circumscribed research program, which has received an increasing interest in the last fifteen years by researchers involved in cultural and social studies, like the Science and Technology Studies field (notably Pickering, Latour) and related disciplines like ethnology, anthropology, sociology and linguistics; in this former case, the term ``turn'' indicates the aim of this research endeavour to investigate an \emph{alternative} way to look at how people interact, work and share knowledge in social settings with respect to more mainstream strands like the pragmatic and realist paradigms, endorsing the claim that people create and recreate meaning and knowledge in social settings \emph{through performance}~\citep{van_house_collocated_2009}, and that even social reality itself is ``created'' while people ``do things''. As~\citet{law_this_2003} put it down:

\begin{quotation}
The differences between realism and pragmatism are important, but neither share the performative
assumption that reality is brought into being in the process of knowing. Or, to
put it more precisely, neither would assume that the object that is known and the subject
that does the knowing are co-produced in the same performance, that the epistemological
problem (what is true) and the ontological question (what is) are both resolved (or not) in
the same moment.
\end{quotation}

This alternative approach shared the critique towards systemic, fully-specified and rationally-conceived abstractions (e.g., with the non-representational theorists\footnote{It should be noted though that ``a performative perspective does not delete the idea of representation, but rather views it as a specific aspect of performativity''~\citep{jensen_experiment_2005}, in that it focuses on the activity of representing, planning, modeling rather than on the material outcome of those practices.}) and drew on the metaphor\footnote{Here and elsewhere we use the term metaphor in the Nietzschean sense, as something that is used to impose order and intelligibility on a world that we cannot access directly.} of ``performance'' to reflect ``a growing discontent with the traditional social sciences and their understanding of practices as texts or representations of genuinely symbolic concepts'', to express ``the reversion from systems of representations to processes of practice and performance'', and to focus on ``the active social construction of reality rather than its representation''~\citep{dirksmeier_time_2008}. 

In an attempt to summarize the recent, and quite protean, discourse about performativity in two lines, after~\citet{bramming_imperfect_2012} we highlight three intertwined aspects: i) reality is understood as incessant creation or practice; ii) matter itself, is understood as ``entangled intra-relation''; and, of course, iii) individuals do not pre-exist their interactions in any essentialist, objectivistic sense.


Yet this recent research line can be said to be grounded in a more \emph{general} idea of performativity. It is this second connotation of the expression ``performative turn'' that we refer to in our proposal. This latter, rather than a specific research program, can be better characterized as a sort of ``sensitivity to specificities of materially heterogeneous events with special reference to differences and relations between performances''~\citep{jensen_performing_2002}. This sensitivity has followed in the last century or so a peculiar karstic trend: it has recurred a number of times by different authors of different cultural milieus and, although each time it was capable to gain a strong interest, this was never sufficient to establish itself as the mainstream thought in any of those milieus, and somehow submerged until a next thinker contributed in its reappraisal. 

In this sense, therefore, the idea of a ``performative turn'' evokes a more a-historical attitude, which was exhibited by individuals that have deliberatively turned away their focus from the allures of representationalism to embrace a more action-oriented and embodied perspective. The term ``turn'' thus indicates the will to reverse the ontological premises that the world is populated with particular objects, entities, configurations that exist in and of themselves and that are endowed with particular essential qualities~\cite[p. 67]{jensen_performing_2002} to consider objects, ``not singular entities, but rather textures of partially coherent and partially co-ordinated performances'' existing through multiple situated practices.

This sensitivity, or will, or discontent with representational/conceptual tenets indicates a sort of ``fil rouge'' that binds together thinkers like Nietzsche, Heidegger, Derrida, Pickering, Latour\footnote{It is typical of fil rouge to binds together unsuspected associates, like Pickering and Latour, for instance. One thing that unites these thinkers, e.g., is that they are both ``happy enough'' to speak of material agency in nature without imputing any intentionality to the word ``agency''~\citep[p. 6]{pickering_mangle_1995}.} and some relevant feminist theorists~\citep{bath_searching_2009} like Judith Butler, Karen Barad and, especially for her involvement in the IT debate, Lucy Suchman (just to mention a few of those authors that influenced our understanding of the performative approach). A common trait among these thinkers seems then the need to find a viable alternative to representationalistic tenets (i.e., stances that could be called as Cartesian, Kantian or simply `modern'\footnote{Yet, we agree with~\citet{jensen_performing_2002} when he points out that ``the performative turn is a way to refuse the choice between the modern and the post-modern. The modern is about order and purity. The post modern is a celebration of fragments and disorder. The performative turn is a series of claims and sensitivities that try to reach a fractional space in between. Something that is beyond the mono-dimensionality of modernity and beyond the free-floating multi-dimensionality of the post-modern. In this sense it has much in common with the parts of the ANT-tradition that claim to be non-modern.''} in some philosophy circles\citep[cf., e.g., ][]{rorty_objectivity_1991}), to shift the focus from questions of correspondence between models/representations and reality, to matters of practices/doings/actions~\citep{barad_posthumanist_2003}.  

In short, a performative approach asks us as observers of social settings to abandon the idea that these are sets of ``object that are'', to embrace the idea they are made of ``events that do''. In doing so, it gives us a ``resource to counter the positivist stance which essentializes categories and naturalizes the qualities of the entities whose stable existence it posits'' (e.g., gender as a fixed attribute of a person)~\citep{licoppe_performative_2010}. The concept of performativity therefore invites us to abandon the Kantian notion of ``thing per se'' (at least in system design) to recognize the relational and manifold nature of any perceived phenomenon, irrespective of its seeming solidity\footnote{Just to back the legitimacy of the turn requested to see action and events where mainstream design sees things, we here recall that our (not too far) ancestors in their language (i.e., Latin, Greek and Old English) used the words `res', `pragma' and `thing' (respectively) in order to denote both an affair, a deed, a business, or assembly~\citep[p. 1]{telier_design_2011}, \emph{as well as} the matters that were discussed and deliberated in such occasions and meetings. In other words, in our past (probably before having become modern - cf. Latour) we could not see as the subject, the act and its motivation or `cause' (cf. `cosa' in Italian) being untwined and disentangled with the material matters involved in such occasions.}, as well as  the \emph{co-constitutive entanglement} of the social and the technological (i.e. material), and ``the performance of the emergent sociomaterial assemblage''~\citep{orlikowski_sociomaterial_2007}. According to this perspective, ``meaning'' is thus seen as an emergent phenomenon (or an epiphenomenon) of interaction~\citep{hug_performativity_2010} but, even beyond this point, as a transient aspect of \emph{embodied interaction}~\citep{dourish_where_2001} that cannot be really decoupled from \emph{situated action}~\citep{suchman_human-machine_2006} nor caught in abstract terms. 

In this vein, researchers adopting a performative turn put first in their research agenda the study of the contingencies of time, space, technology, materiality or discourse, ``the heterogeneous sociomateriality and real-time contingency of performance'', as~\citet{suchman_human-machine_2006} calls them (p. xii), all things that the more classical ``representational'' model of thinking that is typical of ``20th century technoscience''~\citep{suchman_figuring_2004}, i.e., the one assuming a detached observer that studies real objects and their essential properties in an objective world (or that designs and puts new objects into the world), escapes either consciously or unaware with profound consequences also on the conception of the role of technology in society, and of its ``designers''~\citep{orlikowski_sociomaterial_2007}.

\subsection{The performativity fil rouge}

In order to frame how the concept of performativity can influence IT system design in practical ways, we have to briefly outline the fil rouge mentioned above, which binds together influential thinkers of the last 150 years with the foundations of the CSCW approach to system design. To this aim we have first to make a clear distinction between the discourse on performativity we are interested in, and the so called ``performance studies''. These latter are usually at stake where scholars and researchers in the IT literature use expressions like ``designing for performativity''~\citep{wagner_whisperings_2010}, ``the role of performance in design
research''~\citep{jacucci_manifesto_2005} or ``performing design''. These expressions are more related to the traditional meaning of performance~\citep{dirksmeier_time_2008}, as ``showing of a doing'' (cf. Grimes) or ``activity before a particular set of observers'' (cf. Goffman)\footnote{It is nevertheless worthy of note that the meaning of performance as ``performing a play'', ``playing a drama'' is much later than the more general meaning of ``carrying out a promise'', or ``carrying in effect something'' that is from the 16th century approximately.}, and they point all more or less to the ``artistic'' side of the discourse on performativity and as such they tend to ``preserve'', if not enhancing, the creative role of designers instead of contributing in the overturn of the necessity of the idea of design.

Conversely, the concept of performativity we refer to is rooted in the Nietzsche's seminally deconstructive analysis of the relation between words and the world, and in his powerful intuition according to which looking for a specific ``doer'' behind any action is recognized as an arbitrary and unnecessary (and indeed confounding) act\footnote{We are obviously referring to the famous passage in ``The Genealogy of Morals'' where Nietzsche pointed out that  ``there is no `being' behind doing, acting, becoming: `the doer' is merely a fiction added to the `doing'. Doing is all'' (original: es giebt kein `Sein' hinter dem Thun, Wirken, Werden; `der Thaeter' ist zum Thun bloss hinzugedichtet, - das Thun ist Alles).}. This seminal contribution was then retaken by phenomenologist scholars, especially Heidegger, who contributed in the same mould by articulating further the idea that the only way of being of humans (i.e., Dasein) is \textit{engagement in practices} (Existenz)~\citep{riemer_what_2012}, that these latter depend on \emph{equipment}\footnote{Equipment can be seen to denote those things, or artifacts, that the Dasein encounters in fluent use, entangled and experienced in performance, when they are ready-to-hand (Zuhandenheit).} for their performance, and that the relationship between this latter and Dasein is fundamentally co-constitutive~\citep{turner_affordance_2005}. Many affinities can be then found between Heidegger and J. L. Austin~\citep[see e.g., those discussed in ][]{glendinning_being_1998}, the language philosopher who introduced the concept of \emph{performative utterance} to account for the capacity of human speech to act, i.e. have an effect in the material world, rather than just simply describe reality in terms of `true' and `false' statements. In this strand, Law somehow questioned the orderly taxonomy proposed by Austin and claimed that ``all statements are in the slippery space between performative and constative'', thus turning "the question of constative vs. performative [\ldots] into an empirical question, and thus potentially an object for a sociology of performances''\citep{jensen_performing_2002}.

Years later, approximately at the same time when these concepts were taken up in the IT design arena by~\citet{winograd_understanding_1986} in their reappraisal of the Austin's (and Searle's) elaboration of the so called ``speech acts'' , the performative ``fil rouge'' unfolded again in the works of Andrew Pickering, especially in those contributions where he made a clear dinstiction between a ``representational idiom'' and a ``performative idiom'' in scientific and technology-oriented discourses\citep{pickering_mangle_1995}, and in particular for our design-related discourse, when~\citet{pickering_beyond_2008} contrasts the modern technoscientific approach to the design of things with the approach genuinely followed by British cyberneticians, like Beer, Ashby and Pask, i.e., a hand-on experimental, performative and non representational one. At the same time, other authors drew upon the critical reinterpretation by Derrida~\citep{simon_knowing_2010} of the Austin's original differentiation between performatives and constatives, most notably Judith Butler and Karen Barad. These latter elaborated a complex concept of (posthumanist) performativity around the repetitive, or \emph{citational}, aspects of performance, i.e., its ability to produce materiality. In this view, social structures, like rules and categories (such as gender) are not pre-existent attributes of a given object or its behavior, but rather they are continuously produced through processes of repetition and social legitimization. This conviction echoes, but also in some way goes beyond, the views animated by Wittgenstein philosophy that recognize how, due to the intrinsic underspecification of human behaviors~\citep{schmidt_dispelling_2011}, it is the practice that determines the rule rather than the opposite, and that invite to abandon an ``objectified and detached view of rules and procedures as external objects with fixed properties, to a performative view where rule following is characterized as a typically emergent, distributed and artifact-mediated activity''\citep{dadderio_performativity_2008}.

\subsection{Performativity for IT system development}

All that said, one could rightly wonder what the performative turn, as it has been characterized above, has to do with the discourse regarding IT design in socio-technical settings and, above all, if there is anything new. Although some of the performativity tenets, like paying attention to ``the negotiations between actors''~\citep[p. 67]{wagner_exploring_2010} and the question of when design stops and use begins\citep[cf. e.g., ][]{brand_how_1995} ``may seem old to people within the CSCW tradition'' and related ones (i.e., HCI, PD, and the like)\footnote{This was honestly admitted by~\citet{jensen_cscw_2008}, who has nevertheless advocated a better consideration of these ideas within those traditions. Yet, two years \emph{later}~\citet[p. 31]{bratteteig_research_2010} have conversely recognized that ``the performative turn in post-structuralism is perhaps under-articulated in design research.''}, we believe that this perspective can be fruitful along both the practical and conceptual dimension.

\subsubsection{On the practical side: towards new meaningful development cycles}

From the practical point of view, only a few contributions so far refer to the performative tenets explicitly with respect to design;, for instance,~\citet{jensen_cscw_2008} advocates a reorientation of both the understanding and (less clearly, though) the practice of the process of IT design (or more specifically of CSCW design) along the performativity strand; to this aim, he submits recommendations to keep in mind performative aspects in the design process, such as that ``neither humans nor technologies \emph{determine} each other'' deterministically, and that ``materiality might trick us in practice''. Unfortunately, the author falls short of clarifying how a performativity-aware disposition, or ``relativising one's own ontology'' (although certainly a useful exercise) could also ``revitalize design'', and really change the practice of IT design. With a more practical attitude, ~\citet{danholt_prototypes_2005} makes an argument about the performative nature of prototypes, by implying with this expression that prototypes `` affect users in concrete, material, bodily ways in situ''.  Recognizing the performative nature of prototyping is then related to recognizing that this way of designing artifact is ``mutually transformative for users as well as for the technology, a process of co-construction of humans and artifact''; if design ``is considered to be performative'' it is recognized as ``an emergent process where the end result is not predicated by either users or designers, but [is] an outcome of the process. [\ldots] Performativity thus also means that the existing is continuously performed and reiterated in order to persist, which means that the existing is also always under construction and transformation. Slight changes in the way things are done lead to novel existences. Performativity thus imply a continuous possibility of transforming the existing''. 

While we would fully subscribe these conclusions, we notice how user-centered, and even participatory design, approaches (let alone any approach within the more traditional, engineering mythology), in which users are considered to hold important knowledge on their practice and, \emph{in virtue of this competence}, are involved in the design process (in some form), still consider end-users, as well as their target practice, as pre-existing the design process and somehow invariant to the task, \emph{in its essential traits}. Thus, we agree that prototyping and participatory prototyping, especially when prototypes are not merely representational ones (i.e., mock-ups) but rather are working gears~\citep[like in the framework presented in ][]{harel_can_2008}, can make the distance between design and use (and hence designers and users) shorter; but, still, if the prototype is tweaked and co-constructed in a \emph{controlled environment}, in a \emph{design ambit}, that is \emph{off-line} with respect to the flesh and bones of (situated) action, then the performative dimension of the development process is still kept at the margin of the real never-ending (and very aptly depicted as loop-closed) process of the task-artifact cycle~\citep{carroll_task-artifact_1991}, where both coevolve as a whole \emph{and} at a different pace. 

\begin{figure}[tbh]
  \centering
      \includegraphics[width=.6\textwidth]{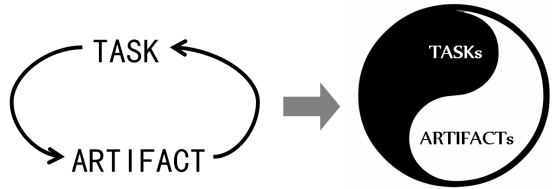}
  \caption{The evolution of the task-artifact cycle in a more co-constitutive vision.}
  \label{fig:yinyang}
\end{figure}

Within a performative strand, such a cycle would likely resemble a more intertwined figure, where \emph{there is no such a thing} like the task \emph{without} the artifact by which it is accomplished, and the artifact alone is just inert accouterment outside the task. Taking seriously that ``the social organisation of work does not pre-exist in any precise or detailed way, but is constituted `in the [artifact-mediated] doing' by practitioners''~\citep{buescher_landscapes_2001} suggests then to consider a variation of the widely known Taijitu symbol (see Figure~\ref{fig:yinyang}) to represent the task-artifact co-evolution: tasks\footnote{We use it the plural form, as it could be difficult, if not arbitrary, to distinguish between different tasks in actual, often multi-tasking practice where means, concerns and ends got mutually mingled.} \emph{occur} only when artifacts are used; artifacts make sense to practitioners only when these are put to work; in other words, there is \emph{anything but situated action}, emerging from the indissoluble \emph{entanglement of tasks and artifacts}. It goes without saying that entanglements cannot really be designed, as ``the take-up, modification and rejection of technology in a work setting, and the [conseguent] accommodation of work practices that take place around a developing technology, are radically unknowable and unpredictable''~\citep{buescher_landscapes_2001} till it actually occurs.

\subsubsection{On the conceptual side: back to the future}

\begin{figure}[tbh]
  \centering
      \includegraphics[width=\textwidth]{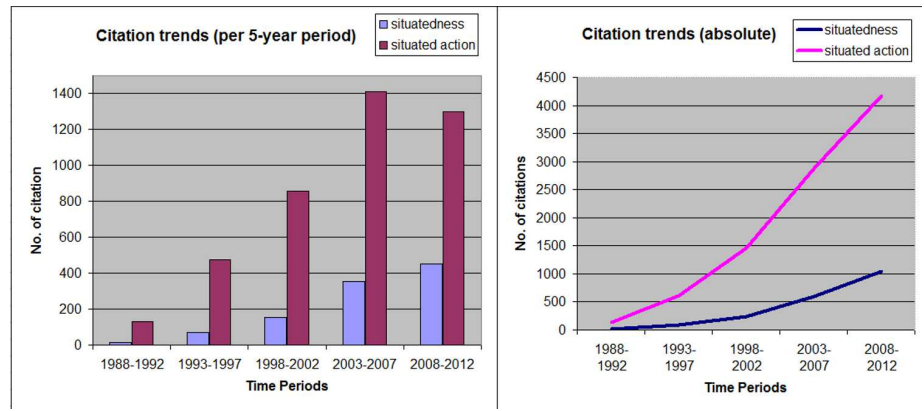}
  \caption{Trends in scholarly citations for the terms `situated action' and `situatedness'.}
  \label{fig:trends}
\end{figure}

The conceptual contribution is no less important if it's true what~\citet{schmidt_maps_1999} once pointed out, i.e., that ``Lucy Suchman's radical critique of cognitive science and the ``situated action'' perspective she proposed has played a significant role in defining the CSCW agenda and has become a shared frame of reference to many, perhaps most, of us''. In the last 25 years, from the publication of ``Plans and Situated Action'' (1987), the number of references to this construct of analysis has ever and ever increased, as well as the related (though not similar at all) concept of situatedness: a brief look at the citation trends (see Figure~\ref{fig:trends}) in the communities that are closest to EUSSET\footnote{The query was performed on the 16th of October 2012 on Google Scholar and was expressed as follows: (`situated action' OR `situatedness') AND (`CSCW' OR `HCI') AND design).} shows how the discourse around the concept of `situation' is central in system design and further proof of that comes from the recent initiative within EUSSET to prepare a ``Situated Computing Manifesto'' and to present such a document to the European Commission as a contribution to the new research program of the European Community called ``Horizon 2020'' with the aim to open a ``new'' research area inside the program focused on research on human practices and design of new systems\footnote{http://www.thinkinnovation.org/it/blog/2012/06/eusset-has-just-engineered-the-manifesto-of-situated-computing/}.

Notwithstanding its relevance, this focus on ``situation'', rather than on performativity, resulted in being problematic, perhaps for its apparent roots in the concept of a (static) place (cf. Latin situatio, site) that laid it open to representationalist drifts (cf., e.g., the connotations acquired by the term `context', among which that of ``container-like''~\citep[p. 19 ]{suchman_human-machine_2006}, in IT-related discourses about ``context-aware systems''). As pointed out by~\citet[p. 23]{clancey_situated_1997}, ``the overwhelming use of the term \emph{situated} [\ldots] since the 1980s has reduced its meaning from something conceptual in form and social in content to merely ``interactive'' or ``located in some time and place''. Even~\citet{suchman_human-machine_2006} herself admitted that the passage where she had written that ``the situation of action \emph{can be defined} as the \emph{full range of resources} that \emph{the actor has available} to convey significance of his or her own actions and to interpret the actions of others'' could be erroneously ``taken to imply that `the situation' exists somehow in advance of action and that it could at least in principle \emph{be fully enumerated and represented} in the form of a model to be referenced'' and drawn by some professional (i.e., the designer) before actually going ``where the action is''~\citep{dourish_where_2001}. Conversely, ``the sense of the situation [Suchman is] after \emph{is a radically performative and interactional one}, such that action's situation is in significant respects constituted through, or stands in a reflexive relationship with, ongoing activity'' (p. 125, our emphasis).  

This remark can not be underestimated. Indeed, when Suchman exposed the main themes pertaining to her decades long research in the field of HCI in the preface of ``Human-Machine-Reconfigurations'' (a reprint of ``Plans and Situated Actions'' that was enriched by new footnotes and additional chapters) she mentions: ``the irreducibility of lived practice, embodied and enacted; the value of empirical investigation over categorical debate; the displacement of reason from a position of supremacy to one among many ways of knowing in acting; the heterogeneous socio-materiality and real-time contingency of performance; and the new agencies and accountabilities effected through reconfigured relations of human and machine''\citep[xii]{suchman_human-machine_2006}. 
It is for us extremely indicative that Suchman did not mention ``situated action''\footnote{Perhaps a sort of semantic pleonasm if it is true that action can not be but situated\ldots}, nor \emph{situatedness}. Here we briefly recall that the former concept was originally chosen ``to underscore the view that every course of action\footnote{including planning itself or ``calling out a plan as a self standing artifact'' cf., respectively p. 17 and 21.} depends in essential ways on its material and social circumstances'' (p. 70); on the other hand, the latter term was actually \emph{never} used by Suchman, although hundreds of scholarly papers associate it to her work\footnote{As a matter of fact, in Human-Machines Reconfigurations Suchman speaks of situatedness only once, and only to challenge the meaning intended for such term by Rodney Brooks, the MIT engineer that questioned symbolic representational approaches in the field of robotics, as she found such meaning ``evacuated of sociality''.} and has been object of some criticisms, among which\footnote{In this number we can also mention people, like~\citet{lave_situated_1991} lamenting the vagueness of the definition itself of situatedness, as well as, from an exactly antithetical stance, other scholars, like~\citet{vera_situated_1993}, contesting what they think to have understood of this concept.} we recall here the point by~\citet{ciborra_mind_2006} regarding the paradoxical and somehow extraordinary lack in such concept of any affective, human, but we would also say performative, element\footnote{~\citet{ciborra_mind_2006} writes: ```Situated' is the translation of the German `befindlich'; situatedness is `befindlichkeit'. [The former term]  not only refers to the circumstances one finds himself or herself in, but also to his or her `inner situation', disposition, mood, affectedness and emotion.''. Moreover incidentally, we could also note that the German courtesy expression ``Wie ist Ihre Befindlichkeit?'' can be translated ``How are you \emph{doing}?''.}. In the same vein, also the current interests on either ``situated software''~\citep{balasubramaniam_situated_2008} or ``situated computing'' can be questioned. As Suchman put it (our emphasis):

\begin{quotation}
I believe that the argument made [in 1987] holds equally well today, across the many developments that have occurred since. The turn to so-called situated computing \emph{notwithstanding}, the basic problems identified previously~--~briefly, the ways in which prescriptive representations presuppose contingent forms of action that they cannot fully specify, and the implications of that for the design of intelligent, interactive interfaces~--~continue to haunt contemporary projects in the design of the ``smart'' machine~\citep[p. 3, our emphasis ]{suchman_human-machine_2006}.
\end{quotation}

Thus, we observe how, periodically in the IT system design discourse, terms and expressions that do have the potential to overturn the traditional oppositions like those of abstraction vs. materiality, representation v. performance, and that of design of systems that ``solidify and stabilize procedures and classifications''\citep{orlikowski_duality_1992} vs. their use in situated performances, although they are ``sensitizing concepts'' ``which draw attention to important features of work and provide guidelines directing research in specific settings''~\citep{crabtree_wild_2001}, nevertheless end up by getting themselves into a sort of seventh room of the Bluebeard castle, where concepts are seemingly kept alive, honored and dolled up, but actually in a state of harmless captivity, with no real influence on actual practices and on the inner convictions of practitioners. This could be just the plain consequence of having ``engineering education had over-invested in analytical technique and scientific understanding at the expense of the practical, `hands-on', the creative, the reflective, the social, the constructive, the ethical, the economic''~\citep[p. 295]{bucciarelli_designing_2003}. Or maybe of just opening the wrong doors. 

In conclusion, we assert the topicality of the performative turn (especially in the sense of the intellectual legacy argued above) and advocate the concept of \emph{performativity} to be taken more seriously in the future for at least two reasons: first it refers to a ``doing'' explicitly, and in that it differs from the keywords like ``situation'', ``situated(ness), and ``context'' which all refer to a ``state of being''; in the hope that this could be enough to avoid getting sucked into the Charybdis of essentialism/representationalism, this could also facilitate the reappropriation of the \emph{original} lesson by Suchman, at least within the CSCW community. Second, we believe that the performative view, in its being anti-conceptual, anti-representational and against surreptitious divides between design and use (i.e., practice), have a potential to bring us on the other side of the river~\citep[cf. the life-raft model mentioned by ][]{buescher_landscapes_2001}) and let us assert that technology-in-practice~\citep{orlikowski_using_2000} can not be really ``designed'' but rather allowed to ``emerge''\footnote{Of course some has still to develop the technological artifact.}, and finally shift with no regrets our concerns  ``from a focus on invention [we would say of design, Ed.], understood as a singular event, to an interest in ongoing practices of assembly, demonstration and performance. The shift from an analysis in terms of form and function to a performative account''~\citep[p. 165]{suchman_working_2002}. We re-propose this resolution within our alternative mythology as a way to bridge the literature contributions mentioned above and the following discourse on bricolage.

\section{From concepts to `\textit{bricolages}'}
\label{sec:bricolage}
\begin{verse}
\begin{flushright}
\scriptsize
I often try out little bits\\
wheresoever they might fit.\\
The sages call this bricolage,\\
the promiscuous prefer menage\ldots 
\footnote{Thomas Erickson, 2000, allegedly written upon reading a commentary for a special issue of CSCW Journal on Theory.}
\end{flushright}
\end{verse}

If the discourse on complexity suggests that complex systems, and in particular socio-technical systems, can not be really designed; and the discourse on the performativity nature of socio-technical systems suggests us to recognize that designing for interaction and action is overambitious for its irreducible distance from the actual performance of the task (which actually, ultimately and yet continually creates meaning and materiality in a situated setting), the last discourse we aim to in order to build our mythology is what gives us ``some hope'', that a way, if not a method, can be taken toward the actual realization of technological scaffoldings~\citep{orlikowski_material_2006} for collaborative complex socio-technical systems: the discourse on the bricolage.

As many other authors before us, in very different fields, we are also fascinated by what the word ``bricolage'' evokes, both in light of its etymology, which differently from what it is usually reported can be traced back to old expressions meaning to `stray', `swerve' but also `rebound', `redound'~\citep{miller_bricoleur_1996}\footnote{so that it is fully justified to claim that bricolage pertains both to an indirect, unexpected, somehow deviating action (cf. The Oxford English Dictionary) but also to something that is essentially redundant and ``inefficient'', at least with respect to some modernist views.}, as well as in light, of course, of the research~\citet{levi-strauss_savage_1966} accomplished on mythical thought and so called ``primitive'' cultures that he reported in his book `The Savage Mind'. The word itself legitimates the fact that this concept has been adopted in different fields and that it happily ``redounded --~itself a bit of bricolage~-- to other discourses: postmodern philosophy, theology, depth psychology, literary theory''~\citep{miller_bricoleur_1996}, cultural studies and IT design-related research. In this latter field, we draw heavily on the concept of bricolage for its ``overall generative effect[, which] seems to be more dependent on \emph{interaction} rather than on some \emph{overriding design rationale}''\citep[our emphasis][p. 347]{lanzara_between_1999} and because ``bricolage privileges combinatory logics, loose coupling, and garbage can processes'' (\textit{ibidem}), right the elements that are suggested by the certainty that any designed thing, no matter how well conceived, will necessarily fall short of avoiding the ``law of unintended consequences''~\citep{mansfield_nature_2010}.

To our knowledge the first authors to draw the rich semantic sphere hovering around the concept of bricolage near to the equally rich sphere regarding design were (almost independently)~\citet{weick_organizational_1993} and~\citet{ciborra_thinking_1992} in the close ambits of organizational design and information systems design, respectively. These studies, although kept themselves mainly abstract and theoretical in analysing the value of adopting the bricolage metaphor in the design research,  provided the conceptual background for many subsequent contributions that leveraged, or simply were inspired, by this metaphor. Among these, we 
also consider the impressive contribution by~\citet{buescher_landscapes_2001} that was one of the first ones to try to give more concreteness to the notion of bricolage within the factual process of the development of computer-based information systems in organizational settings. In their work~\citet{buescher_landscapes_2001} suggestively propose a ``life-raft'' model of systems development~--~a continuously unfolding bricolage of technologies to hand, requiring much patching and baling, with an unknown destination'' (p. 17). In this ``overarching framework within which newly developed technologies are set in place and helped to `work''', they argue that the design process had to become more ``immediate and continuous'' in order ``to cope with the deeply built-in uncertainty of the relationship between technical systems and work practices'' (p. 22). Differently from other authors, they provide a pretty concrete definition of bricolage in a CSCW context:

\begin{quotation}
Bricolage can be described as `designing immediately', using ready-at-hand materials, combinations of already existing pieces of technology~--~hardware, software and facilities (e.g., Internet providers)~--~as well as additional, mostly `off-the self' ones. It therefore also involves design as assembly [and] requires investigation of the process of assemblage as well as designing for it.(p. 23)
\end{quotation} 

While we could substantially agree with the point regarding the \emph{immediacy} of a bricolage-oriented approach (which we interpret as `unmediated spontaneity'', i.e., ``without the mediation of designers or IT specialists'', rather than in terms of ``ad-hoc quickness''), we look with diffidence the subsequent points that bricolage ``is not just an assembly of technical components, but also of appropriate workpractices, skills and training, communications, affordability, legal and contractual arrangements, etc.'' (p. 23)

In fact, if bricolage can at the same time be considered as ``a description of the existing context''; the general activity of bricoler as well as as its ``(unforeseeable) outcome'' (i.e., an assemblage of `things that work', the solution coming out from a particular round of development); and even as a (presumably context-independent) ``method for design'', we would agree to face a very rich concept but, at the same time, we would consider it a far too all-encompassing one to really support a factual approach to system development.

For this reason we propose to focus on the notion of \emph{bricoleur}, i.e., who \emph{performs} bricolage, and therefore especially on the notion of the \emph{bricoleur-in-practice} (to mirror~\citet[][]{orlikowski_using_2000} terminology), i.e., who is actively involved in the activity of \emph{bricoler}, and we submit such an idea of user as the one intended to get value from a bottom-up approach to technology development that tries to do without conceptual design (although not necessarily without ``designers'', as we will see in Section~\ref{subsec:role}). As put in words by~\citet{hartswood_being_2000}:

\begin{quotation}
Users need the opportunity that only their work can offer to explore fully the possibilities for
adopting, and adapting to, new systems and artefacts. When this is allowed to happen, and given the right choice
of technologies, development work can assume the characteristics of `bricolage' -- i.e., the rapid assembly and
configuration of `bits and pieces' of software and hardware~--~led by users acting within their own work settings,
with IT specialists taking on the role of facilitator.
\end{quotation}

Yet, for the detrimental nuance that the term bricoleur has in French as well in its closest English translation `tinker', we propose to denote the expression bricoleur-in-practice in terms of the single term ``bricolant'', i.e., ``who is performing \emph{bricolage}''. This is the meaning that we want to adopt and that we trace back to the specific archetype of bricoleur that Levi-Strauss himself introduced to contrast the opposite archetype of `engineer'. In our view, then, the latter can personify the rational designer that builds systems from scratch after and in virtue of a conceptual effort; while the former denotes the user that fabricates her own tools from available resources, being immersed in situated performances and contingencies.

\begin{quotation}
The bricoleur is adept at performing a large number of diverse tasks; but, in contrast to the engineer, he does not subordinate each one of them to the acquisition of raw materials and tools conceived and procured for the project: his universe of tools is closed, and the rule of his game is to always make do with `what's available', that is, a set, finite at each instance, of tools and materials, heterogeneous to the extreme, because the composition of the set is not related to the current project, or, in any case, to any particular project, but is the contingent result of all the occasions that have occurred to renew or enrich the stock, or to maintain it with the remains of previous constructions or destructions~\citep[p. 17]{levi-strauss_savage_1966}.
\end{quotation}



For our aims, the key points and motivations for focusing on the role of the \emph{bricolant} bricoleur, and hence on the \emph{performance} of bricolage (or \emph{bricolag}-ing), rather than on its outcome or the corresponding a-temporal activity, i.e., the bricolage, is somehow buried in three statements by~\citet{levi-strauss_savage_1966}. In what follows, we will review these three passages to some extent, for their importance in making our position more clear. First, objects ``are not known as a result of their usefulness; they are deemed to be useful or interesting because they are first of all known'' (p. 9). This means that what is ``useful'' or not cannot be pre-determined in terms of functional requirements, irrespectively of the competence of the analyst/designer, as these are necessarily decoupled from the actual availability of the corresponding functionalities in the workspace of users. Conversely, each work item, be it both operator or operand of a computer-based functionality, is meant by users as useful if they have already internalized its function, that is if they already \emph{know} it and have made sense of it. This means that the bricoleur is certainly someone that uses the objects she can find around her, but it is also necessary that she has previously been somehow involved in the creation of those objects, so that she can really know them and know how to arrange them meaningfully at any time. Thus, bricolage is seen as an arrangement of predefined objects, where pre-defined here just means ``defined before'' and not ``from above by someone else''.

This lead to the second passage where a distinction between the engineer/ designer and the bricoleurs is made in virtue of ``the inverse functions which they assign to events and structures as ends and means, [the designer] creating events (changing the world) by means of structures and the `bricoleur' creating structures by means of events.'' (p. 22). This point is particularly important in the vein of how the performative stance sees every event\footnote{That is as ``an autonomic and contingent occurrence with its own conditions and its own time-structure, [in respect to which] the meaning of the past for the present is not fixed but radically ambiguous''~\citep{dirksmeier_time_2008}, i.e., inextricably intertwined with the given situation.}. This means a word of caution regarding any structure that the designer could conceive to either enable or constraint action (i.e., change the world) as these structures may be changed in the process of their enactment, even if such a change is unintentional and unacknowledged~\citep{orlikowski_improvising_1996}\footnote{As Derrida has cunningly observed, there is an inverse relation between play, to which bricolage as an irrational, nonlinear and fragmented activity can be assimilated~\citep{pohn_cosmicplay.net_2007}, and structure. 
If one of the most ambitious purposes of our mythology is to challenge this distinction by highlighting the supremacy of use over design, we owe to Derrida the insight that also this dichotomy is a myth, ``simultaneously meaningless and useful''\citep{pohn_cosmicplay.net_2007}, which is itself made up, and probably fostered by ``tinkerers of abstractions'' who try to sell representational services to potential customers~\citep{robinson_questioning_1991,nandhakumar_fiction_1999}}. It also relates to the more manifest feature of the \emph{bricoler} (that is the activity of bricolage): not only, as said above, to make things out of the materials one has lying about, \emph{but also} to make sense of those materials according to an interpretative act that reinvents the objects (at least their meaning, their function, their value) anew in face of change, and that is hardly anticipatable and mostly unplannable as it is also deeply conditioned by past interactions (we would also say `situated' of course). Thus, we can say that change urges the \emph{bricolant} user to modify her bricolages, as well as its building blocks. 

This leads to the third, and more important, passage to our aims: ``the engineer works by means of concepts, and the bricoleur by means of signs''  (p. 20), in light of the fact that ``signs can be opposed to concepts [in that] whereas concepts aim to be wholly transparent with respect to reality, signs allow and even require the interposing and incorporation of a certain amount of human culture into reality'' (p. 20). 
The idea of transparency from this passage hints suggestively at a clear development recommendation: whereas the engineer aims to hide information\footnote{Cf. the principle of encapsulation, which is defined by Grady Booch as ``the process of compartmentalizing the elements of an abstraction''.} and to make his idea of, say, \emph{Patient} into a number of attributes codified in a relational DBMS, well underneath the application logic, the \emph{bricolant} user needs to have well under his \emph{gaze} what fields will represent the patient in her artifacts, arrange them the way she needs, fill them in on the basis of informal conventions and customs, as well as to disregard and create some new attribute/field at need irrespective of any ideal model of that disembodied entity.
Moreover, the passage mentioned above also clearly requires, first, that a second but by no means less important activity of the \emph{bricolant} user follows the activity itself of having built the bricolage (the artifact), and consists in a continuous and seamless accumulation of any sign that could help her make sense of the bricolage-in-practice: so content can enrich the bricolage-artifact (not just ``be contained'' or ``stored'' therein), as well as any kind of meta-content produced by the users can, like comments, tags, nested threads of conversations that unfold \emph{around} and \emph{about} the tangible artifact. In short: bricolage as a continuous and creative ``playing with signs''\footnote{This passage is strongly influenced by the reading of Nieztsche by Derrida in  ``Structure, Sign, and Play'', where the Nieztschean perspective is related to ``the joyous affirmation of the play of the world and of the innocence of becoming, the affirmation of a world of signs without fault, without truth, and without origin which is offered to an active interpretation.''. Bricolage itself is a concept that urges us considering system development as a game-related social underatake.}. Second, this passage sheds light on the requirement that any computational support of the activity of the \emph{bricolant} user (i.e., the bricoleur \emph{at work}) must be oriented towards this continuous creative and interpretative activity (that can accumulate data as well as coordinate activities~\citep{berg_accumulating_1999}), which nevertheless is completely on her own; towards the reconciliation of multiple, possibly diverging, interpretations; and above all towards the coexistence of these multiple and contextual \emph{incorporations} (see above), both in the local and in the global dimension.

This latter point is what makes us believe that the bricoleur-oriented mythology (as a specific kind of end user, or better yet, of end user enabled by a specific kind of platform that we will outline in the next section) does have the potential to oust the designer-oriented mythology, or at least the mythology where the designer is the proverbial Renaissance figure able to exhibit the competence, in short what~\citet{hirschheim_four_1989} in a still timely contribution denoted as the ``systems expert''. This stereotype, although it has been considered ``unrealistic and an arrant nonsense'' for more than 20 years within the CSCW community, still distorts in professional practice (and not only there) the fragile symmetry of the Janus-like relationship between users and designers~\citep{bowers_janus_1991}. In the next sections, we will speak about how a ``\emph{laissez faire les bricoleurs}'' method can be flanked by a specific ``\emph{logic of bricolage}'', in order to empower end users and have them become the builders of their own artifacts within their daily practices.

\section{Towards environments that support the \emph{bricoleur}}
\label{sec:environments}
\begin{quotation}
Everything that can be said, can be said clearly.\\
(Ludwig Wittgenstein, 1922)\footnote{Tractatus Logico-Philosophicus, 4.116}
\end{quotation}
In this section we would like to address how the three discourses that we have outlined above can converge into a single, coherent and practical proposal for the development of interactive and collaborative information systems. If these strands can actually converge, the related mythology of system development should situate itself among the relatively new research lines that are emerging within the HCI field. As also recently pointed out by~\citet{ardito_composition_2012} these lines focus on concepts such as:

\begin{itemize}
\item Appropriation: i.e., the process by which technologies are understood and used by users in their own ways, possibly subverting the designers' intentions~\citep{orlikowski_learning_1992,dix_designing_2007}.
\item Meta-design~\citep{fischer_meta-design:_2006}: also denoted as ``design for designers'', a design paradigm, which allows various stakeholders, including end users, to act as codesigners even at use time; according to this paradigm, software engineers do not design the final application, as in traditional design, but they create software environments through which different stakeholders can contribute to the \emph{design} of the final application
\item End-User Development~\citep{lieberman_end-user_2006}: a paradigm that focuses on the capability of systems to offer support during run time to
empower users to develop their applications, blurring the distinction between design time and run-time;
\end{itemize}

As we will argue in what follows, the alternative proposal we advocate moves from the first concept of those mentioned above to the last one, in a progressive approaching towards its gist tenets. The term appropriation is indeed representative of a stance we desire to distinguish ourselves from since it is clear that one can appropriate, i.e., take as one's own\footnote{One could notice that the Latin root of the word appropriate, i.e., ad-propriare, hints more at a taking than at a making (cf. the preposition \emph{ad} here indicates a motion toward oneselves.)}, a ``thing'' that has been constructed by someone else. For instance,~\citet{carroll_completing_2004} writes of ``the crucial role played by users' actions in completing the design process'' and that ``[technology appropriation] is actually part of the design process. The design of a technology innovation is completed by users as they appropriate it''. We find then that the notion of appropriation is deeply ingrained in the design-oriented rhetoric. 

In the same mould, also meta-design is a term that explicitly refers to the phase of design, and that was programmatically aimed at investigating ``techniques and processes for creating environments that allow `owners of problems' (or end users) \emph{to act as designers}''~\citep[our emphasis,][]{fischer_meta-design:_2004}. The main contribution that we want to retain from this framework is then the idea of ``underdesign''; this notion relates to design for purposely ``incomplete'' systems that, once deployed, would allow for important modifications by end users themselves, in face of unexpected needs that show up at use time, and that could not be anticipated at design time. Underdesign hints at a conceptual design that does not have the ambition to fully set the system up for its embedding in a complex socio-technical system, but it also hints at a design for the ``under-layers'' i.e., aimed at the construction of environments where applications can be developed with a strong interaction (co-design) between users and professional designers.~\citet{fischer_meta-design:_2004} use the term ``seed'' to denote an underspecified application that users can complete during its use; the authors of this work also report about the action research initiatives that led to the construction of such environments by means of specialized editors (e.g., a Map Editor). The term seed is fully coherent with the idea that applications grow~\citep{truex_growing_1999} and evolve with their environments but, in some way, this latter idea seems to clash with the claim that that end users have to \emph{act as designers}, if this means to reiterate a conceptual, detached and abstract if not formal way to envision how the application ought to be and ought to behave in the unknown future, even if this activity is performed by end users, who play the ``designer'' role.

Moving in the same direction but covering some road further, EUD is the approach the most clearly and explicitly has stated in its agenda (as well as in its name) the involvement of users in the construction of their technology, and without expecting them to act as designers\footnote{As every rose has its thorns, also the expression End-User Development has actually its own; ironically one could find a little thorn for each single word therein contained. \emph{End}: this reiterates the concept of an actor who is the ultimate terminal of a process that she does not control or owns, as indeed she comes at the end of it; \emph{user}: this reflects the idea that an artifact is built and then used, and that some actors design it and some others just use it; \emph{development}: this can hint at the fact that the more the end users are able to develop their own tools in the traditional sense of the term (i.e., by programming it) the better it is, and that EUD research is mainly concerned with filling in the gap between this ideal vision and drab reality. Although we take this exercise in a tongue-in-cheek way, we can nevertheless make a more serious link to our advocacy of more deconstruction-oriented analyses of our given-for-granted categories and discipline language in the quest of (some of) the deep reasons why IT research has such a low impact on IT professional practice (see Section~\ref{sec:nodesign}).}. This shows a strong affinity with the approach we are going to discuss, especially if the meaning at stake for the term `development' is the original one mentioned in Section~\ref{sec:intro}, that is the notion of a continuous and indefinite (not necessarily teleological) ``unfolding'' over time, pruned of its abstraction and differentiation from the actual work practices; only in this way, the end user is left in her natural, or better yet ecological, environment, and she is not eradicated from her work situation to be transplanted, within another well circumscribed and protected environment, into the engineering framework of ``coding for designed structures''. This is the point that resonates more with the passage by Levi-Strauss reported in Section~\ref{sec:bricolage} where the end user, the \emph{bricolant} bricoleur of our mythology, is expected to ``work with signs instead of with concepts''. Thus, \emph{constructing} (or modifying) the artifact should not be seen as radically different than \emph{working with} the artifact. The constructs and structures with which the end users work should be familiar (cf. the other passage highlighted in Section~\ref{sec:bricolage}), like blocks and parts of the artifact itself, are indeed conceived to be rearranged or created by composition from smaller subcomponents that are not ontologically different from their compounds (e.g., big field sections in forms are made of smaller fields groups, and these in their turn are but data fields).

Adopting a fully and coherent EUD approach has a strong impact both on the dimensions of \emph{who} is involved in the development; and \emph{what kind} of system is supposed to support this kind of developing. We will discuss both these dimensions, starting from the more general one, which regards what application macro-classes must be considered for a platform that supports the continuous bricolage-based construction of \emph{convivial tools}\footnote{This expression is taken from Illich. A convivial tool is defined as ``that which gives each person who uses it the greatest opportunity to enrich the environment with the fruits of his or her vision'' and whose ``renewal would be as unpredictable, creative, and lively as the people who use them''~\cite{illich_tools_1973}. This term then evokes a concept that is central to both the way we intend bricolage, as a collaborative and creative activity, and technology, as a tool also for socialization, not only for efficient production, whose building and development should give the opportunity to ``end users' to collaborate and socialize. From Latin \emph{con-vivium}, \emph{living together}, \emph{having a nice time together}, a convivial tool is a tool that unites people in both its use and production, that does not alienate them and indeed give them opportunities to enjoy life together.}. 

\subsection{What meta-system for end users' systems?}

It is possible to distinguish between two main ways a system supports the development of an application at the level of the end users or at least for its tight adaptation to their needs, that is two main ways such a system can act as a sort of meta-system for the development of EUD systems. On the one hand, we can consider systems that primarily (or exclusively) support \emph{configuration}. This regards the so called ``flexibility through control'' of systems that offer ways for people to adjust settings, reprogram the system or otherwise technically adjust it~\citep{dourish_evolution_1999}. Yet, allowing the setting of more or less articulated parameters that affect the application's behavior or its appearance at the interface level entails to give end users little room for intervention, as this is limited to a set of elements that can only take one from a predefined (at design time) set of values and corresponding effects on the application at run time; accordingly, such systems allow for an involvement that is, to our aims, too \emph{superficial} (also literally speaking) and that comes up by being constrained by some model of feasible action, or better yet by some feasible pattern in the ``fitness landscape''~\citep[p. 50]{mansfield_nature_2010} that results from precise configurations in the ``design space''.

On the other hand, other kinds of systems offer an environment that is ``flexible through openness''~\citep{dourish_evolution_1999}, that is a sort of ``meta-system'' by which users are supported in the \emph{creation} of new systems and applications of different complexity, according to their needs and competences: macro programming, visual programming, programming by demonstration are among the solutions that are given to users to ``encourage their participation in the design process''~\citep[p. 170]{dourish_where_2001}. Here the risk may arise that the good motivations and purposes of EUD-oriented researchers may clash with the scope and aims of the actual tools that are made available to the end users: specific features of the environment (or their absence) can introduce, or even impose, rigid models of practice, possibly unaware, and affect how end users build and maintain their equipment. This point relates to an important feature that environments enabling EUD practices should possess: we call this quality \emph{universa\emph{ti}lity}, to hint at something in between the traditional qualities of generality, universality and versatility. While generality is usually defined as ``the degree to which a software product can perform a wide range of functions''\citep{khosravi_quality_2004} and hence serve multiple purposes, universality and versatility (from which \emph{universatility}) refer to the qualities of being both general-purpose, but also easily tailorable to the needs of specific settings and thus able to fit local needs. In other words, where generality refers to the typical quality exhibited by Swiss army knives, that is to have multiple specific functions to serve distinct but anticipated purposes, \emph{universatility} refers to the typical quality of Sardinian Pattada knives\footnote{A pattada is ``an Italian pocketknife [dating] back to the 15th century, [that] is a folding model that opens to a length of 15 to 35 cm [and that] farmers and shepherds always carried it with them to do all sorts of jobs in the fields.''~\citep{de_michelis_swiss_2003}}, that is to allow for an \emph{open} space of possible usages by which users reach their unanticipated purposes with creativity and autonomy. Thus, a powerful environment has to be uni-versatile enough to avoid imposing restrictions on the applications that it allows to construct. Here the core of the problem lies in how this quality is guaranteed and on what conceptual premises (i.e., myths, to recall the vocabulary introduced in Section~\ref{sec:intro}) is grounded on.  

\subparagraph{Universatility based on an ontological approach}

The first way to make an environment general enough to be applied to any cooperative setting but also versatile enough to fit any (in principle) of its situated tasks is what we call the ``ontological approach''. This is expression of the representational and objectivistic approach we discussed in Section~\ref{sec:performance}: the designer of the environment decides how to guarantee wide customization on the basis of a pre-understanding of how actors behave in a number of recurring situations in multiple domains; consequently, on the basis of this understanding (which is based on deep introspection or more interactive and qualitative techniques), the designer conceives a set of ``labels'' that univocally identify the ``things'' that users will handle, associates that classification scheme with intended universal building blocks, and provides users with those elements, all together with specific rules for their composition, so that they can (acknowledge and) make value out of that given model. A paradigmatic example of this approach was the Coordinator~\citep{flores_computer_1988} 25 years ago: there the ontological claim was that actors coordinate their actions in terms of negotiation of commitments, and according to this model the technology offered a universal set of possible categories to characterize setting-specific behaviors and routines. The assumptions underlying this technological proposal have been widely discussed, and contrasted, since then~\citep[e.g.,~][]{suchman_categories_1994} but other examples of this approach still abound, both in daily life, e.g. where reference management softwares force us to univocally associate our academic works or books with a specific category, and in recent academic research, e.g., when users are called to categorize others' comments in public discussion with a system like Reflect~\citep{kriplean_is_2012}. 

In addition to systems where the ontological approach is adopted in an explicit form, we notice that such an approach can also act within an IT system \emph{implicitly} (if not surreptitiously), especially in all those systems that adopt a characteristic or strong metaphor representing ``the'' one way in which human allegedly organize their world and practices: this is the case of the most famous (and nowadays notorious) ``desktop metaphor'', as well of some recent alternative, like the metaphor of ``story'' proposed by~\citet{de_michelis_itsme:_2009}: in both cases, users are called to associate the objects they work with with a concept (i.e., the notion of file, or of resource) and characterize it in terms of a category~--~being it the name of the folder in which the file is virtually stored (as well as the location of this latter in the ``file system''), or the name of a sequence of interactions with someone or about something (i.e., \emph{the} topic of a conversation). The same phenomenon occurs in the ambit of context-aware or situated computing where, as mentioned in Section~\ref{sec:performance}, tools to characterize a context or a situation are part and parcel of the design of the application itself, whenever this task occurs, and they are based, again, on a predefined domain model that the users can only customize (or appropriate); this seems in basic contrast with the idea of context as ``embodied action'' that we share with~\citet{dourish_what_2004}.

All these approaches, either explicitly or implicitly ontological, are grounded on the hypothesis that things could be described univocally or, at least, that the ``name for a thing'' would mean \emph{that} thing irrespective of the setting where such name is used, and for what aim~\citep[cf. e.g.~][]{mark_reconciling_2002,anderson_down_2008}: this is the essence of an ontological stance. However useful this approach may be for ordering and retrieval purposes, any more or less structured ``ontology'' (in the broadest sense, i.e., taxonomy, classification scheme, interaction metaphor and the like), this is conceived at \emph{design time} and it is given to the users so that they make sense of their world in a way that can make some tasks more orderly efficient; yet, this way, counterproductively (cf. Section~\ref{sec:intro}), may also hinder the construction of supports of other, possibly more ``hidden'', tasks: this mirrors platforms that provide users with functionalities that allow for some degree of tailorability but, as the latter is constrained within the boundaries of the metaphor itself, do not encompass functionalities to let the application (and its underlying ontology) evolve towards and align with the idiosyncratic customs of the users: the availability of such functionalities, and their subsequent use, could seriously undermine the consistency of the overall model, and hence the effectiveness of the former tasks~\citep[e.g.,~][]{peters_against_2006}.

\subparagraph{Universatility based on a performative approach}

The main tenet of EUD regards giving users a more substantial role in technology conception, development and evolution. Component design is proposed as an approach that allows users to tailor their applications by enriching them with suitable components offering specific functionalities. This would require the application to be open to such type of tailorization~\citep{lieberman_breaking_2006}. Moreover, while for a task of integration, ``component thinking'' could seem natural, we have observed~\citep{locatelli_community_2010} that in the construction of application from scratch this could be perceived as difficult by users, who usually see their application in a more holistic way than the component-based approach would have them to think about.
A more radical stance is taken by~\citet{fischer_meta-design:_2006} and their notion of meta-design. To this regard we have already raised our perplexity, at least on a purely conceptual level, regarding those platforms that would be aimed at making users act as ``designers'' (in the modern and irresistible connotation of the term), rather than allowing them to construct their tools much alike they already do with their traditional artifacts: i.e., by individual or bottom-up organized initiatives, trials and errors, progressive amendments, patchwork and bricolage attitudes. Indeed, we have observed that actors in their everyday (working) life do not follow a traditional design-based approach to solve their problems and to construct the tools they need to this aim~\citep{cabitza_leveraging_2012}; this perception finds confirmation in a number of field studies~\citep[e.g.,~][]{carstensen_two_1995,morrison_observing_2009,blackwell_logical_2010,handel_working_2011,morrison_multi-disciplinary_2011} that focus on non-yet-digitized settings and that show a continuity in work practices development and paper-based tools construction that the current technology and the approaches used in its construction are still not able to reproduce or guarantee.

Indeed, if we agree that one of the main issues here at stake is that about the gap between users and designers (in the traditional sense) --~which is grounded, as we discussed in Section~\ref{sec:nodesign} on a conceptualization of design that will always prevent users from taking full control and responsibility\footnote{Notably,~\citet{beath_contradictory_1994} notice how most of the user-centered development methodologies that put a strong emphasis on user involvement (they make the case of information engineering) actually relegate users to playing a passive, although present, role during development and, in virtue of this participation, ask a more clear responsibility for project outcomes. We stress here the need to give full control, rather than only responsibility, to the community of users that will host the information system for its construction.} of the development process~-- we believe that this gap can be bridged only if design, and the conceptual modelling activity that design implies, are simply avoided and if the approach toward ``technology co-construction'' takes work practices ``seriously'', by avoiding any sort of compromise at the application level and by deriving the related consequences at the technological infrastructure level. In this line, taking work practices seriously means conceiving of technology construction as part of \emph{work and articulation work}, in the same way as paper-based artifacts are constructed by the actors when they need and use them~\citep[e.g.,][]{morrison_observing_2009}. 

We then call the sort of universatility that platforms must guarantee be based on a \emph{performative approach} for two reasons: first, according to an ontological approach specificity and situatedness are reached by having actors apply a universal model locally: such a model can be both adopted and adapted, but adaptation is here a sort of extension, rather than a tinkering that could undermine its basic assumptions and first-class concepts; conversely, a performative approach guarantees such locality by delegating the users in creating essentially open, underspecified, incomplete and even ambiguous ``models'', which that notwithstanding are totally \emph{theirs}, by which they can make sense of their do-it-yourself tools~\citep{cabitza_leveraging_2012,cabitza_supporting_2012}.

Second, adopting a performative approach calls for the requirement of an environment that limits itself in providing primitives by which users can build their application in a \emph{bottom-up fashion}, that is in an emergent process of tries and errors, and \emph{while they work}, as a way to improve the odds that the application will really reflect and support their situated practices: if this construction were ``extracted'' from those practices and moved to a controlled environment of introspection, modelling and ontological representation, we believe that we would again tap into a less than effective ritual, roped in the tar-baby of thinking that the task-artifact entanglement can be really untangled without losing both (see Section~\ref{sec:performance}). As Lanzara put it down: ``systems do not only operate or change in time, but are literally `made with time'''. Within a performative approach, as we discussed in Section~\ref{sec:performance}, end users can be seen as bricoleurs who build their digital tools tapping into their tacit knowledge and their creative skills to 
build the portion of IT support that comes closer to their work practices to fit their needs better. 

\section{Concrete Steps Towards a Logic of Bricolage}
\label{sec:environments}

In order to make a contribution toward the conceptual foundation of environments supporting the practice of bricolage in EUD terms, we will take inspiration from the point~\citet{lanzara_between_1999} made on the importance of ``transient constructs and persistent structures'' (p. 332), which are seen as the results of ``a practical, situated, context-sensitive mode of design that feeds on the dynamic tension between the requirements of change and stability''. We also think that what he called the ``logic of bricolage'' emerges from the intertwined interplay of structures and constructs, transiency and permanency, universality and locality: this  requires that an environment supporting bricolage is not supposed to provide users with sophisticated (i.e., semantically rich) modelling tools that facilitate the top-down construction of the application (from the conception of the ``entities'' involved, their attributes, their mutual relationships, and of the ``business processes'' where all these latter interact); but rather this logic is supposed to offer to the users a set of ``bricks'' that they can arrange and compose together in a bottom-up fashion within a conceptually consistent environment (the rules of composition).

\begin{figure}[tbh]
  \centering
      \includegraphics[width=\textwidth]{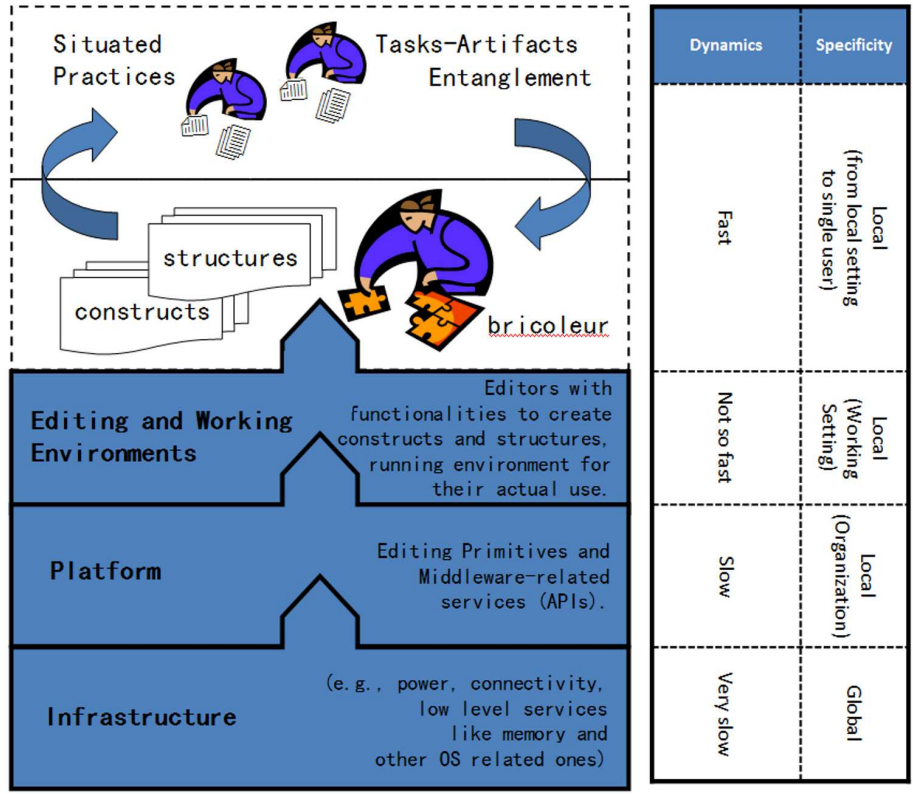}
  \caption{A conceptual architecture for an environment supporting bricolage.}
  \label{fig:architecture}
\end{figure}

In order to envision such an environment, we propose a multi-layered architecture that is inspired by the research accomplished in the COMIC project\footnote{The COmputer-based Mechanisms of Interaction in Cooperative work project was an EC ESPRIT-funded Basic Research
Project No. 6225, from 1992 to 1995}; in such an architecture the layers that are closer to the greater source of uncertainty and unpredictability, that is the layers that are closer to the users or where these act, are those allowed to be changed both faster and to a greater extent (see the column `dynamics' in Figure~\ref{fig:architecture}). With reference to Figure~\ref{fig:architecture}, we distinguish between an infrastructure, which is the set of available services that are used by the computational platform that is specifically designed to support \emph{bricolant} users in building and using their own tools; this platform, in its turn, exposes specific services to make the bricolage-based information system possible and computationally augmented; to this aim, the platform instantiates a working environment where either a persistent storage and a working memory, as well as an execution engine are made available to the users, and it instantiates an EUD environment where users can create their building blocks and edit their tools; while working in this latter environment, users use specific visual editors to build both their constructs and their working structures, that are put in operation when the working environment is ``on line''.

\begin{figure}[tbh]
  \centering
      \includegraphics[width=\textwidth]{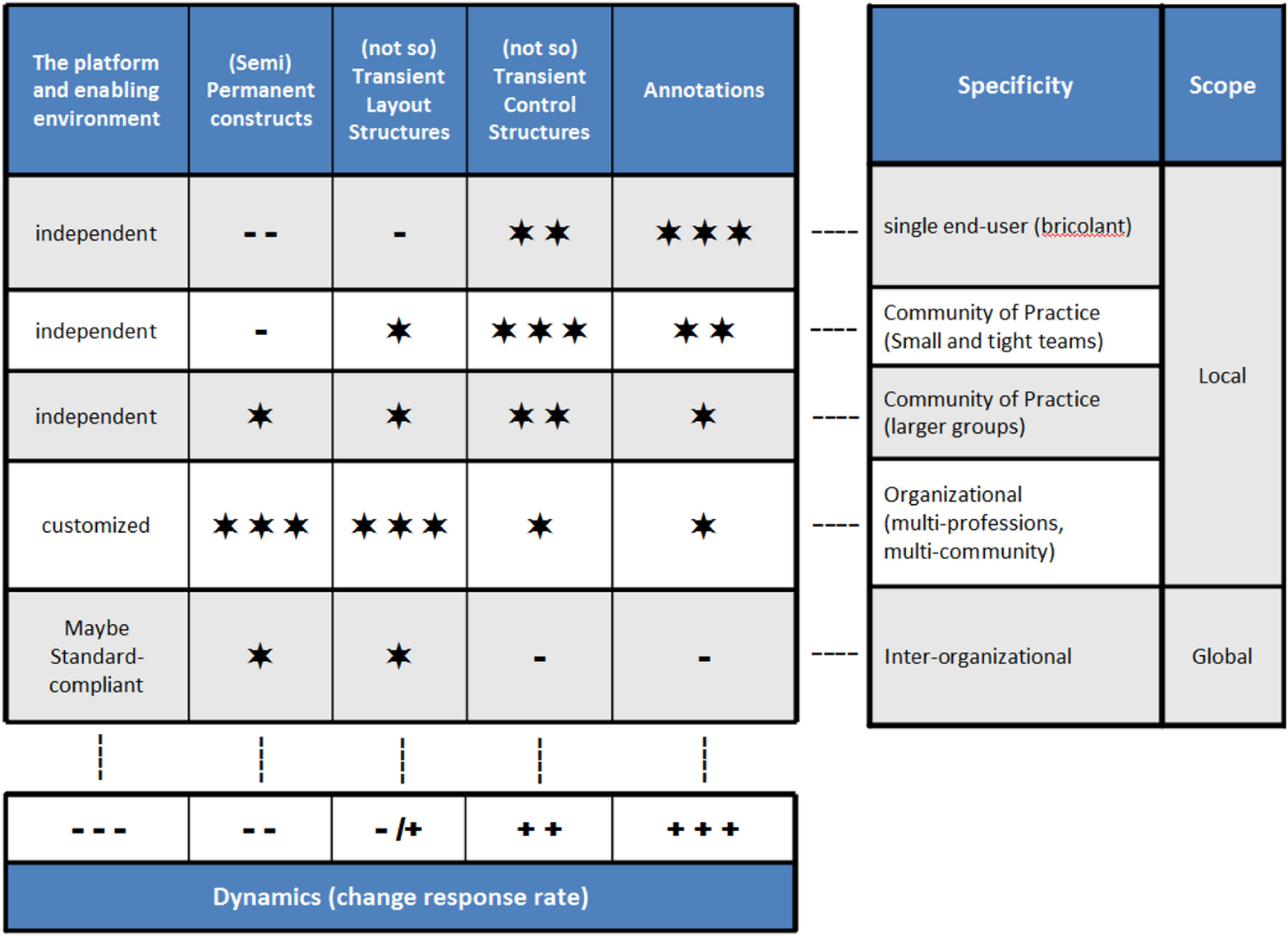}
  \caption{Synopsis of the main concepts encompassed within the proposed ``logic of bricolage''. Specificity, scope and change rate (dynamics) of each concept is indicated; the number of stars and plus signs indicates the extent an instance of each concept is, respectively, specific and fast-rate changing.}
  \label{fig:scope-change}
\end{figure}

We used the technical terms \emph{constructs} and \emph{structures} partly inspired by the original contribution by \citet{lanzara_between_1999}. More specifically, our proposal is based on the following first-class concepts or elements (see the top layers in Figure~\ref{fig:architecture} and the table depicted in Figure~\ref{fig:scope-change}):

\subparagraph{constructing constructs} These are \emph{constructs} that we denote as such because they are both \emph{construct(ed)} during the inception phase of the platform within a cooperative setting or organization as a result of participatory design-do activity; and they are also \emph{constructing}, that is are used as atomic ``building blocks'' by which the bricoleur users can create their working spaces and artifacts. We can further characterize these constructing constructs distinguishing between \emph{operand constructs} and \emph{operator constructs}: operators are all the feasible operations and micro-functions that users deem necessary to be performed over the operands; these latter are the most atomic data structures, components and variables that the platform must make available in both the editing and working environments to be used during situated work practices. Both operands and operators are the ``things'' that are arranged and put together in the bricolage activity in order to, respectively, compose the artifacts and endow these of computational capabilities.

In the same vein of~\citet{simone_computational_1993}, we described a method to recognize and characterize these atomic components from the observation and qualitative research analysis of paper-based artifacts used within a document-intensive work domain~\citep[i.e., two large hospitals ][]{cabitza_remain_2011}. In Table~\ref{table:operations}, we report the list of constructs identified as ``operations'' (i.e., operator constructs) that users agreed with us they could need to apply to the ``data fields'' (i.e., operand constructs) of their documents; these fields were identified as constructs as well, as a number of these fields or groups could be used and inserted in multiple document templates; in the study mentioned above we called these atomic groups of fields ``datoms'', as they were documental \emph{atoms} (i.e., not further decomposable elements) to be arranged into the needed templates. Not all the constructs are easy to build. Indeed, as our subsequent studies show, while datoms can be created with a relatively simple editor~\citep{cabitza_tailorable_2011}, which we realized to allow users both create data fields and their templates, operations clearly need to be associated with specific behaviors exposed by the platform or the infrastructure (like, e.g., printing, or sending as a message, see Table~\ref{table:operations} for a comprehensive list). According to the specific application domain (e.g., information systems, computer-assisted design), the platform can expose a series of elemental Application Programming Interfaces (API) to simplify the work of ``constructing'' the atomic operations to be invoked while using the artifacts conceived for the specific cooperative setting. In general, we intend both kinds of constructing constructs to be semi-permanent (to mirror the Lanzara's suggestive naming) in that their life spans (from creation to discarding) and change rate (i.e., the extent they are supposed to change during their use in the working space) can be put on a low-changing scale (see Dynamics in Figure~\ref{fig:scope-change}), although probably the set of available operators will change less frequently than the set of the operands (e.g., operations and data elements, respectively).

\begin{table}
\scriptsize
\begin{threeparttable}
\begin{tabularx}{\textwidth}{llX}
\toprule 
 & \textbf{Document-based Operations}\tabularnewline
\midrule
1	& create & This operation is akin to picking a new empty sheet of a specific template to insert into the folder. \tabularnewline \midrule
2 & retrieve & This operation is akin to picking a sheet from an archive and make it available for other operations. \tabularnewline \midrule
3	& open/read & This operation is akin to getting explicit access to the content of a sheet or instance of artifact. \tabularnewline \midrule
4	& write & This operation is akin to adding some new content to the artifact and accumulating new inscriptions on it. \tabularnewline \midrule
5	& select	 & This operation is akin to pointing either an artifact (among others) or a specific portion of its content. \tabularnewline \midrule
6	& copy  &  This operation is akin to putting some content into a buffer memory, like a little pocket-sheet. \tabularnewline \midrule
7 & correct  & This operation is to be considered different from regular writing but rather similar to striking through some content and substitute it with a correct one. \tabularnewline \midrule
8 & transmit  &  This operation is akin to sending either the physical artifact or (part of) its content to an external party; \tabularnewline \midrule
9 & print  &  This operation regards the physical printing, or copy, of part/whole content of an artifact  \tabularnewline \midrule
10 & officialize  &  This operation regards the formalization/certification of part/whole content of an artifact; \tabularnewline \midrule
11 & annotate & This operation differs from \texttt{write} in that it is aimed at adding informal or side content: it stands to writing as metadata stands to data. \tabularnewline \midrule
12 & attach & This operation can encompass affixing an external resource to the artifact. \tabularnewline \midrule
13 & cache & This operation regards the saving of part/whole content of an artifact for future use (modifications are still possible). \tabularnewline \midrule
14 & store & This operation regards the storing of the artifact in some repository, where only an operation of \texttt{retrieve} can take it from. \tabularnewline \midrule
15 & protect & This operation regards the preservation of part/whole content of an artifact from further operations. \tabularnewline \midrule
16 & delete & This operation regards the partial/complete elimination of either an artifact or parts of its content. \tabularnewline
\bottomrule 
\end{tabularx}
\end{threeparttable}
\caption{Operator constructs identified in~\citep{cabitza_remain_2011}.}
\label{table:operations}
\end{table}

\subparagraph{Structures} These are what \textit{bricoleur} end users create by composing and arranging \emph{constructing constructs} together. We distinguish between \emph{layout structures}\footnote{We prefer the expression ``layout structure'' instead of ``information structure'' (or ``data structure''), which would perhaps be the traditional mode to indicate those structures, as the latter term would have given the nod to the high level, conceptual element those structures could be referred to by a human user. Conversely, we mean to hint at the material, spatial arrangement of meaningful signs that ``act at the surface'' in promoting cognitive processes of sense making and interpretation.} and \emph{control structures}. The former ones are sort of material (yet non necessarily tangible) and symbolic work spaces that are recognized by members of a community of practitioners as the physically inscribed \emph{technological artifacts}~\citep{orlikowski_using_2000} where and by which to carry their work on. In document-based information systems, layout structures are the document templates of forms and charts that are to be used to both accumulate data and coordinate activities~\citep{berg_accumulating_1999}, endowed of both physical properties (i.e., the topological arrangement of the constructs mentioned above, i.e., data fields and sections) and symbolic properties (the boilerplate texts, any iconic element and visual affordances conveyed through the graphical interface). In the domains of computer-aided design and collaborative drawing/editing, a layout structure can be considered the working space where users arrange command docking bars, symbol stencils and predefined configurations of elements that must be set up before working on them. 

On the other hand, \emph{control structures} specify how the computational engine of the underlying layers of the architecture (see Figure~\ref{fig:architecture}) reacts in response to events generated at artifact (interface) level, how this latter acts on the content inscribed therein, and how it interacts with the users during their use of the tool. 
On a formal level, control structures can be expressed in terms of rewriting systems (see Listing~\ref{here_the_lbl}), a general formalism that can be instantiated, e.g., as rule based control systems, Petri nets, Business Process Modelling Language, that is any sort of declarative control construct. On a concrete level, control structures are generated by bricoleur users by composing constructs together and specifying how application behaviors (that are in their turn composition of operators defined over operands) should be exhibited in response to events and contextual changes. The simplest form of control structure is a type-based constraint defined over some operand; we~\citep{cabitza_leveraging_2009} have described if-then control structures (i.e., rules) by which the platform could convey information to promote collaboration awareness according to the content of a web of coordinative artifacts. More complex control structures are obtained by composing operations and selection points (defined over the construct's content) in a workflow-like manner. These \emph{not so} transient structures (to refer again to the Lanzara's proposal) are intended to be both the outcome and the scaffolding elements for the activity of bricolage: since structures are composed by arrangement and composition of the available constructs, they are supposed to change at a steeper rate than these latter ones (see Figure~\ref{fig:scope-change}): e.g., templates and rules defined on their content will certainly change more often than their building blocks, i.e., data fields and single operations.

\subparagraph{Primitives} Primitives are basic operations that the platform makes available through the editing environment where bricoleurs can create both their constructs and their structures. To adopt a pseudo-formal analogy: if constructs are the elements of an alphabet, the primitives can be seen as the composition rules of a grammar by which end users can generate meaningful sentences (i.e., structures). Primitives are part of the platform and exposed through the enabling environment and, as such, are not expected to change too often, besides the traditional activity of corrective and evolutionary maintenance. Specific primitives allow users to populate these structures with both content and \emph{meta}-content (see Figure~\ref{fig:scope-change}), that is any collaborative annotation.

\subparagraph{Annotations} We consider annotations be part of the first-class concepts of a logic of bricolage for their central role in work articulation, knowledge sharing and mutual understanding~\citep{luff_tasks--interaction:_1992,cadiz_using_2000,bringay_annotations:_2006,cabitza_leveraging_2012}, yet at a more informal level with respect to institutionalized (layout) structures and to the official content that is accumulated therein during situated practice. To this respect, any form of annotation carried out by practitioners \emph{over} and \emph{upon} structures and their content can be seen as a more ephemeral, informal and more user-driven piece of bricolage, which acts at a sort of different layer with respect to primitives, structures and content (see dynamics and specificity in Figure~\ref{fig:scope-change}) but that nevertheless (or right in virtue of this complementarity) plays an equally important role in making the artifacts-in-use flexible enough to support also invisible work and hence fully appropriated by their users. Annotations are then either stigmergic signs and marks attached to the borders of documents, extempore comments, semantic tags from either domain specific taxonomies or setting-specific folksonomies, or nested threads of both, as we described in~\citep{cabitza_supporting_2012}: all pieces of a bricolage that hosts informal communication and handover between practitioners, their silent and ungoverned work of meaning reconciliation, the sedimentations of habits and customs in effective (yet still unsupported computationally) conventions of cooperative work; for these reasons, we believe that any working environment aimed at enabling users in preserving (or even augmenting) their record-keeping conventions in the digitization of their traditional artifacts should support annotation as a first class activity of workers in their natural ``ecosystem''; In particular, then, also annotations should be referrable in control structures as we described in~\citep[p. 232]{cabitza_affording_2012}.

\begin{lstlisting}[basicstyle=\scriptsize,breaklines=true,breakatwhitespace=true,float=tb,caption=A preliminary formalization of the Logic of Bricolage,captionpos=b,label=here_the_lbl,frame=shadowbox]
Generative Productions of the Logical Of Bricolage

1) <web-structure> ::=  <layout-structure> +
2) <layout-structure> ::= <topological-object>+
3) <topological-object> ::= <operand-construct> <coordinates>?
4) <operand-construct> ::=  <constant> | <typed-variable> | <operator-construct>(<operand-construct>+) 
5) <operator-construct> ::= <functional-operator-construct> | <actional-operator-construct> 
6) <annotation> ::= <style> <target-ref>+ | <constant> <target-ref>+
7) <target-ref> ::=  <functional-operator-construct> (<target>)
8) <constant> ::= <domain-values> | <multimedia-text> 
9) <target> ::= <control-structure> | <topological-object> | <annotation>
10) <style> ::= <conventional-symbol>+ | operator-construct(<target>)
11) <control-structure> ::=  <rewriting-rule> |   <connector>  
12) <connector> ::  <functional-operator-construct> (<rewriting-rule>+)
13) <rewriting-rule> ::=  <condition>* <action>+
14) <condition> ::= <functional-operator-construct> (<state>)
15) <action> ::=  <actional-operator-construct>(<state>)
16) <state> ::= <operand-construct>+

---
Legenda: The LOB grammar is expressed in EBNF-like notation; therefore, the symbol `|'  means  `alternative'  ; `<>' means  'variable'; `+'  means `one or more occurrences';  `*' means 'zero or more occurrences'; `?' means `zero or one occurrences'; ``domain values'' are not specified and are the terminal symbols of the Grammar (e.g. True and False).




\end{lstlisting}

\subsection{Some first implications on research}
\label{subsec:implications}

The three-layered architecture described above and depicted in Figure~\ref{fig:architecture} is aimed at addressing the user-centered requirement to provide shop-floor practitioners involved in a digitization program with (at least) the same \emph{space of possibility} they have when they work with non-digitized artifacts, mainly by decoupling such a requirement from those of who could actually initiate such a program: top management (the buyer) would get the services they pay for, i.e., rationalizing the bureaucratic administration of their firm through the informated control of communication and knowledge~\citep{zuboff_age_1988,yates_control_1993}, from the infrastructure-platform stack; on the other hand, end users would get the opportunity to transition from their paper-based artifacts to computationally augmented ones by means of the editing and working environments, so that the layout structures that scaffold their activities~\citep{orlikowski_material_2006} would change with the necessary gradualness (or don't differ at all). At least in theory.

To this \emph{win-win} aim, the model is left purposely flat, general and simple: we do not want to introduce surreptitious entities, like the concept of artifact, activity, task, role and the like, which conversely traditional methods of software production employ as either scaffolding or means for the phase of design. We have already recalled how any design of IT technologies either produces or adopts a model, sooner or later. These models, irrespective of the layers at which they manifest or are adopted, will necessarily end up by conflicting with work practices~\citep[for a recent account on this phenomenon see, e.g.~][]{morrison_multi-disciplinary_2011}, because these latter ``by definition'' change over time and make sense only in their doing (see Section~\ref{sec:performance}), while models as much ``by definition'' introduce the level of representational stiffness that is necessary for their role in requirement elicitation and formal specifications~\citep[e.g., ][]{bowers_janus_1991,robinson_questioning_1991,bannon_cscw_1994}.

Yet, the architecture depicted in Figure~\ref{fig:architecture}, irrespective of its simplicity, requires a radical change of perspective for all the stakeholders involved in technology conception and construction. In particular, this proposal requires to focus on the idiosyncratic and fine grained ways in which users cope with unexpected change in their work environment~\citep{bannon_human_1992}. For IT professionals this means to focus on how constraints have to be dynamically expressed to support the definition of the appropriate ordering of action and interaction in cooperative work in any circumstance~\citep[e.g. ][]{pesic_constraint-based_2007,van_der_aalst_declarative_2009}; which pieces of information are used to support articulation and cooperative work and how these are arranged in suitable artifacts~\citep[e.g. ][]{nemeth_master_2003,cabitza_remain_2011}; what habits, customs and conventions are at stake and silently inform the exchange of information and the sense making occurring \emph{within} and \emph{across} communities of cooperating actors~\citep{mark_conventions_2002,cabitza_leveraging_2009}. In sum, conceiving artifacts (and entangled tasks) as more or less transient ``entities'' emerging from the composition of constructs requires to understand what elementary bricks users \emph{already} have on hand to flexibly compose their artifacts (remember the requirement that bricoleurs already \emph{know} the available pieces) and to conceive of ways to make those bricks computational, that is, associated to specific system behaviors (or functionalities), so that the performative and entangled nature of tasks and artifacts can be preserved and supported.

With respect to a research-oriented agenda, this requires further studies and meta-studies in the same vein of that by~\citet{martin_patterns_2004,cabitza_remain_2011}, which aim to identify recurring and ``universal'' elemental operations/behaviors. Their identification would facilitate the reuse of ways to map either domain- or setting-specific (operator) constructs with the APIs that the common platform has to expose to make the execution of operator constructs possible.
These studies would share the assumption that leveraging general \emph{operator constructs} will not impose users any specific practice or way to treat information (which is mainly represented in terms of \emph{operand} constructs); this assumption seems reasonable first because the identified operations would be intended to be as ``atomic'' and elementary as possible; second, because what could change in any specific setting would be the practices users are familiar to and hence the way users would make sense of and articulate together those basic elements within their work. 

\subsection{For whom tolls the bell?}
\label{subsec:role}

In the previous section we hinted at the fact that an architecture that enables bricolage requires all the stakeholders to reconfigure their traditional roles to make the best use of it within the win-win game that both motivates and pays off the effort of such reconfiguration. But who are the stakeholders that are involved on a practical level? In this section we will just limit ourselves to the ones that are closer to the task-artifact entanglement\footnote{We are aware that buyers, top management executives, middle management officers, more or less official and institutionalized representatives of business units and their employees have always been part and parcel of the development process of a Corporate Information System. Yet, articulating the reconfiguration of the larger actor network that encompasses all these levels of involvement and accountability would be out of the chapter's scope, although a topic of compelling interest for sure.}: traditional frameworks usually denote these as end users, key users, actors from one side of the divide; designers, analysts, programmers and developers, from the other\footnote{We already recalled in Section~\ref{sec:nodesign} that this divide has historical roots, and hence it is contingent and not ontological by all means. In particular, there was a time in which the concept of ``user'' stemmed from that of ``programmer'' (and not vice versa): this happened approximately in the second half of the 1950s (notably when Licklider and Engelbart were writing their seminal essays on the future role of IT) when the computer, which had been that far intended only as a mathematical instrument for which each of its users had to write her own code to executed when it was her turn, became a full-fledged time-sharing equipment and established itself as a business machine, or better yet an electronic data-processing machine~\citep{oneill_evolution_1992,campbell-kelly_computer:_2004}.}.

The former ones are those that are supposed to invest an important effort in \emph{bricolaging} with the system, in the hope that this could be paid off in terms of a better fit between the resulting system and their needs, and of a smaller impact on their traditional coordinative practices and accepted power relationships. To this regard, it is often argued that one should distinguish at least between the regular end user, and the so called ``power user''. This distinction, which can be of some value for purely analytical purposes, should yet be taken with caution if it is to drive the decision of ``what to offer/allow to whom''. The conventional label of power user, far from being used~--~as often is~--~to indicate a role having special rights in modifying the technology (like a sort of administrator, who is distinct from regular users for her ``powers''), should be rather interpreted as originally intended by~\citet{bandini_eud_2006}, i.e., as an organizational or even more informal category that allows to distinguish end users on the basis of their motivations in improving the artifacts \emph{they also use}, and for the competences they have acquired in understanding how things could be changed and why; then, power-users are not the ``chosen ones'' that receive the right to modify the application from above; but rather who, in virtue of their motivations and competencies, are either formally or informally delegated by their colleagues to the aim of taking personal care that the tool continuously evolves and fits the current needs of the community where it is put to work. Therefore, within our perspective, the difference between power and regular user fades into that of \emph{bricolant} user: someone that \emph{can} be factually involved in constructing and developing the bricolage, but that \emph{anyway} also uses it, and hence contributes in building and consolidating related habits and conventions of usage and interpretation; in this light, also the slackest employee, with her implicit resistance or explicit complaint of the inadequacy of her tools, can be said to be an active part of the internalization of a bricolage tool within a cooperative setting, and hence its continuous \emph{accommodation}, for the mutual interdependence between every node of the socio-technical network. For this reason, access to the editing environment should be purposely left to be regulated according to local and socially relevant conventions and initiatives that are just outside the scope of the technology itself.

In regard to the IT practitioners: obviously abandoning the traditional view of rational design does not entail to get rid of designers at all; besides being socially impossible, this sounds also undesirable. Rather, it requires designers, business analysts and IT analysts to focus on different first class purposes and services to supply, as we hinted at in Section~\ref{subsec:implications}. In Section~\ref{sec:bricolage}, we recalled the work by~\citet{hartswood_being_2000} and hinted at the role of designers in terms of \emph{facilitators} of the process of co-construction of both tasks and artifacts. A similar role has been identified, not occasionally, within the approach that proposes Participatory Evolutionary Design as a virtuous integration of EUD and Participatory Design~\citep{sumner_evolution_1997}. The term ``facilitator'' yet must be taken more in the connotation first discussed by~\citet{hirschheim_four_1989}, that is more as that of a \emph{catalyst} within a chemical reaction that really follows unpredictable, and above all, uncontrollable, dynamics; otherwise, the risk is to conceive them as professionals supposed to facilitate the process in which computer-based systems are finally accepted as ``perfect bureaucratic tools''~\citep{harris_better_1999} and adopted within a community of practice or organizational setting. 

In this view, we can detect two main roles involved in IT development from the IT perspective, which are characterized by specific and complementary competences. One, who acts as the catalyst/facilitator mentioned above, could be referred as a \emph{maieuta}-designer\footnote{The pronunciation of this term is quite similar to that of \emph{meta}-designer, not completely by chance (mee'yootah vs. 'mee-tah).}. Although such a word would be pronounced quite similarly to that of meta-designer, its meaning would refer to a quite different thing. \emph{Maieuta} is who performs the art of \emph{maieutics}, the Socratic approach where someone helps \emph{bring out} implicit notions in the interlocutors' beliefs, mainly through a dialogic and narrative way encompassing a series of open questions that do not necessarily require an answer, or just help them further refine their understanding and become more autonomous in their expression. Such a designer is primarily concerned with the front-end of the enabling technology, i.e., with the graphical and semiotic aspects of the artifacts ``to be built and to be used'' through it; in some way, also, the \emph{maieuta}-designer is who is supposed to ``close'' the design process of the merely technological part and to \emph{pass the baton on} the end users, i.e., (the bricoleurs), by helping them in finding the ways and motivations for the ``in vivo'' development of their artifacts by their own. As such, the \emph{maieuta}-designer does not have to possess strong programming or architectural skills: rather she has to be a domain expert, a connoisseur of how the users of a particular domain (if not particular setting) are used to conceiving their tools and tinkering them over time (to what ends, on the basis of what political and cultural drives and constraints, and the like) in order to help users in exploiting the available editing environment in such a way that their bricolage does not become an erratic process but rather it is \emph{sustainable} over time. 

Typical technology-oriented competencies must be conversely mastered by the IT professionals working on and developing the platform itself, or who we could denote as back-end designers and programmers: these are called to the role of guaranteeing that the artifacts built on top of the platform can evolve over time, that is that the enabling environments are easy to use for as many \emph{actual} users as possible (and not just for few management officers). This means to guarantee that the best software engineering techniques (e.g, modularity, integration of data and routines) are employed to make the platform and exposed environments powerful and flexible enough to allow for the bricolage activities at the higher level layers of the overall architecture, besides guaranteeing also that the platform is modular and robust enough to cope (and align) with (low rate) changes in the underlying infrastructure. We could say that ``designing for unanticipated construction'' could flank (or perhaps substitute) the old claim for ``designing for unanticipated use'' by~\citet{robinson_design_1993}.

\subsection{What's outside like this?}
\label{subsec:rw}

We have claimed that our proposal is not original in any strong sense, but rather it tries to reinvigorate a discourse among the current mythologies of system design that conceives the progressive delegation of power and control to end users as a feasible way to cope with increasingly complex socio-technical systems, supported by increasing complicated IT systems~\citep{latour_interobjectivity_1996}. 

In the recent years a small number of frameworks have been developed to exhibit at least some of the more relevant characteristics of the architecture presented above. Among these we obviously mention the particular document-based information system platform that we are developing in the last few years, called Web of Active Documents~\citep[WOAD, ][]{cabitza_woad:_2010,cabitza_web_2011}. This is a platform endowed with two editors, one for the construction of rule-based control structures~\citep[therein called mechanisms, ][]{cabitza_providing_2012} and one for the construction of operand constructs~\cite[atomic data structures called datoms, ][]{cabitza_tailorable_2011} and their spatial arrangement in document templates, which provides also an execution environment that has been currently under experimentation in the healthcare domain~\citep{cabitza_woad_2011,cabitza_rule-based_2012}. 

The core concepts of WOAD can be summarized as follows in terms of: i) the information system is parcellized in a set of hyperlinked active documents that can be annotated in every parts and sections and be associated with any other document, comment and computational behavior; ii) there is no rational and unified data model: users define their forms in a bottom up manner and, in so doing, the platform instantiates the underlying flat data structures that are necessary to store the content these forms will contain and to retrieve the full history of the process of filling in them; iii) the presentation layer is in full control of end users, who are called to both generate their own templates and specify how their appearance should change later in use under particular conditions; iv) execution control is rule-based. Users can define local rules that act on the documents' content and, as hinted above, change how documents look like (i.e., their physical affordances), to make themselves aware of pertinent conditions according to some cooperative convention or business rule like, e.g., the need to revise the content of a form, or to consider it provisional, or to carefully consider some contextual condition\footnote{see e.g.~\citet{cabitza_leveraging_2009} for other examples of such conventions.}.

Although WOAD has been natively conceived to allow for the degree of flexibility and user autonomy that we described at length above, the specialist literature reports also other platforms and frameworks that exhibit similar features. For instance, Placeless Documents~\citep{dourish_extending_2000} introduced the idea of document properties that are attached by single end users and, above all, properties that represent active ways to operate with documents (called active properties): users can add these properties to documents to make them carry executable code that can be invoked to control or augment their functionalities. This work, to our knowledge, was among the first ones to carry into the HCI scientific arena notions from the prototype-based object-oriented programming and operating system programming, like that to attach code to documents as a means to control their behavior and the idea to let users develop some bunches of runnable code to extend the system functionalities. Also in this case, users have a relatively small set of operations to endow their documents with to make them active, but are forced to conceive for any document all those operations that could be invoked about and upon its content. WOAD, instead, aims to decouple layout structures from control structures, although these can be related to each other by means of if-then mechanisms defined over the document's content. 

Enabling end users to build their own documents is a common trait of recent initiatives of visual data-driven form generation; these projects are usually aimed at allowing users to generate even complex forms, intended as data-entry points to an underlying flat data structure, without particular programming skills, e.g., by means of a visual editor like Microsoft\textsuperscript{\textregistered} InfoPath\textsuperscript{\textregistered} as described in~\citep{mamlin_cooking_2006}, or by means of the Layout Mode in FileMaker Pro\textsuperscript{\textregistered} as used in~\citep{chen_developing_2011}. This allows to take ``form design out of the programmers' hands and put it into the realm of content management, much as form-generation tools (like Ruby on Rails or Plone's Archetypes) aid the developer in rapidly generating forms''~\citep{mamlin_cooking_2006} and hence to address a specific need that so far has been raised especially by practitioners in the hospital care domain~\citep[e.g.,][]{mamlin_cooking_2006,morrison_observing_2009,chen_developing_2011,cabitza_woad_2011}. The system described by~\citet{mamlin_cooking_2006} presents also the feature to associate form elements with specific rules (expressed in the Arden Syntax), making this similar to WOAD, although the editor defined in this latter framework allows for the reuse of form components (i.e., the datoms) and for the above mentioned decoupling between layout and the logic-based control flow of execution (datoms vs. mechanisms). 

In the same vein, some effort has been paid by researchers to address the requirement of making end user really autonomous in creating their document-specific rules, and this is usually enabled by means of visual and user-friendly tools~\citep[e.g., ][]{cabitza_rule-based_2012,krebs_combining_2012}. An even more comprehensive approach to this general aim has been recently proposed by~\citet{harel_come_2003} in the Play framework: this latter allows users to build reactive systems by \emph{playing}, so to say, their specifications in a performative way, that is through scenarios that are subsequently implemented by means of a Play-Engine that ``plays out'' the corresponding models of interaction; these are explicitly represented in terms of multi-step control structures, what the authors call ``live sequence charts''. These are hence way more complex and articulated interaction structures than simple rules are, and we have discussed above how this is not necessarily a good thing to cope with unknown emergent behaviors. Yet, in Play it is how users can specify these models of human-computer interaction that is innovative and peculiarly aligned with some of the tenets we discussed above: end users can interact with prototype user interfaces and have the system build the corresponding structures, or write the intended behavior and its main exceptions handling procedures in brief sentences expressed in natural (yet structured) language, or even tell it directly to the system, in a sort of versatile multi-mode way to teach the system what to do if some events occur (typically at interface level). 

The Play framework has been specifically proposed as a concrete first step toward what~\citet{harel_can_2008} suggestively calls the \emph{liberation of programming from its three straightjackets}: these are the ``1) need to write down a program as a symbolic, textual, or graphical artifact; 2) the need to specify requirements (the what) separately from the program (the how) and to pit one against the other; 3) the need to structure behavior according to the system's structure, providing each piece or object with its full behavior'' (p. 29). The aim to liberate programming from these representational straightjackets resonates in very close affinity with the tenets of a EUD approach to supporting end users perform bricolage in their situated practices like the one we described in this chapter and seems to go in the direction of drawing a common agenda where practitioners from different disciplines like the software engineering field, Human Computer Interaction and CSCW can perhaps meet together and inform their own research and development initiatives in a positive manner.

%

\section{Some final remarks for future discussions}
\label{sec:remarks}

In this section we will just outline two important aspects that should be object of future research (or discussion) about the concrete applicability of the \emph{laissez-faire} method and of the \emph{logic of bricolage} (see also Listing~\ref{here_the_lbl}) in the development and evolution of socially embedded systems. These two topics regards important strands of research that are receiving a crescent interest from diverse communities of researchers involved in studying the impact of IT in social settings in the last years. We broadly denote these two topics as ``concerns about risk'' and ``concerns about interoperability'' (and hence standardization).

\subsection{How risky is a different development strategy?}
\label{subsec:risk}

One could detect a harsh irony in our advocacy of a \emph{laissez-faire} approach (that is, of a \emph{no method}) for the construction of an efficient, effective and safe technology, whereas it has been the need to guarantee such qualities that motivated the consolidation of engineering methods and methodology in IT system construction (see Section~\ref{sec:nodesign}). This feeling could be indeed reinforced by our explicit confidence that an environment enabling and supporting \emph{ateleological} bricolage by end users could be a feasible alternative to any *-design of those systems. Indeed, in the specialist literature there is a consolidated tendency in considering bricolage akin to improvisation~\citep{weick_organizational_1993,lanzara_between_1999} and the bricoleur as someone, \emph{at best}, who draws on the materials at hand to create a response to a task \emph{on the spot}~\citep{levi-strauss_savage_1966}; in the Lanzara's words: ``in a broadly diffused \emph{engineering ideology}, bricolage is usually associated with second- best solutions, maladaptation, imperfection, inefficiency, incompleteness, slowness''. This prejudice is renovated several times within the system development discourse.

One reason for that is to be found in the misunderstanding coming from understating the radical novelty that the myth of the bricolant/bricoleur carries with itself. In fact, as long as bricolage is evoked in discourses that still refer to a traditional way to build computational artifacts or that even just spare the traditional wording (e.g., design \emph{with} users, \emph{meta-}design), that is as long as bricolage is ingrained in a traditional way to \emph{think} of IT design (even in the light of contributions coming from participatory design, action research, and ethnography), we keep undermining the \emph{original} sense of this concept in some (important) way and, worse yet for our aims, weakening its full potential to move into a new more useful mythology. 

From what we argued at length in Section~\ref{sec:bricolage} and then in this section, it should be clear that we stress the ability of the bricoleur to ``work and play with the stock [with] parts that are not standardized or invented, [but rather] appropriated for new uses''~\citep[pp. 161-162]{weinstein_george_1991}, and that are taken from ``an inventory of semi-defined elements [that] are at the same time abstract and concrete [and that] carry a meaning, given to them by their past uses and the bricoleur's experience, knowledge and skill, a meaning which can be modified, up to a point, by the requirements of the project and the bricoleur's intentions''~\citep{louridas_design_1999}: exactly what we above called constructs. In this compositional mythology then, the concept of \emph{bricolant} end user should not be seen any longer as an ``improviser'', a tinker, hobbyist or hacker in the negative connotation of these terms\footnote{ This latter connotation has been unfortunately preserved, if not even reinforced (probably against their intentions) by who introduced such a concept in the IT discourse, often for some sort of patronizing empathy with the revolutionary and yet ill-organized attitude that is typical of bricoleurs in front of very well structured work contexts. As paradigmatic of this risk, we can mention Ciborra, who is certainly one of the most inspired researchers that in his works referred to and discussed several times the idea of embedding bricolage strategies in organizations;
\begin{quotation}
As soon we leave the realm of method, procedure, and systematic ways of organizing and executing work according to rational study, planning, and control we enter the \emph{murky world of informal}, worldy, and everyday modes of operation and practices. It is the real of \emph{hacking}; practical intelligence; the \emph{artistic embroidery} of the prescribed procedure; the \emph{shortcut and the trasgression }of the established organizational order as embedded in systems and formalized routines. Bricolage, \emph{improvisation} and hacking [...] all seem to share the same way of operating: small forces, tiny interventions, and on-the-fly add-ons lead, when performed skilfully and with close attention to the local context, to momentous consequences~\citep[p. 47-48]{ciborra_labyrinths_2002} (our emphasis).
\end{quotation}
Thus, notwithstanding his good intentions, and acknowledging the context in which Ciborra provided his argumentations, i.e., a context in which information systems were rationally designed, digitization programs strictly planned and work activities tightly ordered (!), we find that his purposely ill-concealed sympathies for this guerrilla-like and spontaneous, istinctive, almost extempore resistance of revolutionary users that sabotage the megamachine~\citep{latouche_megamachine_2004} with smart workarounds and interventions ``that diverge from the formalized, pre-planned ways of operating''~\citep{ciborra_labyrinths_2002}, and his sympathizing and benevolent gaze to something that Ciborra himself acknowledged being perceived derogatorily and ``sanctioned as marginal, belonging to the red light districts of the organization'', all that ended up by undermining the actually revolutionary potential of the concept of bricolaging, certainly against his intentions.}; but rather as a creative actor who is \emph{exploited} (however harsh this term could seem) by the sponsors of the digitization initiative to reach their purposes in the awareness that \emph{only} end users are competent \emph{enough} to reach a sustainable balance between effectiveness and efficiency in cooperative ambits and, above all, to meaningfully improvise in face of the unexpected, in virtue of the fact that users, communities and organizations \emph{already} exist (i.e., they pre-exist digitization and informating initiatives) and \emph{already} possess ``a disposition towards their environment [...,] already [are] committed in a self-meaningful manner towards [their] own survival and prosperity''~\citep{angell_dispositioning_2009}.

Supporting bricolage then is not the slothful retreat of the blas\'{e} researcher that releases responsibility advocating a more empowered and active role of end users in virtue of her democratic feelings. Rather it is the ultimate strategy to make the complex socio-technical systems that our IT solutions contribute to set up more \emph{resilient} in face of the \emph{normality of accidents}.
More than this, we submit that an architecture that adopts a \emph{laissez-faire method} and the \emph{logic of bricolage} described above could be said to be both intrinsically resilient and evolutive, two terms that are attracting more and more interest by researchers involved in the safety of IT systems~\citep{hanseth_risk_2007} especially in critical or delicate settings, like, e.g., healthcare~\citep{hollnagel_resilience_2008}. These two concepts regards the capability of a system to react or change to unexpected events, changes or conditions at different time scales, respectively short and long with respect to the occurrence of unexpected event. Being both \emph{resilient}, that is able to reach a stable and safe working state after that some unexpected event has occurred, and \emph{evolutive}, that is able to grow and increase one's own fit with respect to the surrounding environment, is a fundamental characteristic of socio-technical systems to properly face the increasing odds of failure of some of their multiple components as also a function of the increasing complexity of their interrelated parts and corresponding links. This characteristic also constitutes a progress within the rhetoric around safety with respect to the concept of robustness; this refers more to the capability of a system to resist and withstand adverse events and change, also in virtue of design and analysis phases usually aimed at identifying, prioritizing and handling specific exceptions, or at reducing the opportunity for their occurrence.

The architecture we envision above is intrinsically resilient as it, paradoxically as it can seem, delegates to end users the burden and responsibility to react to unexpected events by leveraging their innate creativity and the invaluable, and often irremediably tacit, knowledge of the overall system dynamics, sometimes much more based on intuition than on rationality~\citep{mark_resilience_2008}; and because such an architecture purposely avoids to provide users with information and execution structures that are constrained and articulated on the basis of strong assumptions on how the system will behave under certain conditions: a bricolage-based information system is just an environment for both the human and automated manipulation and processing of signs\footnote{If the worst thing occurs, e.g., if power goes down, the overall socio-technical system is made more resilient simply by printing down some layout structures on paper and have the users work as usual, just without the computational augmentation of those structures.}. Moreover, interactive systems that are built on top of such an architecture are naturally and concretely open to evolution, even more than many other ones as they, differently from those systems that are built by someone else than their actual users, are changed opportunistically by end users on their own when (or in short time after that) they feel this necessary to \emph{accomodate} their artifacts and tools to emerging conditions and newly recurring situations.

Again in the Lanzara's words:
\begin{quotation}
[\dots], systems assembled by bricolage have an evolutionary advantage: being loosely connected and incoherent assemblies of mixed components, they can be partially reworked without much investment effort. Bricolage is a design strategy that makes sunk costs recoverable. In case of system's depletion, obsolescence or low performance, regeneration can be done without having to throw away the whole structure. [\dots]
As a consequence, systems are persistent and robust because cannot be changed or moved easily, but at the same time keep structural plasticity and exhibit some self-correcting properties. Innovation can be accommodated locally. [\dots] As [bricolage] exploits the properties of existing structures for interactive and generative purposes, it successfully mediates the dilemmas of change and stability, innovation and conservation. On the one hand, by experimenting with transient constructs it allows for some variability and improvisation without incurring in the possible disruptions caused by excessive instability and radical change; on the other hand, by assembling robust but furtherly manipulable structures, it allows for some order and reliability without curbing the chances for system improvement and innovation. In short, it makes both radical innovation and complete unraveling unlikely. (p. 347)
\end{quotation}

We subscribe this understanding of the role that an active, conscious and responsible \emph{bricoler} can play in system development, but also associate this with the awareness that new platforms must be built supporting this role, and new professionals must be educated to mediate between the possibly conflicting stances of such an empowered actor and other roles that are in a hierarchical relationship with it.

\subsection{Interoperability Concerns}
\label{subsec:inter}

The second issue we propose for future research and discussion regards the tension between the global dimension and the local one in the construction of a technology aimed at supporting whole organizations and/or networks of these latter; this is therefore the extension of what we have discussed so far in terms of collaboration within groups of limited size, what we call the local dimension, in the new terms of interoperability ``in the large'', i.e., across multiple settings and organizations. The separation between global and local has been already shown as illusory by who have recently proposed the concept of information infrastructure~\citep[e.g., ][]{ellingsen_integration_2012}, as it is recognized that any local event or practice can have the potential to affect the overall system, sooner or later (cf. the butterfly effect that is often associated with complex systems).

In Section~\ref{sec:bricolage} we have very briefly hinted at how this separation can be just seen as one way by which the interests of someone (typically institutional authorities of control and high-level regulatory bodies at Regional or National level) are imposed on the practices of others (typically the producers of inscribed data and their first consumers for articulative reasons~\citep{berg_accumulating_1999,winthereik_ict_2005,cabitza_whatever_2012}). Consequently, addressing the tension between local vs. global requirements means to also address one of the main root causes behind the pervasive (and still relatively neglected) phenomenon usually denoted as ``work around''; such a circumvention of the system and its intended (and designed) uses which users perform to overcome the rigidity that the global view of ISs imposes on their local practices can indeed undermine any serious attempt to engineer and deliver safe and robust technologies in socio-technical systems~\citep{niazkhani_evaluating_2011,handel_working_2011}. In this section, we will outline how a \emph{laissez-faire} method can be compatible with the increasing need for global interoperability between local systems and, in its little own, can contribute in going beyond the above mentioned tension by interpreting such seeming opposition (see Section~\ref{sec:nodesign}) in a more dialectical way towards the development of information systems that can be flexible enough to meet the local needs of data producers', i.e., the \emph{primary} users of any information~\citep{cabitza_whatever_2012}, as well as the needs/expectations of some relevant data consumers (i.e., \emph{secondary} users) at the same time.

A typical solution adopted in the organizational domain is to solve the tension between global and local needs by building technologies that are conceived to be part of a \emph{global infrastructure} aimed at becoming the backbone enabling an integrated/federated (sometime also called cooperative) management of data and business processes across multiple systems: the aim of these initiatives is to rationalize data schemes and business processes under a unified program that guarantees that the requirements that really counts at global level are met thanks to a centralized control. If the process requires a progressive construction of smaller components, then the main issue at stake becomes how to guarantee interoperability among these components through the definition and enactment of standards of both data classifications and protocols for the exchange of data and control structures, respectively. This trend characterizes recent attention to the so called semantic Web and ontology-driven orchestration of services and processes.

Yet, as aptly recalled by~\citet{lanzara_between_1999}, ``anthropologist Clifford Geertz has pointed out that the more we try to make the world ``global'' the more the world responds with the emergence of multiple ``local worlds'' and identities that seem to be irreducible to one another (Geertz, 1996). In the same line, but with a more technological perspective,~\citet{ciborra_thinking_1992} suggests that ``Top management needs to appreciate local fluctuations in system practices as a repository of unique innovations and commit adequate resources to their development, even if the systems go against traditional approaches. Rather than looking for standard models in the business strategy literature, [strategic information systems] should be sought in the theory and practice of organizational learning and innovation, both incremental and radical''.

We are not interested here in considering the innovation issue mentioned in this latter passage. Nevertheless, we believe that this passage offers an interesting stimulus to go beyond the way in which firms conceive their (strategic) information systems.  Actually, letting information systems ``[\ldots] emerge from the grass roots of the organization, out of end-user hacking, computing, and tinkering''  asks for a significant change of perspective in IT design that is closely related to the performative and bricolage-oriented stance we advocated in this chapter. Since ``organizational learning and innovation'' occur \emph{where} practices are and action is~\citep{dourish_where_2001}, we subscribe the suggestion by~\citet{ciborra_thinking_1992} and ~\citet{lanzara_between_1999} to start from this ``local dimension'' to build a technological support that promotes firms' vitality not only for the sake of innovation but also for an effective every day performance. Moreover, since ``learning and innovation''  are ``both  incremental and radical'', the requirement of having different layer dynamics (or change rate) that was discussed in Section~\ref{sec:environments} can be put in relation with the need to cope with incremental and radical changes in the organization itself and in the co-evolving technology.

The idea we would like to be considered, discussed and above all experimented regards the ``whole'' not as a centralized and monolithic entity (to some extent); but rather as the composition of ``small'' entities, highly specialized to the local needs, tightly connected in a web of loose connections interconnecting ``local spaces'', i.e., peer nodes that are each characterized by local structure and semantics~\citep{bandini_www_2007} and that are easily adaptable to unexpected contingencies since they are concerned with (and hence control) a limited and well known ``piece'' of the world. Obviously, soon a need to make this new kind of ``whole'' coherent and consistent with the overall good performance of the network would arise, which is the main concern of any higher-level management unit. Yet, instead of having the management promote (and enforce) coherence in terms of ``obtrusive'' control of local behaviors and top-down ontologies to order resources, the management unit itself should behave as a ``local entity'' with specific goals and expectations on data produced by other units, expressed in terms of its own local quality level constraints. These latter can be exposed as public \emph{expectations} that other units can (or have to) comply with to interoperate, once either the producer or the consumer have selected what data structures are involved by these constraints and how these data have to be arranged and aggregated to respect the consumer's needs. 

\begin{figure}[tbh]
  \centering
      \includegraphics[width=\textwidth]{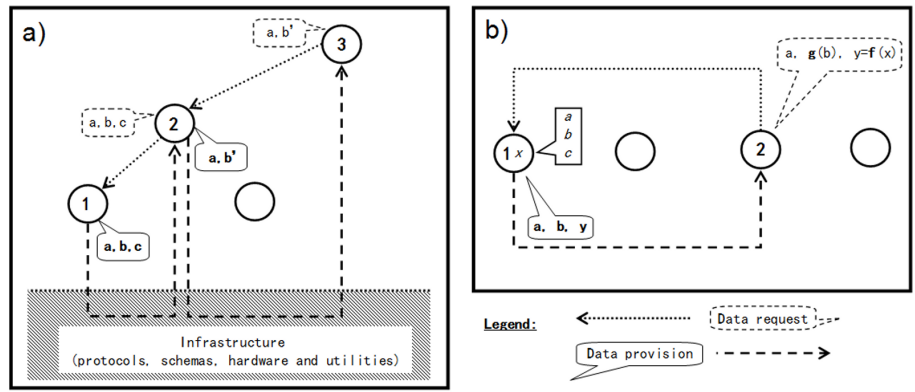}
  \caption{Two ways to reach interoperability between data producers and consumers: \textit{a)} through standardizing infrastructures; \textit{b)} through peer-to-peer negotiations and ad-hoc specifications.}
  \label{img:global}
\end{figure}

As we discussed in~\citep{cabitza_whatever_2012}, interoperability can be achieved in two ways: either by having higher level entities require (and obtain) data from data producers (e.g., node 2 asking for data a, b and c to node 1 in Figure~\ref{img:global}.\textit{a}, and node 3 asking for data a and b' to node 2 in the same figure), which is the traditional way where the global dimension of a data provision is obtained through a sort of trans-lation of locally produced locally data into a standardizing infrastructure (by means of imposed protocols, schemas and formats); or having consumer entities (which need data for secondary purposes) declaring interest in specific data sets that are being produced by some producer unit (for their primary purposes, see the letter sequences in the dashed balloons in Figure~\ref{img:global}, in both box \textit{a} and \textit{b}), obviously in the assumption that a list of available data are shared by these latter ones (see the sequence of letters in the square balloon in box \textit{b} of Figure~\ref{img:global}); the point-to-point provision between consumer and producer can be then enriched by two kinds of mechanisms: \emph{presentation mechanisms} (see the function \textbf{f}($\cdot$) in Figure\ref{img:global}.\textit{b}); and \emph{affording mechanisms} (see the function \textbf{g}($\cdot$) in Figure\ref{img:global}.\textit{b}). The former kind of mechanism is a functional specification of the consumer about how this would like to receive the raw data packed up by the producer, in terms of specific aggregation, basic processing and reporting, in order to have those data more meaningful (or convenient) at its side. Obviously, this function can be applied by the consumer itself once the producer has supplied the input data, but most of the times it will be processed by the producer directly on the basis of data that are not necessarily exposed publicly (in Figure~\ref{img:global}.\textit{b} see data x, which is not published by node 1 but in some sense requested by node 2 in terms of a functional specification of data y, \textbf{f}($\cdot$).)
Affording mechanisms, as presented in~\citep{cabitza_whatever_2012}, are mechanisms dispatched by requesters in order to convey minimal expected quality levels to producers and make them aware of quality constraints that, if not complied with, could seriously undermine the secondary use of the requested data, e.g., to make unit 1 (in Figure~\ref{img:global}.\textit{b} understand when archival or statistical purposes at unit 2 are partly or totally undermined by their otherwise perfectly reasonable (yet idiosyncratic) ways to produce and handle data locally. In so doing, local units can maintain the full autonomy, as well as being endorsed by an explicit accountability and liability contract, with respect to how to fulfil data requests/requirements. This approach is naturally scalable with respect to higher-level management units, even outside the organization itself, whereas compliance with presentation mechanisms can be more or less mandatory at political level (still, not embedded at technological level).

In other words, the bottom up construction of (information) structures that we have mentioned in Section~\ref{sec:environments} are then one way to deal with interoperability, as the requests from one local entity (i.e., the consumer) can be expressed as a sort of unstructured lists of requested items from those structures~\citep[as depicted in the squared balloons in Figure~\ref{img:global}.\textit{b}][]{simone_adaptability_2001}, as well as of sets of ordering/aggregating policies on these latter~\citep[see dashed balloons in Figure~\ref{img:global}.\textit{b}][]{cabitza_whatever_2012} that another local entity (i.e., the producer, node 1 in Figure~\ref{img:global}.\textit{b}) can employ to process/package those items and send them to the requester (node 2 in Figure~\ref{img:global}.\textit{b}). These data (or reports obtained by applying the presentation mechanisms) can be then reinterpreted by the receiving entity without the need to know anything (i.e., their situated meaning, let alone their upper-level meaning) about the sender' structures. 

In this scenario, interoperability is achieved not by semantically enriching data at the producer's side, i.e., where those data are not natively attached with that semantics, but rather by having consumers pragmatically express what data they need, subscribe to a set of data that are exposed by producers and, possibly, express also \emph{how} they need those data be presented (i.e., reported, typically in aggregated form). In this view, the definition of ``standard structures''~\citep[which play the role of boundary objects][]{star_sorting_1999}) end up by regarding much simpler pieces of information, i.e., a subset of the operand constructs of one unit that this latter exposes to the outside world (or to specific correspondent units, typically of higher level in a social hierarchy); as said above, these latter items can be considered as changing with a low speed rate, since they play a sort of \emph{minimum data set} that characterizes the unit at hand, but they are in any case easily modifiable since they are not universally (and hence not rigidly) incorporated in more complex structures. In this view, consistency has to be obtained by an effective monitoring of the received structures, instead of a prescriptive way on how to achieve interoperability about them.


\section{Conclusions}
\label{sec:conclusions}
\begin{quotation}
Let every careful man be very far from writing about things truly worthy of care.\\
(Plato, 352 BC)\footnote{Seventh Letter}
\end{quotation}
The main claim of this paper is that, in order to bridge the gap between what users need and what is given to them as solutions to those needs, the concept of design has to be substantially challenged and its role in IT development reformulated. To this aim, we submit that an old mythology of design, which is based on the separation between conceptual design and situated use, and consequently on the modelling activity that entails this separation (see Section~\ref{sec:nodesign}), should be abandoned in favour of a new mythology; we advocate this new mythology be grounded on both the notion of \emph{performativity}, from the conceptual perspective, and on the notion of the \emph{bricolant} end user, from the more practical perspective. Reviewing the main tenets of this mythology has brought us to introducing a lean method for the development of socially embedded technologies, epitomized by the motto ``\emph{laissez faire les bricoleurs}'' and the preliminary proposal of a ``\emph{logic of bricolage}'' that specific environments should enact to empower end-users in the process of development of their tools.
Quite contrarianly with respect to whom welcomes the increasing blurring between the roles of designers and users~\citep[e.g.,][]{fischer_meta-design:_2004,johannessen_users_2012}, we do not advocate the idea by which users should increasingly ``act as designers'' (and researchers work to that aim), as such an idea would foster the approach by which users adopt a spurious attitude and end by putting themselves out of the practice, even if temporarily: in so doing, it is a short step that users become people who think to ``design her own practices'' as well~\citep[as it is claimed even recently by][]{johannessen_users_2012}. Conversely, the role of IT professionals and of end-users has to be characterized by a clear separation of concerns in the development of computer-based supports of cooperative, organizational work: hard engineering-based \emph{design of meta-systems} for the former ones, \emph{bricolag}-ing for the latter ones. Indeed, we submit that this separation, that is at least conceptually (if not also pragmatically) opposite to proposals that advocate a tight integration between, if not a unification of, conceptual design and end-user practices (e.g., the participatory design and the meta-design frameworks) can have the advantage to make the relationships among these two roles not only less harmfully ambiguous, but also and above all, more productive with respect to the timely and effective deployment and maintenance of computational artifacts, since end-users are in full control of this process, which many contributions in the specialist literature recognize as highly situated and intertwined to social aspects that can not really be untangled from the technical ones.

As we have mentioned in Section~\ref{sec:environments}, there are examples of technologies that can be interpreted as steps toward the goal of letting users be in true control of the technology they think to need and wish to use. At a cursory glance, these technologies implement different kinds of platforms that enable users to perform bricolage-like activities in which the pieces that these platforms make available are composed and arranged in meaningful ways. However, there is still a long way to go, in order to collect findings that would confirm that there is an alternative way to design than the conceptual and representational one that is currently ruling in our development methodologies and professional practices. In order to make this journey effective, we believe that the main future steps should reconsider some aspects that still characterize the robust stronghold of the mainstream mythology. 

Apart form any technical consideration, this approach would require a substantial change in the way young IT-related students are educated to information system design: this is traditionally based on a plethora of data models, from the business-oriented one (the conceptual model) up to the more machine-related one (the physical model), and on a collection of business process models and notations by which to describe how work is (or should be) carried out on those data. However, from a broader perspective, the main role that we have advocated for the development of collaborative applications and information systems, namely the role of the facilitating \emph{maieuta-designer}, would instead require an educational agenda that is quite different and way much less consolidated and agreed upon than the one so far conceived for the role of the system developer and programmer: this would encompass, for instance, teaching the basics of social informatics~\citep{kling_understanding_2005} and semiotics~\citep{de_souza_semiotic_2009}, some qualitative research methods adapted to the IT domain~\citep{kling_understanding_2005}, insights on current theories on IT impact and risk management~\citep{hanseth_risk_2007}, as well as notions of socially-informed history of technological evolution~\citep{akera_using_2004}. The point is that all three of the roles we discussed in Section~\ref{subsec:role}, namely the user\footnote{We consider also the user in this point, because also potential end-users should be invited to see technology as a convivial and creative tool to build and maintain over time, in their active responsibility, rather than as an immutable (yet mobile) commodity to passively use or as a consumerist gadget to possess and use well below its full potential.} (as expert of the setting), the facilitator (as domain expert) and the IT developer (as expert of infrastructural concerns) should receive a newly formulated or seriously revisited educational program so that an effective way to take ``human actors'' seriously can be promoted again~\citep{bannon_human_1992}. 

On the other hand, the layered conceptual architecture that we have illustrated in Section~\ref{sec:environments} has still to prove its practical value, feasibility and efficacy in a reasonable range of application domains (or settings): our personal research experience makes us confident that such an architecture is promising for the case of document-based, knowledge-intensive collaborative (information) systems; although many such systems can be found in the world out there, we are aware that this macro-class of applications simply does not cover all IT-based supports.  In any case, in order to go a step further in this direction, better platforms and environments supporting an effective and reliable EUD approach are needed; we have proposed some basic principle on which these systems could all be based on: decoupled modularity of information and control structures; loose integration of the latter ones in terms of recomposition of elemental common constructs according to local needs (as opposed to the construction of unifying general schemes); full homogeneity across the layers for the construction of aggregated functionalities so that users can access them and operate with them with the same high-level language; and finally, tools supporting the technology development and managing its intrinsic complexity that are based on users' building practices (vs. the introduction of more or less ``hard'' engineering tools in EUD).



The three brief (and necessarily partial and unbalanced) outlines of current discourses on complexity, performativity and bricolage, as well as the novel (or not so novel) contributions of ours, like the little tale on the historicity of conceptual design, the notion of task-artifact entanglement, the requirement of universatility for EUD environments, the concept of the bricolant/bricoleur end-user, the foundation of a logic of bricolage, the distinction between \emph{maieuta}-designers and traditional designers (also on an educational level), the laissez-faire method as conscious way to cope with socio-technical complexity, the peek to the local/global illusion, these are all provided as pieces of a bricolage. Like any bricolage, we do not see a particular truth in any of the pieces we brought together in this chapter; rather we have argued about the potential of the resulting jigsaw puzzle, in our opinion coherently kept together by the mythology of the performative end-users, as a whole to come out being use-ful. Obviously our last hope is that the EUSSET forum will host many similar discourses, among which those picked up and assembled in this chapter, and give them some sort of legitimacy to inform future common initiatives of research, education and IT professional practice.





\begin{thebibliography}{193}
\providecommand{\natexlab}[1]{#1}
\providecommand{\url}[1]{{#1}}
\providecommand{\urlprefix}{URL }
\expandafter\ifx\csname urlstyle\endcsname\relax
  \providecommand{\doi}[1]{DOI~\discretionary{}{}{}#1}\else
  \providecommand{\doi}{DOI~\discretionary{}{}{}\begingroup
  \urlstyle{rm}\Url}\fi
\providecommand{\eprint}[2][]{\url{#2}}

\bibitem[{{AA.}(2001)}]{vv._aa._extreme_2001}
{AA} V (2001) Extreme chaos. Tech. rep., The Standish Group International,
  Inc., \urlprefix\url{www.cin.ufpe.br/~gmp/docs/papers/extreme_chaos2001.pdf}

\bibitem[{van~der Aalst et~al(2009)van~der Aalst, Pesic, and
  Schonenberg}]{van_der_aalst_declarative_2009}
van~der Aalst WMP, Pesic M, Schonenberg H (2009) Declarative workflows:
  Balancing between flexibility and support. Computer Science - Research and
  Development 23(2):99--113, \doi{10.1007/s00450-009-0057-9},
  \urlprefix\url{http://www.springerlink.com/index/10.1007/s00450-009-0057-9}

\bibitem[{Akera and Aspray(2004)}]{akera_using_2004}
Akera A, Aspray W (eds)  (2004) Using History To Teach Computer Science and
  Related Disciplines. Computing Research Association

\bibitem[{Anderson et~al(2008)Anderson, Hardstone, Procter, and
  Williams}]{anderson_down_2008}
Anderson S, Hardstone G, Procter R, Williams R (2008) Down in the
  {(Data)base(ment):} supporting configuration in organizational information
  systems. In: Resources, Co-Evolution and Artifacts, Computer Supported
  Cooperative Work, Springer London, pp 221--253,
  \urlprefix\url{dx.doi.org/10.1007/978-1-84628-901-9_9}

\bibitem[{Angell and Ilharco(2009)}]{angell_dispositioning_2009}
Angell I, Ilharco F (2009) Dispositioning {IT} all: A theory for thriving
  without models. In: Bricolage, Care and Information Claudio Ciborra's Legacy
  in Information Systems Research, Palgrave Macmillan, pp 401--422

\bibitem[{Ardito et~al(2012)Ardito, Costabile, Matera, Piccinno, Desolda, and
  Picozzi}]{ardito_composition_2012}
Ardito C, Costabile MF, Matera M, Piccinno A, Desolda G, Picozzi M (2012)
  Composition of situational interactive spaces by end users. In: {NordiCHI}
  2012: proceedings of the 7th nordic conference on human-computer interaction:
  making sense through design, October 14th-17th, Copenhagen, Dk, {ACM} Press

\bibitem[{Ash et~al(2004)Ash, Berg, and Coiera}]{ash_unintended_2004}
Ash JS, Berg M, Coiera E (2004) Some unintended consequences of information
  technology in health care: The nature of patient care information
  system-related errors. Journal of the American Medical Informatics
  Association 11(2):104--112, \doi{10.1197/jamia.M1471},
  \urlprefix\url{http://www.sciencedirect.com/science/article/B7CPS-4BRRG2S-6/2/7acb3f613cd174d18c99d2610662807b}

\bibitem[{Axelrod and Cohen(1999)}]{axelrod_harnessing_1999}
Axelrod RM, Cohen MD (1999) Harnessing complexity : organizational implications
  of a scientific frontier. Free Press, New York

\bibitem[{Balasubramaniam et~al(2008)Balasubramaniam, Lewis, Simanta, and
  Smith}]{balasubramaniam_situated_2008}
Balasubramaniam S, Lewis G, Simanta S, Smith D (2008) Situated software:
  Concepts, motivation, technology, and the future. Software 25(6):50 --55,
  \doi{10.1109/MS.2008.159}

\bibitem[{Bandini and Simone(2006)}]{bandini_eud_2006}
Bandini S, Simone C (2006) {EUD} as integration of components off-the-shelf:
  The role of software professionals knowledge artifacts. In: End User
  Development, vol~9, Kluwer Academic Publishers, pp 347---369

\bibitem[{Bandini et~al(2007)Bandini, Sarini, Simone, and
  Vizzari}]{bandini_www_2007}
Bandini S, Sarini M, Simone C, Vizzari G (2007) {WWW} in the small towards
  sustainable adaptivity. World Wide Web 10(4):471--501,
  \doi{10.1007/s11280-007-0024-y}

\bibitem[{Bannon(1992)}]{bannon_human_1992}
Bannon L (1992) From human factors to human actors: the role of psychology and
  human-computer interaction studies in system design. In: Greenbaum J, Kyng M
  (eds) Design at work, L. Erlbaum Associates Inc., Hillsdale, {NJ}, {USA}, pp
  25--44, \urlprefix\url{dl.acm.org/citation.cfm?id=125470.125458}

\bibitem[{Bannon(1994)}]{bannon_cscw_1994}
Bannon L (1994) {CSCW}, challenging perspectives on work and technology. In:
  Proceedings of the {''Information} Technology {\textbackslash}\&
  Organisational Change'' Nijenrode Business School, The Netherlands, 28-29
  April, 1994

\bibitem[{Barad(2003)}]{barad_posthumanist_2003}
Barad K (2003) Posthumanist performativity: Toward an understanding of how
  matter comes to matter. Signs: Journal of Women in Culture and Society
  28(3):801--831, \doi{10.1086/345321},
  \urlprefix\url{http://www.jstor.org/stable/10.1086/345321}

\bibitem[{Bath(2009)}]{bath_searching_2009}
Bath C (2009) Searching for methodology: Feminist technology design in computer
  science. In: Proceedings of the 5th European Symposium on Gender
  {\textbackslash}\& {ICT} Digital Cultures: Participation - Empowerment -
  Diversity March 5 - 7, 2009 - University of Bremen, Bremen, D

\bibitem[{Beath and Orlikowski(1994)}]{beath_contradictory_1994}
Beath CM, Orlikowski WJ (1994) The contradictory structure of systems
  development methodologies: Deconstructing the {IS-User} relationship in
  information engineering. Information Systems Research 5(4):350--377,
  \doi{10.1287/isre.5.4.350},
  \urlprefix\url{http://isr.journal.informs.org/cgi/doi/10.1287/isre.5.4.350}

\bibitem[{Berg(1999)}]{berg_accumulating_1999}
Berg M (1999) Accumulating and coordinating: Occasions for information
  technologies in medical work. Computer Supported Cooperative Work, The
  Journal of Collaborative Computing 8(4):373--401

\bibitem[{Blackwell and Green(2008)}]{blackwell_abstract_2008}
Blackwell A, Green A (2008) The abstract is 'an enemy'. lternative perspectives
  to computational thinking. In: {PPIG} 08: Proceedings of the 20th workshop of
  the Psychology of Programming Interest Group

\bibitem[{Blackwell and Morrison(2010)}]{blackwell_logical_2010}
Blackwell AF, Morrison C (2010) A logical mind, not a programming mind:
  Psychology of a professional end-user. In: {PPIG} 2010: Proceedings of the
  22nd Annual Workshop of the Psychology of Programming Interest Group.
  September 19-22, 2010. Madrid, Spain., pp 175--184

\bibitem[{Bowers(1992)}]{bowers_politics_1992}
Bowers J (1992) The politics of formalism. In: Lea M (ed) Context of Computer
  Mediated Communications, Harvester, Hassocks, {UK}

\bibitem[{Bowers(1991)}]{bowers_janus_1991}
Bowers JM (1991) The janus faces of design: some critical questions for {CSCW}.
  In: Bowers JM, Benford SD (eds) Studies in computer supported cooperative
  work, North-Holland Publishing Co., Amsterdam, The Netherlands, pp 333--350,
  \urlprefix\url{dl.acm.org/citation.cfm?id=117730.117753}

\bibitem[{Bramming et~al(2012)Bramming, Hansen, Bojesen, and
  Olesen}]{bramming_imperfect_2012}
Bramming P, Hansen BG, Bojesen A, Olesen KG (2012) {(Im)perfect} pictures:
  snaplogs in performativity research. Qualitative Research in Organizations
  and Management: An International Journal 7(1):54--71,
  \doi{10.1108/17465641211223465},
  \urlprefix\url{http://www.emeraldinsight.com/10.1108/17465641211223465}

\bibitem[{Brand(1995)}]{brand_how_1995}
Brand S (1995) How buildings learn : what happens after they're built. Penguin
  Books, New York

\bibitem[{Bratteteig et~al(2010)Bratteteig, Morrison, and
  Wagner}]{bratteteig_research_2010}
Bratteteig T, Morrison A, Wagner I (2010) Research practices in digital design.
  In: Exploring digital design : multi-disciplinary design practices, Springer
  Berlin, D, pp 17--54

\bibitem[{Bringay et~al(2006)Bringay, Barry, and
  Charlet}]{bringay_annotations:_2006}
Bringay S, Barry C, Charlet J (2006) Annotations: A functionality to support
  cooperation, coordination and awareness in the electronic medical record. In:
  {COOP'06:} Proceedings of the 7th International Conference on the Design of
  Cooperative Systems, France, Provence

\bibitem[{Bryant(2000)}]{bryant_its_2000}
Bryant A (2000) It's engineering jim... but not as we know it: software
  engineering; solution to the software crisis, or part of the problem? In:
  {ICSE'00:} Proceedings of the 22nd international conference on Software
  engineering, {ACM}, New York, {NY}, {USA}, pp 78--87,
  \doi{10.1145/337180.337191},
  \urlprefix\url{doi.acm.org/10.1145/337180.337191}

\bibitem[{Bucciarelli(2003)}]{bucciarelli_designing_2003}
Bucciarelli L (2003) Designing and learning: a disjunction in contexts. Design
  Studies 24(3):295--311, \doi{10.1016/S0142-694X(02)00057-1},
  \urlprefix\url{http://linkinghub.elsevier.com/retrieve/pii/S0142694X02000571}

\bibitem[{Buescher et~al(2001)Buescher, Gill, Mogensen, and
  Shapiro}]{buescher_landscapes_2001}
Buescher M, Gill S, Mogensen P, Shapiro D (2001) Landscapes of practice:
  Bricolage as a method for situated design. {CSCW}, Computer Supported
  Cooperative Work 10(1):1--28, \doi{10.1023/A:1011293210539}

\bibitem[{Cabitza(2011)}]{cabitza_remain_2011}
Cabitza F (2011) {"Remain} faithful to the earth!": Reporting experiences of
  artifact-centered design in healthcare. Computer Supported Cooperative Work
  {(CSCW)} 20(4):231---263, \doi{10.1007/s10606-011-9143-1},
  \urlprefix\url{http://www.springerlink.com/index/10.1007/s10606-011-9143-1}

\bibitem[{Cabitza and Gesso(2011)}]{cabitza_web_2011}
Cabitza F, Gesso I (2011) Web of active documents: an architecture for flexible
  electronic patient records. In: Fred A, Filipe J, Gamboa H (eds) Biomedical
  Engineering Systems and Technologies. Third International Joint Conference,
  {BIOSTEC} 2010, Valencia, Spain, January 2010. Revised Selected Papers,
  Communications in Computer and Information Science, vol 127, Springer, pp
  44---56

\bibitem[{Cabitza and Gesso(2012)}]{cabitza_rule-based_2012}
Cabitza F, Gesso I (2012) Rule-based programming as easy as a child's play. a
  user study on active documents. In: {IHCI'12:} {IADIS} International
  Conference Interfaces and Human Computer Interaction 2012 Lisbon, Portugal 21
  - 23 July 2012

\bibitem[{Cabitza and Simone(2010)}]{cabitza_woad:_2010}
Cabitza F, Simone C (2010) {WOAD:} a framework to enable the end-user
  development of coordination oriented functionalities. Journal of
  Organizational and End User Computing {(JOEUC)} 22(2),
  \doi{10.4018/joeuc.2010101905}

\bibitem[{Cabitza and Simone(2012{\natexlab{a}})}]{cabitza_affording_2012}
Cabitza F, Simone C (2012{\natexlab{a}}) Affording mechanisms: an integrated
  view of coordination and knowledge management. Computer Supported Cooperative
  Work {(CSCW)} 21(2):227---260, \doi{10.1007/s10606-011-9153-z}

\bibitem[{Cabitza and Simone(2012{\natexlab{b}})}]{cabitza_whatever_2012}
Cabitza F, Simone C (2012{\natexlab{b}}) {''Whatever} works'': Making sense of
  information quality on information system artifacts. In: Viscusi G,
  Campagnolo GM, Curzi Y (eds) Phenomenology, Organizational Politics, and {IT}
  Design: The Social Study of Information Systems, {IGI} Global, pp 79--110,
  \urlprefix\url{10.4018/978-1-4666-0303-5.ch006}

\bibitem[{Cabitza and Simone(2013)}]{cabitza_computational_2013}
Cabitza F, Simone C (2013) Computational coordination mechanisms: A tale of a
  struggle for flexibility. Computer Supported Cooperative Work {(CSCW)}
  forthcoming

\bibitem[{Cabitza et~al(2009)Cabitza, Simone, and
  Sarini}]{cabitza_leveraging_2009}
Cabitza F, Simone C, Sarini M (2009) Leveraging coordinative conventions to
  promote collaboration awareness. Computer Supported Cooperative Work {(CSCW)}
  18(4):301---330

\bibitem[{Cabitza et~al(2011{\natexlab{a}})Cabitza, Corna, Gesso, and
  Simone}]{cabitza_woad_2011}
Cabitza F, Corna S, Gesso I, Simone C (2011{\natexlab{a}}) {WOAD}, a platform
  to deploy flexible {EPRs} in full control of end-users. In: Blandford A,
  De~Pietro G, Gallo L, Gimblett A, Oladimeji P, Thimbleby H (eds) {EICS4Med}
  2011: Proceedings of the 1st International Workshop on Engineering
  Interactive Computing Systems for Medicine and Health Care, co-located with
  the {ACM} {SIGCHI} Symposium on Engineering Interactive Computing Systems
  {(EICS} 2011) Pisa, Italy, June 13, 2011, {CEUR-WS.org}, vol 727, pp 7---12

\bibitem[{Cabitza et~al(2011{\natexlab{b}})Cabitza, Gesso, and
  Corna}]{cabitza_tailorable_2011}
Cabitza F, Gesso I, Corna S (2011{\natexlab{b}}) Tailorable flexibility: Making
  end-users autonomous in the design of active interfaces. In: Blashki K (ed)
  {MCCSIS} 2011: {IADIS} Multi Conference on Computer Science and Information
  Systems, Rome, Italy, July 20--26, 2011, {IADIS}

\bibitem[{Cabitza et~al(2012{\natexlab{a}})Cabitza, Colombo, and
  Simone}]{cabitza_leveraging_2012}
Cabitza F, Colombo G, Simone C (2012{\natexlab{a}}) Leveraging
  underspecification in knowledge artifacts to foster collaborative activities
  in professional communities. International Journal of Human - Computer
  Studies In press, \doi{10.1016/j.ijhcs.2012.02.005}

\bibitem[{Cabitza et~al(2012{\natexlab{b}})Cabitza, Gesso, and
  Simone}]{cabitza_providing_2012}
Cabitza F, Gesso I, Simone C (2012{\natexlab{b}}) Providing end-users with a
  visual editor to make their electronic documents active. In: {VL/HCC} 2012:
  Proceedings of Short papers of the {IEEE} Symposium on Visual Languages and
  Human-Centric Computing September 30 - October 4, 2012 - Innsbruck, Austria,
  {IEEE} Computer Press, Innsbruck, Austria

\bibitem[{Cabitza et~al(2012{\natexlab{c}})Cabitza, Simone, and
  Locatelli}]{cabitza_supporting_2012}
Cabitza F, Simone C, Locatelli MP (2012{\natexlab{c}}) Supporting
  artifact-mediated discourses through a recursive annotation tool. In:
  {GROUP'12:} Proceedings of the 17th {ACM} international conference on
  Supporting group work, {ACM}, New York, {NY}, {USA}, pp 253--262,
  \doi{10.1145/2389176.2389215}

\bibitem[{Cadiz et~al(2000)Cadiz, Gupta, and Grudin}]{cadiz_using_2000}
Cadiz JJ, Gupta A, Grudin J (2000) Using web annotations for asynchronous
  collaboration around documents. In: {CSCW} 2000: Proceedings of the 2000
  {ACM} conference on Computer supported cooperative work, {ACM}, New York,
  {NY}, {USA}, pp 309--318, \doi{10.1145/358916.359002}

\bibitem[{Campbell-kelly and Aspray(2004)}]{campbell-kelly_computer:_2004}
Campbell-kelly M, Aspray W (eds)  (2004) Computer: A History Of The Information
  Machine. Westview Press

\bibitem[{Carr(2004)}]{carr_does_2004}
Carr N (2004) Does {IT} Matter? Information Technology and the Corrosion of
  Competitive Advantage. Harvard Business School Press

\bibitem[{Carroll(2004)}]{carroll_completing_2004}
Carroll J (2004) Completing design in use: closing the appropriation cycle. In:
  European Conference on Information Systems

\bibitem[{Carroll et~al(1991)Carroll, Kellogg, and
  Rosson}]{carroll_task-artifact_1991}
Carroll JM, Kellogg WA, Rosson MB (1991) The task-artifact cycle. In: Carroll
  JM (ed) Designing Interaction: Psychology at the Human-Computer Interface,
  Cambridge University Press, New York, {NY}, {USA}, pp 74---102,
  \urlprefix\url{http://portal.acm.org/citation.cfm?id=120352.120358}

\bibitem[{Carstensen et~al(1995)Carstensen, Sorensen, and
  Borstrom}]{carstensen_two_1995}
Carstensen PH, Sorensen C, Borstrom H (1995) Two is fine, four is a mess:
  Reducing complexity of articulation work in manufacturing. In: {COOP'95},
  Proceedings of the International workshop on the design of cooperative
  systems. Sophia Antipolis, {FR}, pp 314--333

\bibitem[{Chen and Akay(2011)}]{chen_developing_2011}
Chen W, Akay M (2011) Developing {EMRs} in developing countries. Information
  Technology in Biomedicine, {IEEE} Transactions on 15(1):62---65,
  \doi{10.1109/TITB.2010.2091509}

\bibitem[{Christensen(2012)}]{dugdale_trouble_2012}
Christensen LR (2012) The trouble with {'Knowledge} transfer': On conduit
  metaphors and semantic pathologies in our understanding of didactic practice.
  In: Dugdale J, Masclet C, Grasso MA, Boujut JF, Hassanaly P (eds) From
  Research to Practice in the Design of Cooperative Systems: Results and Open
  Challenges, Springer London, pp 111--121,
  \urlprefix\url{dx.doi.org/10.1007/978-1-4471-4093-1_8}

\bibitem[{Ciborra(2002)}]{ciborra_labyrinths_2002}
Ciborra C (2002) The labyrinths of Information challenging the wisdom of
  systems. Oxford University Press, Oxford; New York, {N.Y.}, {USA}

\bibitem[{Ciborra(2006)}]{ciborra_mind_2006}
Ciborra C (2006) The mind or the heart? it depends on the (definition of)
  situation. Journal of Information Technology 21:129--139

\bibitem[{Ciborra(1992)}]{ciborra_thinking_1992}
Ciborra CU (1992) From thinking to tinkering. Information Society 8:297--309

\bibitem[{Clancey(1997)}]{clancey_situated_1997}
Clancey WJ (1997) Situated cognition : on human knowledge and computer
  representations. Cambridge University Press, Cambridge, {U.K.;} New York,
  {NY}, {USA}

\bibitem[{Crabtree et~al(2001)Crabtree, Rodden, and Bb}]{crabtree_wild_2001}
Crabtree A, Rodden T, Bb NN (2001) Wild sociology: Ethnography and design.
  Tech. rep., Department of Sociology for the Degree of Doctor of Philosophy,
  Lancaster University, Lancaster, {UK}

\bibitem[{{D'Adderio}(2008)}]{dadderio_performativity_2008}
{D'Adderio} L (2008) The performativity of routines: theorising the influence
  of artefacts and distributed agencies on routines dynamics. Research Policy
  37:769--89

\bibitem[{Danholt(2005)}]{danholt_prototypes_2005}
Danholt P (2005) Prototypes as performative. In: {CC'05:} Proceedings of the
  4th decennial conference on Critical computing, between sense and
  sensibility, {ACM} Press, p~1, \doi{10.1145/1094562.1094564},
  \urlprefix\url{http://portal.acm.org/citation.cfm?doid=1094562.1094564}

\bibitem[{De~Michelis(2003)}]{de_michelis_swiss_2003}
De~Michelis G (2003) The {''Swiss} pattada''. interactions 10(3):44--53,
  \doi{10.1145/769759.769760},
  \urlprefix\url{doi.acm.org/10.1145/769759.769760}

\bibitem[{De~Michelis et~al(2009)De~Michelis, Loregian, and
  Moderini}]{de_michelis_itsme:_2009}
De~Michelis G, Loregian M, Moderini C (2009) itsme: Interaction design
  innovating workstations. Knowledge, Technology \& Policy 22(1):71--78,
  \urlprefix\url{dx.doi.org/10.1007/s12130-009-9069-9},
  10.1007/s12130-009-9069-9

\bibitem[{Derrida(1981)}]{derrida_positions_1981}
Derrida J (1981) Positions. The University of Chicago Press

\bibitem[{Dirksmeier and Helbrecht(2008)}]{dirksmeier_time_2008}
Dirksmeier P, Helbrecht I (2008) Time, non-representational theory and the
  {`Performative} turn' - towards a new methodology in qualitative social
  research. Forum Qualitative Social Research 9(2):Art. 55,
  \urlprefix\url{nbn-resolving.de/urn:nbn:de:0114-fqs0802558}

\bibitem[{Dix(2007)}]{dix_designing_2007}
Dix A (2007) Designing for appropriation. In: Proceedings of the 21st British
  {HCI} Group Annual Conference on People and Computers: {HCI...} but not as we
  know it, British Computer Society, Swinton, {UK}, {UK}, {BCS-HCI} '07, pp
  27--30, \urlprefix\url{dl.acm.org/citation.cfm?id=1531407.1531415}

\bibitem[{Dourish(1999)}]{dourish_evolution_1999}
Dourish P (1999) Evolution in the adoption and use of collaborative
  technologies. In: Proceedings of the {ECSCW'99} Workshop on Evolving Use of
  Groupware; 1999 September 16; Copenhagen, Denmark.

\bibitem[{Dourish(2001)}]{dourish_where_2001}
Dourish P (2001) Where the Action Is: The Foundations of Embodied Interaction.
  {MIT} Press, Cambridge, {USA}

\bibitem[{Dourish(2004)}]{dourish_what_2004}
Dourish P (2004) What we talk about when we talk about context. Personal and
  Ubiquitous Computing 8(1):19--30

\bibitem[{Dourish et~al(2000)Dourish, Edwards, {LaMarca}, Lamping, Petersen,
  Salisbury, Terry, and Thornton}]{dourish_extending_2000}
Dourish P, Edwards WK, {LaMarca} A, Lamping J, Petersen K, Salisbury M, Terry
  DB, Thornton J (2000) Extending document management systems with
  user-specific active properties. {ACM} Transactions on Information Systems
  18(2):140--170, \doi{http://doi.acm.org/10.1145/348751.348758}

\bibitem[{Eder(1966)}]{eder_definitions_1966}
Eder W (1966) Definitions and methodologies. In: Gregory SA (ed) The Design
  Method, Butterworths, London, {UK}, pp 19--31

\bibitem[{Ellingsen et~al(2012)Ellingsen, Monteiro, and
  Røed}]{ellingsen_integration_2012}
Ellingsen G, Monteiro E, Røed K (2012) Integration as interdependent
  workaround. International Journal of Medical Informatics
  \doi{10.1016/j.ijmedinf.2012.09.004},
  \urlprefix\url{http://linkinghub.elsevier.com/retrieve/pii/S1386505612001785}

\bibitem[{Ensmenger(2001)}]{ensmenger_black_2001}
Ensmenger NL (2001) From ''black art'' to industrial discipline: the software
  crisis and the management of programmers. PhD thesis, University of
  Pennsylvania, {USA}, {AAI3015310}

\bibitem[{Fischer and Giaccardi(2006)}]{fischer_meta-design:_2006}
Fischer G, Giaccardi E (2006) Meta-design: A framework for the future of
  end-user development. In: Lieberman H (ed) End User Development -- Empowering
  people to flexibly employ advanced information and communication technology,
  Kluwer Academic Publishers, Dordrecht, The Netherlands, {NL}, pp 427--457,
  \urlprefix\url{http://dx.doi.org/10.1007/1-4020-5386-X_19}

\bibitem[{Fischer et~al(2004)Fischer, Giaccardi, Ye, Sutcliffe, and
  Mehandjiev}]{fischer_meta-design:_2004}
Fischer G, Giaccardi E, Ye Y, Sutcliffe AG, Mehandjiev N (2004) Meta-design: a
  manifesto for end-user development. Communications of the {ACM} 47(9):33--37,
  \doi{10.1145/1015864.1015884},
  \urlprefix\url{dx.doi.org/10.1145/1015864.1015884}

\bibitem[{Fitzpatrick(2003)}]{fitzpatrick_locales_2003}
Fitzpatrick G (2003) The locales framework : understanding and designing for
  wicked problems. Kluwer Academic Publishers, Dordrecht; Boston

\bibitem[{Fitzpatrick and Ellingsen(2012)}]{fitzpatrick_review_2012}
Fitzpatrick G, Ellingsen G (2012) A review of 25 years of {CSCW} research in
  healthcare: Contributions, challenges and future agendas. Computer Supported
  Cooperative Work {(CSCW)} \doi{10.1007/s10606-012-9168-0},
  \urlprefix\url{http://www.springerlink.com/index/10.1007/s10606-012-9168-0}

\bibitem[{Flores et~al(1988)Flores, Graves, Hartfield, and
  Winograd}]{flores_computer_1988}
Flores F, Graves M, Hartfield B, Winograd T (1988) Computer systems and the
  design of organizational interaction. {ACM} Trans Inf Syst 6(2):153--172,
  \doi{10.1145/45941.45943}, \urlprefix\url{doi.acm.org/10.1145/45941.45943}

\bibitem[{Glendinning(1998)}]{glendinning_being_1998}
Glendinning S (1998) On being with others Heidegger, Derrida, Wittgenstein.
  Routledge, London; New York,
  \urlprefix\url{http://site.ebrary.com/id/10017732}

\bibitem[{Greenhalgh et~al(2010)Greenhalgh, Plsek, Wilson, Fraser, and
  Holt}]{greenhalgh_response_2010}
Greenhalgh T, Plsek P, Wilson T, Fraser S, Holt T (2010) Response to {'The}
  appropriation of complexity theory in health care'. Journal of Health
  Services Research \& Policy 15(2):115--117, \doi{10.1258/jhsrp.2010.009158},
  \urlprefix\url{http://jhsrp.rsmjournals.com/cgi/doi/10.1258/jhsrp.2010.009158}

\bibitem[{Haigh(2002)}]{haigh_software_2002}
Haigh T (2002) Software in the 1960s as concept, service, and product. {IEEE}
  Ann Hist Comput 24(1):5--13, \doi{10.1109/85.988574},
  \urlprefix\url{http://dx.doi.org/10.1109/85.988574}

\bibitem[{Haigh(2010)}]{haigh_crisis_2010}
Haigh T (2010) Crisis what crisis? reconsidering the software crisis of the
  1960s and the origins of software engineering. Tech. rep., School of
  Information Studies, University of Wisconsin, Milwaukee, {USA}

\bibitem[{Haigh(2011)}]{haigh_inventing_2011}
Haigh T (2011) Inventing information systems: The systems men and the computer,
  1950-1968. Business History Review 75(01):15--61, \doi{10.2307/3116556},
  \urlprefix\url{http://www.journals.cambridge.org/abstract_S000768050007536X}

\bibitem[{Handel and Poltrock(2011)}]{handel_working_2011}
Handel MJ, Poltrock SE (2011) Working around official applications: experiences
  from a large engineering project. In: {CSCW'11}, Proceedings of the 2011
  {ACM} Conference on Computer Supported Cooperative Work, Hangzhou, China,
  March 19-23, 2011, pp 309--312

\bibitem[{Hanseth and Ciborra(2007)}]{hanseth_risk_2007}
Hanseth O, Ciborra C (2007) Risk, complexity and {ICT}. E. Elgar, Cheltenham,
  {UK;} Northampton, {MA}

\bibitem[{Harel(2008)}]{harel_can_2008}
Harel D (2008) Can programming be liberated, period? Computer 41(1):28--37,
  \doi{10.1109/MC.2008.10},
  \urlprefix\url{http://ieeexplore.ieee.org/lpdocs/epic03/wrapper.htm?arnumber=4445599}

\bibitem[{Harel and Marelly(2003)}]{harel_come_2003}
Harel D, Marelly R (2003) Come, let's play : scenario-based programming using
  {LSCs} and the play-engine. Springer, Berlin; New York

\bibitem[{Harris and Henderson(1999)}]{harris_better_1999}
Harris J, Henderson A (1999) A better mythology for system design. In:
  {CHI'99:} Proceedings of the {SIGCHI} conference on Human factors in
  computing systems: the {CHI} is the limit, {ACM} Press, New York, {NY},
  {USA}, pp 88--95, \doi{10.1145/302979.303003}

\bibitem[{Hartswood et~al(2000)Hartswood, Procter, Rouncefield, and
  Slack}]{hartswood_being_2000}
Hartswood M, Procter R, Rouncefield M, Slack R (2000) Being there and doing
  {IT} in the workplace: a case study of a co-development approach in
  healthcare. In: Cherkasky T (ed) Proceedings of the {CPSR/IFIP} {WG} 9.1
  Participatory Design Conference., pp 96--105.

\bibitem[{Hirschheim and Klein(1989)}]{hirschheim_four_1989}
Hirschheim R, Klein HK (1989) Four paradigms of information systems
  development. Commun {ACM} 32(10):1199--1216, \doi{10.1145/67933.67937},
  \urlprefix\url{doi.acm.org/10.1145/67933.67937}

\bibitem[{Hollnagel et~al(2008)Hollnagel, Nemeth, and
  Dekker}]{hollnagel_resilience_2008}
Hollnagel E, Nemeth CP, Dekker S (2008) Resilience engineering perspectives.
  Ashgate, Aldershot, Hampshire, England; Burlington, {VT}

\bibitem[{Hug(2010)}]{hug_performativity_2010}
Hug D (2010) Performativity in design and evaluation of sounding interactive
  commodities. In: {AM'10:} Proceedings of the 5th Audio Mostly Conference: A
  Conference on Interaction with Sound, {ACM} Press, pp 1--8,
  \doi{10.1145/1859799.1859806},
  \urlprefix\url{http://portal.acm.org/citation.cfm?doid=1859799.1859806}

\bibitem[{Illich(1973)}]{illich_tools_1973}
Illich I (1973) Tools for conviviality. Harper {\textbackslash}\& Row, New
  York, {NY}, {USA}

\bibitem[{Illich(1977)}]{illich_disabling_1977}
Illich I (1977) Disabling Professions. Marion Boyars, London, {UK}

\bibitem[{Jacucci et~al(2005)Jacucci, Jacucci, Wagner, and
  Psik}]{jacucci_manifesto_2005}
Jacucci C, Jacucci G, Wagner I, Psik T (2005) A manifesto for the performative
  development of ubiquitous media. In: {CC'05:} Proceedings of the 4th
  decennial conference on Critical computing: between sense and sensibility,
  {ACM} Press, pp 19---28, \doi{10.1145/1094562.1094566},
  \urlprefix\url{http://portal.acm.org/citation.cfm?doid=1094562.1094566}

\bibitem[{Jensen(2005)}]{jensen_experiment_2005}
Jensen CB (2005) An experiment in performative history: The electronic patient
  record as a future-generating device. Social Studies of Science 25(2):241--67

\bibitem[{Jensen(2008)}]{jensen_cscw_2008}
Jensen CB (2008) {CSCW} design reconceptualized through science studies. In:
  Cognition, Communication and Interaction: Transdisciplinary Perspectives on
  Interactive Technology., Springer Verlag, London, {UK}, pp 132--148

\bibitem[{Jensen(2002)}]{jensen_performing_2002}
Jensen TE (2002) Performing social work, competence, orderings, spaces and
  objects. {PhD}, University of Copenhagen, Department of Psychology,
  \urlprefix\url{http://www.dasts.dk/wp-content/uploads/Elgaard-2001.pdf}

\bibitem[{Johannessen et~al(2012)Johannessen, Gammon, and
  Ellingsen}]{johannessen_users_2012}
Johannessen LK, Gammon D, Ellingsen G (2012) Users as designers of information
  infrastructures and the role of generativity. {AIS} Transactions on
  Human-Computer Interaction 4(2):72--91

\bibitem[{Jones(1996)}]{jones_patterns_1996}
Jones C (1996) Patterns of software system failure and success. International
  Thomson Computer Press, London; Boston

\bibitem[{Jones(1970)}]{jones_design_1970}
Jones J (1970) Design Methods. Wiley-Interscience, London, {UK}

\bibitem[{Kaghan and Bowker(2001)}]{kaghan_out_2001}
Kaghan WN, Bowker GC (2001) Out of machine age? complexity, sociotechnical
  systems and actor network theory. Journal of Engineering and Technology
  Management 18(3-4):253--269, \doi{10.1016/S0923-4748(01)00037-6},
  \urlprefix\url{linkinghub.elsevier.com/retrieve/pii/S0923474801000376}

\bibitem[{Kaplan and Harris-Salamone(2009)}]{kaplan_health_2009}
Kaplan B, Harris-Salamone KD (2009) Health {IT} success and failure:
  Recommendations from literature and an {AMIA} workshop. Journal of the
  American Medical Informatics Association 16(3):291--299,
  \doi{10.1197/jamia.M2997}

\bibitem[{Khosravi and Gueheneuc(2004)}]{khosravi_quality_2004}
Khosravi K, Gueheneuc Y (2004) A quality model for design patterns. Tech. rep.,
  Universite de Montreal, Montreal, {CA}

\bibitem[{Kim and Kaplan(2006)}]{kim_interpreting_2006}
Kim RM, Kaplan SM (2006) Interpreting socio-technical co-evolution: Applying
  complex adaptive systems to {IS} engagement. Information Technology \& People
  19(1):35--54, \doi{10.1108/09593840610700800},
  \urlprefix\url{http://www.emeraldinsight.com/10.1108/09593840610700800}

\bibitem[{Kling et~al(2005)Kling, Rosenbaum, and
  Sawyer}]{kling_understanding_2005}
Kling R, Rosenbaum H, Sawyer S (2005) Understanding and communicating social
  informatics : a framework for studying and teaching the human contexts of
  information and communication technologies. Information Today, Inc., Medford,
  {N.J.}

\bibitem[{Krebs et~al(2012)Krebs, Conrad, and Wang}]{krebs_combining_2012}
Krebs D, Conrad A, Wang J (2012) Combining visual block programming and graph
  manipulation for clinical alert rule building. In: Proceedings of the 2012
  {ACM} annual conference extended abstracts on Human Factors in Computing
  Systems Extended Abstracts, {ACM}, New York, {NY}, {USA}, {CHI} {EA} '12, pp
  2453--2458, \doi{10.1145/2212776.2223818},
  \urlprefix\url{http://doi.acm.org/10.1145/2212776.2223818}

\bibitem[{Kriplean et~al(2012)Kriplean, Toomin, Morgan, Borming, and
  Ko}]{kriplean_is_2012}
Kriplean T, Toomin M, Morgan J, Borming A, Ko AJ (2012) Is this what you meant?
  promoting listening on the web with reflect. In: {CHI} 2012: Proceedings of
  the International conference on Human Computer Interaction, May 5-10, 2012,
  Austin, Texas, {USA}, pp 1559--1568

\bibitem[{Lanzara(1999)}]{lanzara_between_1999}
Lanzara G (1999) Between transient constructs and persistent structures:
  designing systems in action. Journal of Strategic Information Systems
  8(331-349)

\bibitem[{Latouche(2004)}]{latouche_megamachine_2004}
Latouche S (2004) La megamachine : raison technoscientifique, raison economique
  et mythe du progres : essais a la memoire de Jacques Ellul. La Decouverte :
  {M.A.U.S.S.}, Paris

\bibitem[{Latour(1996)}]{latour_interobjectivity_1996}
Latour B (1996) On interobjectivity. Mind, Culture, and Activity 3(4):228--245,
  \doi{10.1207/s15327884mca0304_2}

\bibitem[{Lave and Wenger(1991)}]{lave_situated_1991}
Lave J, Wenger E (1991) Situated Learning: Legitimate Peripheral Participation.
  Cambridge University Press., Cambridge, {UK}

\bibitem[{Law and Singleton(2003)}]{law_this_2003}
Law J, Singleton V (2003) {'This} is not an object'. Tech. rep., Centre for
  Science Studies, Lancaster University, Lancaster {LA1} {4YN}, {UK},
  \urlprefix\url{www.comp.lancs.ac.uk/sociology/papers/Law-Singleton-This-is-Not-an-Object.pdf}

\bibitem[{Levi-Strauss(1966)}]{levi-strauss_savage_1966}
Levi-Strauss C (1966) The savage mind {(La} pensee suavage). Weidenfeld and
  Nilson, London, {UK}

\bibitem[{Licoppe(2010)}]{licoppe_performative_2010}
Licoppe C (2010) the performative turn in science and technology studies.
  Journal of Cultural Economy 3(2):181--188,
  \doi{10.1080/17530350.2010.494122},
  \urlprefix\url{http://www.tandfonline.com/doi/abs/10.1080/17530350.2010.494122}

\bibitem[{Lieberman et~al(2006)Lieberman, Patern{\textbackslash}'{o}, Klann,
  and Wulf}]{lieberman_end-user_2006}
Lieberman H, Patern{\textbackslash}'{o} F, Klann M, Wulf V (2006) End-user
  development: An emerging paradigm. In: End User Development, vol~9, Kluwer
  Academic Publishers, pp 1--8

\bibitem[{Locatelli and Simone(2010)}]{locatelli_community_2010}
Locatelli MP, Simone C (2010) A community based metaphor supporting {EUD}
  within communities. In: Santucci G (ed) Proceedings of the International
  Conference on Advanced Visual Interfaces, {AVI} 2010, Roma, Italy, May 26-28,
  2010, {ACM} Press, {AVI} '10, p 406,
  \doi{http://doi.acm.org/10.1145/1842993.1843084}

\bibitem[{Louridas(1999)}]{louridas_design_1999}
Louridas P (1999) Design as bricolage: anthropology meets design thinking.
  Design Studies 20(6):517--535, \doi{10.1016/S0142-694X(98)00044-1},
  \urlprefix\url{linkinghub.elsevier.com/retrieve/pii/S0142694X98000441}

\bibitem[{Love(2007)}]{love_social_2007}
Love T (2007) Social, environmental and ethical factors in engineering design
  theory : a post-positivist approach. Praxis Education, Western Australia

\bibitem[{Luff et~al(1992)Luff, Heath, and
  Greatbatch}]{luff_tasks--interaction:_1992}
Luff P, Heath C, Greatbatch D (1992) Tasks-in-interaction: paper and screen
  based documentation in collaborative activity. In: {CSCW'92:} Proceedings of
  the 1992 {ACM} conference on Computer-supported cooperative work, {ACM}
  Press, New York, {NY}, {USA}, pp 163--170, \doi{10.1145/143457.143475}

\bibitem[{Lyotard(1986)}]{lyotard_postmodern_1986}
Lyotard JF (1986) The postmodern condition : a report on knowledge. Manchester
  University Press, Manchester, {UK}

\bibitem[{Lyytinen and Robey(1999)}]{lyytinen_learning_1999}
Lyytinen K, Robey D (1999) Learning failure in information systems development.
  Information Systems Journal 9(2):85--101,
  \doi{10.1046/j.1365-2575.1999.00051.x},
  \urlprefix\url{http://doi.wiley.com/10.1046/j.1365-2575.1999.00051.x}

\bibitem[{Maguire and {McKelvey}(1999)}]{maguire_complexity_1999}
Maguire S, {McKelvey} B (1999) Complexity and management: Moving from fad to
  firm foundations. Emergence 1:19--61

\bibitem[{Mamlin et~al(2006)Mamlin, Biondich, Wolfe, Fraser, Jazayeri, Allen,
  Miranda, and Tierney}]{mamlin_cooking_2006}
Mamlin B, Biondich P, Wolfe B, Fraser H, Jazayeri D, Allen C, Miranda J,
  Tierney W (2006) Cooking up an open source {EMR} for developing countries:
  {OpenMRS} - a recipe for successful collaboration. In: {AMIA} Annual
  Symposium Proceedings, vol 2006, p 529

\bibitem[{Mansfield(2010)}]{mansfield_nature_2010}
Mansfield J (2010) The nature of change or the law of unintended consequences.
  Imperial College Press, London, {UK}

\bibitem[{Mark(2002)}]{mark_conventions_2002}
Mark G (2002) Conventions and commitments in distributed {CSCW} groups.
  Computer Supported Cooperative Work {(CSCW)} 11(3):349--387,
  \urlprefix\url{http://dx.doi.org/10.1023/A:1021289427473},
  {10.1023/A:1021289427473}

\bibitem[{Mark and Semaan(2008)}]{mark_resilience_2008}
Mark G, Semaan B (2008) Resilience in collaboration. In: {CSCW} '08:
  Proceedings of the 2008 {ACM} conference on Computer supported cooperative
  work, {ACM} Press, pp 137--146, \doi{10.1145/1460563.1460585}

\bibitem[{Mark et~al(2002)Mark, Gonzalez, Sarini, and
  Simone}]{mark_reconciling_2002}
Mark G, Gonzalez VM, Sarini M, Simone C (2002) Reconciling different
  perspectives: An experiment on technology support for articulation. In:
  {COOP'02:} Proceedings of the International Conference on the Design of
  Cooperative systems. Saint Raphael {(FR)}, 4-7 June., pp 23--37

\bibitem[{Martin and Sommerville(2004)}]{martin_patterns_2004}
Martin D, Sommerville I (2004) Patterns of cooperative interaction: Linking
  ethnomethodology and design. {ACM} Trans Comput-Hum Interact 11(1):59--89,
  \doi{10.1145/972648.972651},
  \urlprefix\url{http://doi.acm.org/10.1145/972648.972651}

\bibitem[{{McLeod} and Doolin(2010)}]{mcleod_documents_2010}
{McLeod} L, Doolin B (2010) Documents as mediating artifacts in contemporary
  {IS} development. In: System Sciences {(HICSS)}, 2010 43rd Hawaii
  International Conference on, pp 1--10, \doi{10.1109/HICSS.2010.155}

\bibitem[{Merton(1936)}]{merton_unanticipated_1936}
Merton RK (1936) The unanticipated consequences of purposive social action.
  American Sociological Review 1(6):894--904

\bibitem[{Miller(1996)}]{miller_bricoleur_1996}
Miller D (1996) The bricoleur in the tennis court: pedagogy in postmodern
  context. In: Proceedings of the Conference on values in higher education.,
  \urlprefix\url{web.utk.edu/~unistudy/ethics96/dlm1.html}

\bibitem[{Morrison et~al(2010)Morrison, Westvang, and
  Skogsrud}]{wagner_whisperings_2010}
Morrison A, Westvang E, Skogsrud SS (2010) Whisperings in the undergrowth:
  Communication design, online social networking and discursive performativity.
  In: Wagner I, Bratteteig T, Stuedahl D (eds) Exploring digital design :
  multi-disciplinary design practices, Springer London, London, pp 221--259,
  \urlprefix\url{http://www.springerlink.com/index/10.1007/978-1-84996-223-0_8}

\bibitem[{Morrison and Blackwell(2009)}]{morrison_observing_2009}
Morrison C, Blackwell A (2009) Observing end-user customisation of electronic
  patient records. In: Pipek V, Rosson M, de~Ruyter B, Wulf V (eds) End-User
  Development, Springer Berlin / Heidelberg, Lecture Notes in Computer Science,
  vol 5435, pp 275--284, \doi{10.1007/978-3-642-00427-8_16},
  \urlprefix\url{dx.doi.org/10.1007/978-3-642-00427-8_16},
  10.1007/978-3-642-00427-8\_16

\bibitem[{Morrison et~al(2011)Morrison, Fitzpatrick, and
  Blackwell}]{morrison_multi-disciplinary_2011}
Morrison C, Fitzpatrick G, Blackwell AF (2011) Multi-disciplinary collaboration
  during ward rounds: Embodied aspects of electronic medical record usage. I J
  Medical Informatics 80(8):96--111

\bibitem[{Nandhakumar and Avison(1999)}]{nandhakumar_fiction_1999}
Nandhakumar J, Avison DE (1999) The fiction of methodological development: a
  field study of information systems development. Information Technology \&
  People 12(2):176--191, \doi{10.1108/09593849910267224},
  \urlprefix\url{http://www.emeraldinsight.com/10.1108/09593849910267224}

\bibitem[{Nemeth(2003)}]{nemeth_master_2003}
Nemeth C (2003) The master schedule how cognitive artifacts affect distributed
  cognition in acute care. Tech. rep., Cognitive Technologies Laboratory

\bibitem[{Niazkhani et~al(2011)Niazkhani, Pirnejad, Sijs, and
  Aarts}]{niazkhani_evaluating_2011}
Niazkhani Z, Pirnejad H, Sijs Hvd, Aarts J (2011) Evaluating the medication
  process in the context of {CPOE} use: The significance of working around the
  system. I J Medical Informatics 80(7):490--506

\bibitem[{Noble(1979)}]{noble_america_1979}
Noble DF (1979) America by design : science, technology, and the rise of
  corporate capitalism. Oxford University Press, Oxford

\bibitem[{Norman and Kuras(2004)}]{norman_engineering_2004}
Norman D, Kuras M (2004) Engineering complex systems. Tech. rep., The {MITRE}
  corporation

\bibitem[{{O'Neill}(1992)}]{oneill_evolution_1992}
{O'Neill} JE (1992) The evolution of interactive computing through time-sharing
  and networking. {Ph.D.} dissertation, University of Minnesota, {USA}

\bibitem[{Orlikowski(1992{\natexlab{a}})}]{orlikowski_duality_1992}
Orlikowski WJ (1992{\natexlab{a}}) The duality of technology: Rethinking the
  concept of technology in organizations. Organization Science 3(3):398--427,
  \doi{10.1287/orsc.3.3.398},
  \urlprefix\url{orgsci.journal.informs.org/cgi/doi/10.1287/orsc.3.3.398}

\bibitem[{Orlikowski(1992{\natexlab{b}})}]{orlikowski_learning_1992}
Orlikowski WJ (1992{\natexlab{b}}) Learning from notes: Organizational issues
  in groupware implementation. In: {CSCW'92:} Proceedings of the Conference on
  Computer Supported Cooperative Work, October 31 - November 4, 1992, Toronto,
  Canada., {ACM}, pp 362---369

\bibitem[{Orlikowski(1996)}]{orlikowski_improvising_1996}
Orlikowski WJ (1996) Improvising organizational transformation over time: A
  situated change perspective. Information Systems Research 7(1):63--92,
  \doi{10.1287/isre.7.1.63},
  \urlprefix\url{http://isr.journal.informs.org/cgi/doi/10.1287/isre.7.1.63}

\bibitem[{Orlikowski(2000)}]{orlikowski_using_2000}
Orlikowski WJ (2000) Using technology and constituting structures: A practice
  lens for studying technology in organizations. Organization Science
  11(4):404--428

\bibitem[{Orlikowski(2006)}]{orlikowski_material_2006}
Orlikowski WJ (2006) Material knowing: The scaffolding of human
  knowledgeability. European Journal of Information Systems 15(5):460---466

\bibitem[{Orlikowski(2007)}]{orlikowski_sociomaterial_2007}
Orlikowski WJ (2007) Sociomaterial practices: Exploring technology at work.
  Organization Studies 28(9):1435--1448, \doi{10.1177/0170840607081138},
  \urlprefix\url{http://oss.sagepub.com/cgi/doi/10.1177/0170840607081138}

\bibitem[{Paley(2007)}]{paley_complex_2007}
Paley J (2007) Complex adaptive systems and nursing. Nursing Inquiry
  14(3):233--242, \doi{10.1111/j.1440-1800.2007.00359.x},
  \urlprefix\url{http://doi.wiley.com/10.1111/j.1440-1800.2007.00359.x}

\bibitem[{Paley and Eva(2011)}]{paley_complexity_2011}
Paley J, Eva G (2011) Complexity theory as an approach to explanation in
  healthcare: A critical discussion. International Journal of Nursing Studies
  48(2):269 -- 279, \doi{10.1016/j.ijnurstu.2010.09.012},
  \urlprefix\url{http://www.sciencedirect.com/science/article/pii/S0020748910003135}

\bibitem[{Pan et~al(2008)Pan, Hackney, and Pan}]{pan_information_2008}
Pan G, Hackney R, Pan SL (2008) Information systems implementation failure:
  Insights from prism. International Journal of Information Management
  28(4):259--269, \doi{10.1016/j.ijinfomgt.2007.07.001},
  \urlprefix\url{http://linkinghub.elsevier.com/retrieve/pii/S0268401207000886}

\bibitem[{Pavard and Dugdale(2006)}]{pavard_contribution_2006}
Pavard B, Dugdale J (2006) The contribution of complexity theory to the study
  of socio-technical cooperative systems. In: Minai AA, Bar-Yam Y (eds)
  Unifying Themes in Complex Systems, Springer Berlin Heidelberg, pp 39--48,
  \urlprefix\url{http://dx.doi.org/10.1007/978-3-540-35866-4_4},
  10.1007/978-3-540-35866-4\_4

\bibitem[{Pesic et~al(2007)Pesic, Schonenberg, Sidorova, and Van
  Der~Aalst}]{pesic_constraint-based_2007}
Pesic M, Schonenberg MH, Sidorova N, Van Der~Aalst WMP (2007) Constraint-based
  workflow models: change made easy. In: {OTM'07:} Proceedings of the 2007
  {OTM} Confederated international conference on On the move to meaningful
  internet systems: {CoopIS}, {DOA}, {ODBASE}, {GADA}, and {IS} - Volume Part
  I, Springer-Verlag, Berlin, Heidelberg, pp 77--94,
  \urlprefix\url{http://dl.acm.org/citation.cfm?id=1784607.1784618}

\bibitem[{Peters(2006)}]{peters_against_2006}
Peters I (2006) Against folksonomies - indexing blogs and podcasts for
  corporate knowledge management. In: Proceedings of Online Information,
  London, {UK}, Learned Information Europe, pp 93--97

\bibitem[{Pickering(1995)}]{pickering_mangle_1995}
Pickering A (1995) The Mangle of Practice: Time, Agency and Science. University
  Of Chicago Press

\bibitem[{Pickering(2004)}]{pickering_science_2004}
Pickering A (2004) The science of the unknowable: Stafford beer's cybernetic
  informatics. Kybernetes 33(3/4):499--521

\bibitem[{Pickering(2008)}]{pickering_beyond_2008}
Pickering A (2008) Beyond design: cybernetics, biological computers and
  hylozoism. Synthese 168(3):469--491, \doi{10.1007/s11229-008-9446-z},
  \urlprefix\url{http://www.springerlink.com/index/10.1007/s11229-008-9446-z}

\bibitem[{Pickering(2010)}]{pickering_cybernetic_2010}
Pickering A (2010) The Cybernetic Brain: sketches of another future. The
  University of Chicago Press, Chicago, Illinois, {USA}

\bibitem[{Pohn(2007)}]{pohn_cosmicplay.net_2007}
Pohn K (2007) Cosmicplay.net. {PhD}, Depth Psycology Dept., Pacifica Graduate
  Institute, Carpinteria, California, {USA},
  \urlprefix\url{www.cosmicplay.net/}

\bibitem[{Poltrock and Handel(2009)}]{poltrock_modeling_2009}
Poltrock S, Handel M (2009) Modeling collaborative behavior: Foundations for
  collaboration technologies. In: Proceedings of the 42nd Hawaii International
  Conference on System Sciences

\bibitem[{Riemer and Johnston(2012)}]{riemer_what_2012}
Riemer K, Johnston RB (2012) What is {IT} in use and why does it matter for
  {IS} design? In: Proceedings of the {IT} Artefact Design {\textbackslash}\&
  Workpractice Intervention, A Pre-{ECIS} and {AIS} {SIG} Prag Workshop, June
  10, 2012, Barcelona, E, Forskningsnaetverket {VITS},
  \urlprefix\url{www.vits.org/uploads/IT_Artifact/What_is_IT_and_why_does_it_matter.pdf}

\bibitem[{Robey and Markus(1984)}]{robey_rituals_1984}
Robey D, Markus ML (1984) Rituals in information system design. {MIS} Quarterly
  8(1):5--15, \urlprefix\url{www.jstor.org/stable/249240}

\bibitem[{Robinson(1993)}]{robinson_design_1993}
Robinson M (1993) Design for unanticipated use... In: {ECSCW'93:} Proceedings
  of the third conference on European Conference on Computer-Supported
  Cooperative Work, Kluwer Academic Publishers, Norwell, {MA}, {USA}, pp
  187--202

\bibitem[{Robinson and Bannon(1991)}]{robinson_questioning_1991}
Robinson M, Bannon L (1991) Questioning representations. In: {ECSCW'91:}
  Proceedings of the Second European Conference on Computer-Supported
  Cooperative Work, Amsterdam, The Netherlands

\bibitem[{Rochlin(1998)}]{rochlin_trapped_1998}
Rochlin G (1998) Trapped in the net : the unanticipated consequences of
  computerization. Princeton University Press, Princeton, {NJ}, {USA}

\bibitem[{Rolland and Monteiro(2002)}]{rolland_balancing_2002}
Rolland KH, Monteiro E (2002) Balancing the local and the global in
  infrastructural information systems. The Information Society 18(2):87--100,
  \doi{10.1080/01972240290075020},
  \urlprefix\url{http://www.tandfonline.com/doi/abs/10.1080/01972240290075020}

\bibitem[{Rorty(1991)}]{rorty_objectivity_1991}
Rorty R (1991) Objectivity, relativism, and truth. Cambridge University Press,
  Cambridge; New York

\bibitem[{Schmidt(1999)}]{schmidt_maps_1999}
Schmidt K (1999) Of maps and scripts: the status of formal constructs in
  cooperative work. Information and Software Technology 41(6):319---329

\bibitem[{Schmidt(2011)}]{schmidt_dispelling_2011}
Schmidt K (2011) Dispelling the mythology of computational artifacts. In:
  Cooperative Work and Coordinative Practices Contributions to the Conceptual
  Foundations of Computer-Supported Cooperative Work {(CSCW)}, Springer,
  Berlin, D, pp 391--413

\bibitem[{Schmidt and Wagner(2004)}]{schmidt_ordering_2004}
Schmidt K, Wagner I (2004) Ordering systems: Coordinative practices and
  artifacts in architectural design and planning. Computer Supported
  Cooperative Work {(CSCW)} 13(5-6):349--408,
  \doi{http://dx.doi.org/10.1007/s10606-004-5059-3}

\bibitem[{Schneberger and {McLean}(2003)}]{schneberger_complexity_2003}
Schneberger SL, {McLean} ER (2003) The complexity cross: Implications for
  practice. Communications of the {ACM} 46(9):216--225

\bibitem[{Shapiro(2005)}]{shapiro_participatory_2005}
Shapiro D (2005) Participatory design: the will to succeed. In: {CC'05:}
  Proceedings of the 4th decennial conference on Critical computing: between
  sense and sensibility, {ACM} Press, pp 29--38, \doi{10.1145/1094562.1094567},
  \urlprefix\url{portal.acm.org/citation.cfm?doid=1094562.1094567}

\bibitem[{Shipman and Marshall(1999)}]{shipman_formality_1999}
Shipman FM, Marshall CC (1999) Formality considered harmful: Experiences,
  emerging themes, and directions on the use of formal representations in
  interactive systems. Computer Supported Cooperative Work 8(4):333---352

\bibitem[{Simon(1981)}]{simon_sciences_1981}
Simon H (1981) The Sciences of the Artificial. {MIT} Press, Cambridge, {USA}

\bibitem[{Simon(2010)}]{simon_knowing_2010}
Simon J (2010) Knowing together: A social epistemology for socio-technical
  epistemic systems. {PhD}, Universitat Wien

\bibitem[{Simone and Sarini(2001)}]{simone_adaptability_2001}
Simone C, Sarini M (2001) Adaptability of classification schemes in
  cooperation: What does it mean? In: {ECSCW'01:} Proceedings of European
  Conference on Computer-Supported Cooperative Work, Kluwer Academic
  Publishers, pp 19--38

\bibitem[{Simone and Schmidt(1993)}]{simone_computational_1993}
Simone C, Schmidt K (1993) Computational mechanisms of interaction for {CSCW},
  {COMIC} deliverable d3.1, esprit basic research project. Tech. rep.,
  Computing Department, Lancaster University, Lancaster, {U.K.}

\bibitem[{de~Souza and Leitao(2009)}]{de_souza_semiotic_2009}
de~Souza CS, Leitao CF (2009) Semiotic engineering methods for scientific
  research in {HCI}. Synthesis Lectures on Human-Centered Informatics
  2(1):1--122, \doi{10.2200/S00173ED1V01Y200901HCI002}

\bibitem[{Star and Bowker(1999)}]{star_sorting_1999}
Star SL, Bowker G (1999) Sorting Things Out: Classification and its
  Consequences. {MIT} Press, London, {UK}

\bibitem[{Stevens et~al(2006)Stevens, Quaisser, and
  Klann}]{lieberman_breaking_2006}
Stevens G, Quaisser G, Klann M (2006) Breaking it up: An industrial case study
  of component-based tailorable software design. In: Lieberman H,
  Patern{\textbackslash}'{o} F, Wulf V (eds) End User Development,
  Human-Computer Interaction Series, vol~9, Springer Netherlands, pp 269--294,
  \urlprefix\url{dx.doi.org/10.1007/1-4020-5386-X_13}

\bibitem[{Suchman(1994)}]{suchman_categories_1994}
Suchman L (1994) Do categories have politics? Computer Supported Cooperative
  Work {(CSCW)} 2(3):177--190

\bibitem[{Suchman(2006)}]{suchman_human-machine_2006}
Suchman L (2006) Human-Machine Reconfigurations: Plans and Situated Actions.
  Cambridge University Press

\bibitem[{Suchman et~al(2002)Suchman, Trigg, and
  Blomberg}]{suchman_working_2002}
Suchman L, Trigg R, Blomberg J (2002) Working artefacts: ethnomethods of the
  prototype. British Journal of Sociology 53(2):163--179,
  \doi{10.1080/00071310220133287},
  \urlprefix\url{http://doi.wiley.com/10.1080/00071310220133287}

\bibitem[{Suchman(2004)}]{suchman_figuring_2004}
Suchman LA (2004) Figuring personhood in sciences of the artificial. Tech.
  rep., Department of Sociology, Lancaster University, Lancaster {LA1} {4YL},
  {UK},
  \urlprefix\url{www.comp.lancs.ac.uk/sociology/papers/suchman-figuring-personhood.pdf}

\bibitem[{Sumner and Stolze(1997)}]{sumner_evolution_1997}
Sumner T, Stolze M (1997) Evolution, not revolution: Participatory design in
  the toolbelt era. In: Kyng M, Mathiassen L (eds) Computers and design in
  context, {MIT} Press, Cambridge, {MA}, {USA}, pp 1--26,
  \urlprefix\url{http://dl.acm.org/citation.cfm?id=270318.270319}

\bibitem[{Telier(2011)}]{telier_design_2011}
Telier A (2011) Design Things. Design Thinking, Design Theory, The {MIT} Press

\bibitem[{Tenner(1997)}]{tenner_why_1997}
Tenner E (1997) Why Things Bite Back: Technology and the Revenge of Unintended
  Consequences. Vintage

\bibitem[{Truex et~al(1999)Truex, Baskerville, and Klein}]{truex_growing_1999}
Truex DP, Baskerville R, Klein HK (1999) Growing systems in emergent
  organizations. Communications of {ACM} 42(8):117--123

\bibitem[{Turner(2005)}]{turner_affordance_2005}
Turner P (2005) Affordance as context. Interacting with Computers
  17(6):787--800, \doi{10.1016/j.intcom.2005.04.003}

\bibitem[{Van~House(2009)}]{van_house_collocated_2009}
Van~House NA (2009) Collocated photo sharing, story-telling, and the
  performance of self. International Journal of Human-Computer Studies
  67(12):1073--1086, \doi{10.1016/j.ijhcs.2009.09.003},
  \urlprefix\url{http://linkinghub.elsevier.com/retrieve/pii/S1071581909001256}

\bibitem[{Vera and Simon(1993)}]{vera_situated_1993}
Vera AH, Simon HA (1993) Situated action: A symbolic interpretation. Cognitive
  Science 17:7--48

\bibitem[{Wagner et~al(2010)Wagner, Stuedahl, and
  Bratteteig}]{wagner_exploring_2010}
Wagner I, Stuedahl D, Bratteteig T (2010) Exploring Digital Design:
  Multi-Disciplinary Design Practices. Springer-Verlag London Limited, London,
  \urlprefix\url{dx.doi.org/10.1007/978-1-84996-223-0}

\bibitem[{Warkentin et~al(2009)Warkentin, Moore, Bekkering, and
  Johnston}]{warkentin_analysis_2009}
Warkentin M, Moore RS, Bekkering E, Johnston AC (2009) Analysis of systems
  development project risks: an integrative framework. {ACM} {SIGMIS} Database
  40(2):8, \doi{10.1145/1531817.1531821},
  \urlprefix\url{http://portal.acm.org/citation.cfm?doid=1531817.1531821}

\bibitem[{Weick(1993)}]{weick_organizational_1993}
Weick KE (1993) Organizational redesign as improvisation. In: Huber G, Glick W
  (eds) Organizational change and redesign: Ideas and insights for improving
  performance, Oxford University Press, New York, {NY}, {USA}, pp 346--379

\bibitem[{Weinstein and Weinstein(1991)}]{weinstein_george_1991}
Weinstein D, Weinstein M (1991) George simmel: Sociological flaneur bricoleur.
  Theory, Culture Society 8:151--168

\bibitem[{Winograd and Flores(1986)}]{winograd_understanding_1986}
Winograd T, Flores F (1986) Understanding Computers and Cognition: a new
  foundation for design. Addison Wesley, Reading {MA}

\bibitem[{Winthereik and Vikkelso(2005)}]{winthereik_ict_2005}
Winthereik BR, Vikkelso S (2005) {ICT} and integrated care: Some dilemmas of
  standardising inter-organisational communication. Computer Supported
  Cooperative Work {(CSCW)} 14(1):43--67,
  \doi{http://dx.doi.org/10.1007/s10606-004-6442-9}

\bibitem[{Yates(1993)}]{yates_control_1993}
Yates J (1993) Control through Communication. The Rise of System in American
  Management. The Johns Hopkins University Press, Baltimore and London.

\bibitem[{Zuboff(1988)}]{zuboff_age_1988}
Zuboff S (1988) In the age of the smart machine : the future of work and power.
  Basic Books, New York

\end{thebibliography}

\end{document}